\documentclass[prd,preprintnumbers,amssymb,amsmath,12pt,nofootinbib,longbibliography]{revtex4}

\usepackage{graphicx,multirow,subfigure}
\usepackage{bm}
\usepackage{amsmath}
\usepackage{amssymb}
\usepackage{amscd}
\usepackage{latexsym}
\usepackage{slashed}
\usepackage{color}
\usepackage{graphicx}
\usepackage{ulem}
\usepackage{color}

\begin{document}

\title{
{Forward-Backward Asymmetry as a Discovery Tool \\
for $Z'$ Bosons at the LHC}
}

\date{\today}

\author{Elena Accomando}%
 \email{E.Accomando@soton.ac.uk}
\affiliation{School of Physics and Astronomy, University of Southampton, Highfield, Southampton SO17 1BJ, UK}%
\author{Alexander Belyaev}%
 \email{A.Belyaev@soton.ac.uk}
\affiliation{School of Physics and Astronomy, University of Southampton, Highfield, Southampton SO17 1BJ, UK}%
\author{Juri Fiaschi}%
 \email{Juri.Fiaschi@soton.ac.uk}
 \affiliation{School of Physics and Astronomy, University of Southampton, Highfield, Southampton SO17 1BJ, UK}%
\author{Ken Mimasu}%
 \email{K.Mimasu@sussex.ac.uk}
\affiliation{School of Physics and Astronomy, University of Sussex, Falmer, Brighton, BN1 9RH, 
UK}%
\author{Stefano Moretti}%
 \email{S.Moretti@soton.ac.uk}
\affiliation{School of Physics and Astronomy, University of Southampton, Highfield, Southampton SO17 1BJ, UK}%
\author{Claire Shepherd-Themistocleous}%
 \email{claire.shepherd@stfc.ac.uk}
\affiliation{Particle Physics Department,
STFC, Rutherford Appleton Laboratory, Harwell Science and Innovation Campus,
Didcot, Oxfordshire, OX11 0QX, UK}%

\begin{abstract}

\noindent
The  Forward-Backward Asymmetry (AFB) in $Z^\prime$ physics is commonly only perceived as 
the observable which possibly allows one to interpret a $Z^\prime$ signal 
by distinguishing different models of such (heavy) spin-1 bosons.
In this paper, we examine the potential of AFB in setting bounds on or even 
discovering a $Z'$ at the Large Hadron Collider (LHC) and show that it might be a powerful tool for this purpose.
We analyse two different scenarios: 
$Z^\prime$-bosons with a narrow and wide width, respectively. We find that, in the first case, the significance of the AFB search can be comparable with that of the bump search usually adopted by the experimental collaborations; however, being  a ratio of (differential) cross sections the AFB 
has the advantage of reducing systematical errors. 
In the second case, the AFB search can win over the bump search in terms of event shape, as the structure of the AFB distribution as a function of the invariant mass of the reconstructed $Z^\prime$-boson could nail down the new broad resonance much better than the event counting strategy usually adopted in such cases.

\end{abstract}

\maketitle

\section{Introduction}

\noindent
Heavy neutral $Z^\prime$-bosons arise in a number of theories that extend the Standard Model (SM) gauge group by adding an extra $U(1)$ {symmetry}. 
The most common
$Z^\prime$-boson benchmark models can be divided in three main classes: $E_6$ models, Generalized Left-Right (GLR) symmetric models and Generalized Standard Models (GSM), {see, e.g., the review in  \cite{Accomando:2010fz} and references therein}. All these models predict 
a {relatively narrow width for the $Z^\prime$-bosons, so that $\Gamma_{Z^\prime}/M_{Z^\prime}$ varies in the $0.5-12\%$ range}.
The {lowest $\Gamma_{Z^\prime}/M_{Z^\prime}$ value
is realised  in the  $E_\psi$ model from the $E_6$ class while the biggest value
appears in the $Q$-model belonging to the GSM class.}
\par\noindent
Experimental searches for a heavy $Z^\prime$-boson at the LHC are usually interpreted in the context of the Sequential Standard Model (SSM), which is part of the GSM class~\cite{Erler:1999ub}.
This benchmark scenario just includes one extra neutral  vector boson with couplings to fermions identical to those of the corresponding SM $Z$-boson and no mixing with the neutral Electro-Weak (EW) SM boson. Being nothing but a heavier copy of the SM $Z$-boson, this $Z^\prime$-boson is characterized by a narrow width: $\Gamma_{Z^\prime}/M_{Z^\prime}\simeq 2.8\%$, including the $Z^\prime$-decay into top-antitop pairs above threshold.
Dedicated search strategies therefore assumes that the new heavy resonance is narrow and can be described by a Breit-Wigner line-shape, standing over the SM background, when looking at the distribution in the invariant mass of the $Z^\prime$-boson decay products. In this way, the new physics signal is thought to have a well defined peaking structure, concentrated in a small interval centred around its mass. 
On the basis of this assumption, the 95\% Confidence Level (C.L.) upper bound on the cross section is derived and limits on the mass of the $Z^\prime$-boson are extracted within the above mentioned benchmark models. In the case of a 
narrow width $Z^\prime$ scenario, even interference effects can be accounted for without substantially altering
the described experimental approach \cite{Accomando:2013sfa}.

However, the narrow width hypothesis is quite strong, even if well motivated. There exist in fact counter examples of theories where the predicted $Z^\prime$-boson is characterized by a large width. For example, this can be  
realized in Technicolor \cite{Belyaev:2008yj} scenarios,  Composite Higgs Models \cite{Barducci:2012kk}
or in more generic setups where the $Z^\prime$-boson couples differently to the first two fermion generations 
with respect to the third one \cite{Kim:2014afa, Malkawi:1999sa} or else interacts with the SM gauge bosons in presence of mixing \cite{Altarelli:1989ff},
so that  large $\Gamma_{Z^\prime}/M_{Z^\prime}$ values are induced by the additional $Z^\prime$ decay channels 
which then onset in all such cases. As a consequence, the resulting resonance is wide and, instead of having a well defined Breit-Wigner 
line shape, it will appear as a broad shoulder spreading over the SM background. 
In such models, the ratio $\Gamma_{Z^\prime}/M_{Z^\prime}$ can easily reach the {50\% value or higher} making a Breit-Wigner line shape {based analysis definitely inappropriate.}
The experimental collaborations have not tackled this kind of {scenarios
yet and  respective dedicated search strategies for wide $Z^\prime$ particles are  absent at the moment.}
 The only frameworks which have been analyzed in order to interpret possible non-resonant deviations from the SM are the graviton production within the ADD model \cite{ArkaniHamed:1998rs, ArkaniHamed:1998nn} and the contact interaction within the left-left isoscalar model \cite{Eichten:1983hw, Eichten:1984eu}. For details on the parametrization of the differential cross section within these two scenarios we refer to Ref. \cite{Khachatryan:2014fba} and references therein. Both processes might give rise to an excess of events spread over the SM background. In this `effectively'
non-resonant case, the experimental analyses are essentially counting experiments: an excess of events searched out of an estimated SM background. To make the analyses more robust, the same background is often estimated with multiple data-driven methods. Kinematical cuts are then optimized in order to maximize the discovery/exclusion potential at the LHC.
Despite this, as one can understand, this analysis can be fragile. The experimental results heavily rely on the good understanding and control of the SM background, as the new physics signal is not expected to have a definite shape. 
The choice of the control region, needed to define the functional form of the SM background to be used in the regions where there 
might be some signal, is not trivial. Interference effects between new physics signal and SM background can indeed affect the low scale region of the distribution in the invariant mass of the $Z^\prime$-boson decay products, proving the assumption that the control region is new physics free to be simply false. Under these premises the experimental analyses could get quite complicated. On the one hand, the possible presence of wide objects could in fact shrink the new physics free region, owing to an increase of the interference effects driven by the large width. {On the other hand, the wide $Z'$-bosons could easily escape detection in the bump searches due to the same interference effects combined with the absence of a resonant peaking structure in the di-lepton invariant mass distribution. All this then conspires to make the discovery of a wide $Z^\prime$ very problematic.

In this paper, we study {the potential of a} complementary observable in order 
to {probe} new massive $Z'$ objects either narrow or broad. The variable we analyse is the Forward-Backward Asymmetry (AFB). In the literature, this observable is usually advocated in the second stage of the data analysis process to interpret experimental results after a possible discovery of a new spin-1 particle using a standard bump hunt. The AFB is indeed used to disentangle different { models with $Z'$-bosons and nail down the underlying theory}. This procedure relies on the assumption that the new heavy $Z^\prime$-boson is characterized by a narrow width and it would be discovered via the bump search. Our purpose is to show that the AFB can be used not only for interpreting a possible discovery but also in the very same search process. We show that the AFB observable  can be associated to the default resonance search to improve and/or extend the discovery potential for both narrow and wide $Z^\prime$s.  Focussing on AFB, we aim at establishing the methods needed to  study $Z^\prime$-boson production at the CERN LHC in the Drell-Yan channel,  giving rise to di-lepton pairs in the final state: $pp\rightarrow l^+l^-$ with $l=e, \mu$. This production process is particularly clean and thus represents the golden channel for $Z^\prime$-boson discovery at the LHC. 

The paper is organized as follows. In Section \ref{sec:bounds}, we derive current and projected bounds for  $Z'$ model benchmarks  for the LHC at 8 and 13 TeV, respectively, for the  models presented in Ref.  \cite{Accomando:2010fz}.
In Section \ref{sec:AFB}, we discuss the role of the forward-backward asymmetry within the Z' physics, its reconstruction and statistical uncertainty. We also
discuss the effect of a rapidity cut on the di-lepton system upon the signal significance. This cut is commonly implemented in order to increase the efficiency in guessing the quark direction in $pp$ collisions, which is needed to reconstruct the AFB observable. 
{The drawback of applying it is a decrease of the number of signal events with the consequent depletion of the AFB significance}.  The outcome is that this stringent cut can be relaxed for the range of $Z^\prime$-boson masses {which we will be looking for in the next LHC run.} 
In Section \ref{sec:narrowZ}, we discuss the role of the AFB in searches for narrow width $Z^\prime$-bosons and  systematically analyse $Z'$ model benchmarks confronting AFB and bump search. We will show that the significance from the AFB can be comparable with that obtained from the cross section studies over the same invariant mass distribution of the di-lepton system. In this scenario, the advantage of using the AFB observable would consist in minimizing the systematics, as the AFB is a ratio of (differential) cross sections.  Moreover, the two observables (cross section and asymmetry) could be of mutual support to make the claim of a possible new physics discovery more robust, if the bump search itself would  provide only a mild evidence for a $Z^\prime$ state.

In Section \ref{sec:wideZ}, we analyse the role of AFB in searches for wide $Z^\prime$ particles. We consider two benchmark models which predict a wide $Z^\prime$-boson with ratio $\Gamma_{Z^\prime}/M_{Z^\prime}$ of the order of several tens  of percent. In this case, again, the  AFB can be complementary to the resonance search and have a distinctive line shape contrary to the invariant mass distribution  of the cross section, which could well mimic the background shape. That is, the latter would only give rise to an excess of events evenly spread over the SM background, which is of difficult interpretation and in many cases of difficult measurement owing to uncertainties in the background modelling. Finally, in Section~\ref{sec:conclusions} we summarize and conclude.


\section{Bounds on the Z'-boson mass}
\label{sec:bounds}

In this section, we re-obtain the existing bounds on the mass of the $Z^\prime$-boson from the aforementioned benchmarks models as obtained by, e.g., the CMS collaboration, after the 7, 8 TeV run with about 20 $fb^{-1}$ of accumulated luminosity, assuming a narrow width: these can be found in Ref.~\cite{Khachatryan:2014fba}. After confirming current CMS limits obtained under the above assumption, we further present those obtained by taking into account the full width effect as well the intereference corrections. Finally,  we produce projection limits for LHC Run 2.

We therefore start by scanning over the thirteen benchmark models predicting a $Z^\prime$-boson characterized by a narrow width ($\Gamma_{Z^\prime}/M_{Z^\prime}\le 5\%$) summarized in Ref. \cite{Accomando:2010fz}, and extract the limits on $M_{Z^\prime}$ by making use of the 95\% C.L. upper bound on the $Z^\prime$-boson production cross section in Drell-Yan, $\sigma ( pp\rightarrow Z^\prime \rightarrow e^+e^-, \mu^+\mu^-)$. In order to reduce systematic uncertainties, the experimental analysis normalizes the $Z^\prime$-boson production cross section in Drell-Yan to the SM $Z$-boson cross section on peak. As shown in Fig. 1, the 95\% C.L. upper bound is indeed given on the ratio $R_\sigma =\sigma ( pp\rightarrow Z^\prime \rightarrow e^+e^-, \mu^+\mu^-)/\sigma ( pp\rightarrow Z, \gamma\rightarrow e^+e^-, \mu^+\mu^-)$. The use of this ratio $R_\sigma$ in fact cancels the uncertainty in the integrated luminosity and reduces the dependence on the experimental acceptance and  trigger efficiency.

\begin{figure}[t]
\centering
\subfigure[]{
\includegraphics[width=7.7cm]{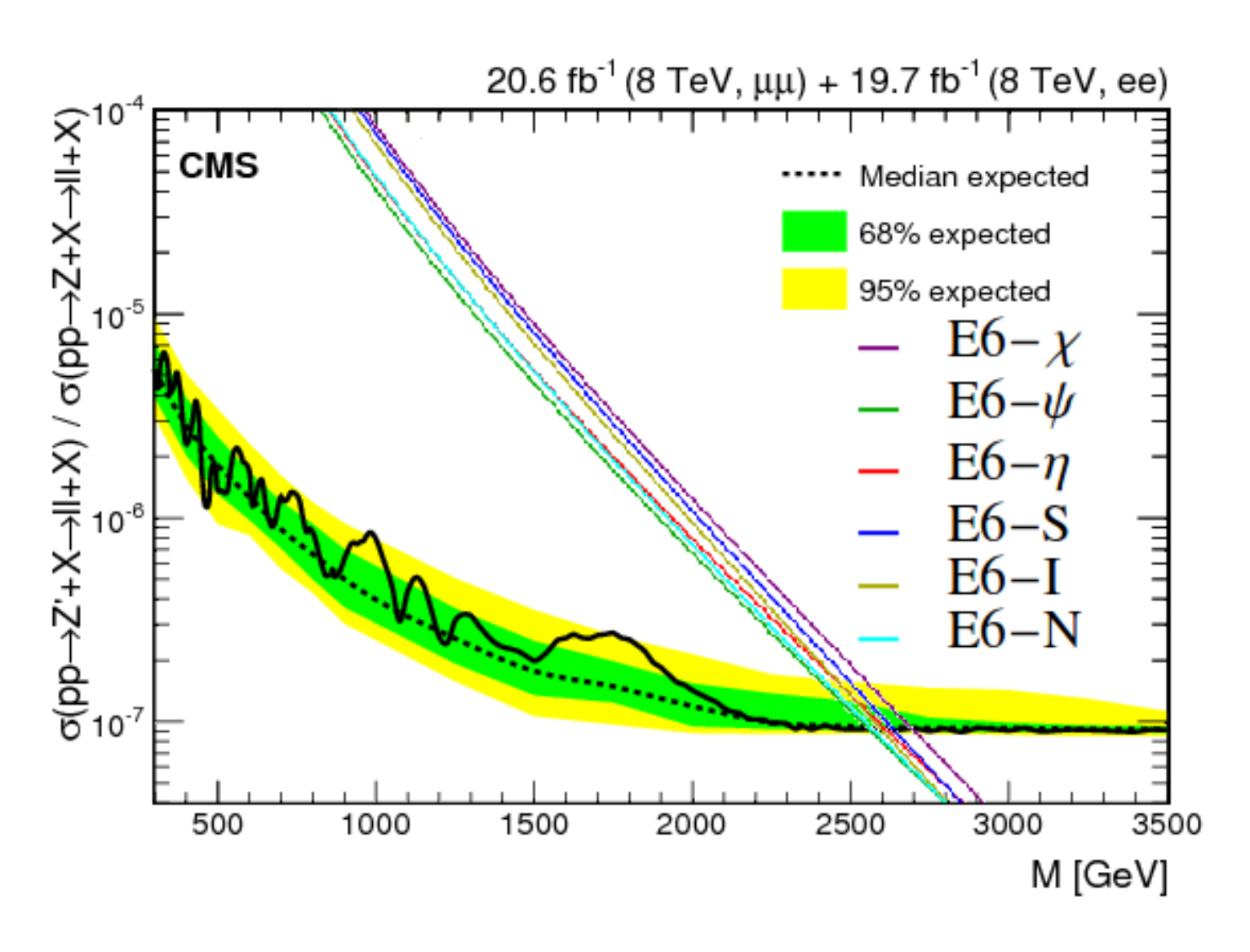}
\label{fig:ratiosa}
}
\subfigure[]{
\includegraphics[width=7.7cm]{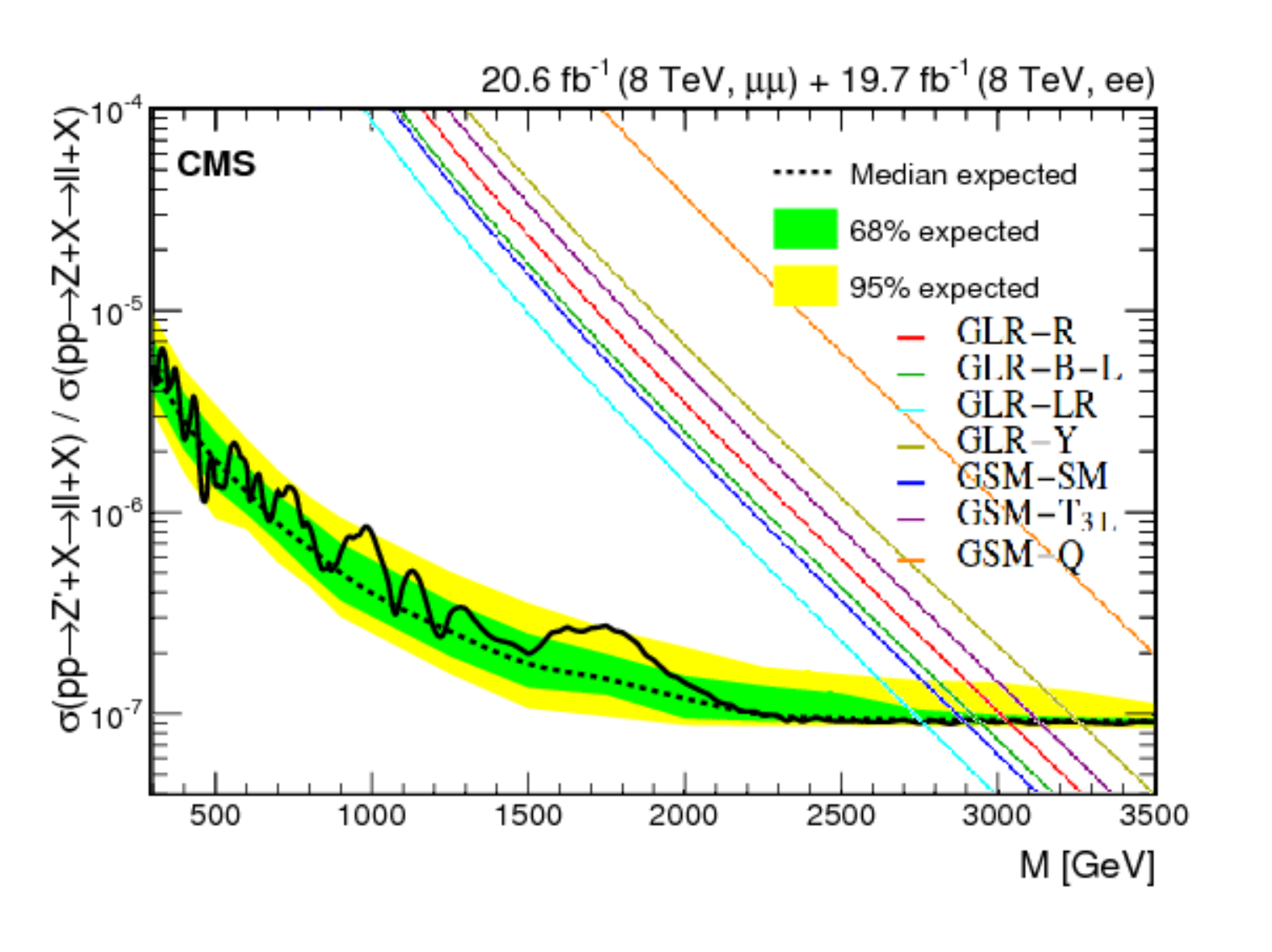}
\label{fig:ratiosb}
}
\caption{(Colour online)
\subref{fig:ratiosa} 95\% C.L. upper bound on the $Z^\prime$-boson production cross section in Drell-Yan normalized to the SM cross section on the $Z$-boson peak: $R_\sigma = \sigma (pp\rightarrow Z^\prime\rightarrow l^+l^-)/\sigma (pp\rightarrow Z, \gamma \rightarrow l^+l^-)$ with $l = e, \mu$. The combined analysis of the di-muon and di-electron channels has been produced by the CMS collaboration with a data sample collected at the 8 TeV LHC, corresponding to an integrated luminosity of 20.6 and 19.7 $fb^{-1}$ respectively \cite{Khachatryan:2014fba}. Theoretical predictions for the class of the $E_6$ models are superimposed to extract the corresponding $Z^\prime$-boson mass limits. As described in the text, in order to match theoretical predictions and experimental results, the optimal cut on the invariant mass of the di-lepton pairs has been implemented: $\Delta M = |M_{ll}-M_{Z^\prime}|\le 0.05\ \ E_{LHC}$ for $E_{LHC} = 8$ TeV.
\subref{fig:ratiosb} Same for the other two classes of GSM and GLR models.
}
\label{fig:CMS_bounds}
\end{figure}

We calculate this ratio, $R_\sigma$, at the Next-to-Next-to-Leading Order (NNLO) in QCD using 
the {WZPROD program \cite{Hamberg:1990np,vanNeerven:1991gh,ZWPROD}
(which we have adapted for $Z'$ models and new Parton Distribution Function (PDF) sets~\cite{Accomando:2010fz})}
and the CTEQ6.6 package \cite{Kretzer:2003it}. 

\begin{table}[t]
\begin{center}

\begin{tabular}{|c||c|c|c|c|c|c|c|c|c|c|c|c|c|}
\hline
Class 
& \multicolumn{6}{|c|}{$E_6$}
& \multicolumn{4}{|c|}{GLR}
& \multicolumn{3}{|c|}{GSM}
\\ \hline
$U^\prime (1)$ Models 
& $\chi$
& $\psi$
& $\eta$
& $S$
& $I$
& $N$
& $R$
& ${B-L}$
& ${LR}$
& $Y$
& ${SM}$
& ${T_{3L}}$
& $Q$
\\ \hline
$M_{Z^\prime}$ [GeV]
& 2700 
& 2560 
& 2620  
& 2640  
& 2600  
& 2570  
& 3040 
& 2950 
& 2765  
& 3260  
& 2900  
& 3135  
& 3720
 \\  \hline
\end{tabular}

\end{center}
\caption{
Bounds on the $Z^\prime$-boson mass derived from the latest direct searches performed by CMS at the 8 TeV LHC with integrated luminosited $L=20 fb^{-1}$. We consider thirteen different models with an extra $U^\prime (1)$ gauge group predicting a new heavy neutral boson characterized by a narrow width. From left to right, the columns indicate the $M_{Z^\prime}$ limit in $GeV$ within the $E_6$, GLR and GSM class of models.
}
\label{tab:events_8tev}
\end{table}

  The NNLO QCD contributions give rise to a $K$-factor which depends on the energy scale, thus we fully take into account such a dependence. The NNLO prediction for the SM $Z, \gamma$ production cross section, $\sigma (pp\rightarrow Z, \gamma\rightarrow l^+l^-)$ with $l = e, \mu$, in the mass window of 60 to 120 GeV is 1.117 nb. With all these ingredients at hand, we compute $R_\sigma$ as a function of  the mass of the new heavy $Z^\prime$-boson, $M_{Z^\prime}$, and derive the corresponding limits for all benchmark models. Fig. 1a shows the bounds on all $E_6$ models, while Fig. 1b displays the results for the remaining two classes of models, GLR and GSM.  As previously mentioned,
traditional experimental analyses work under the hypothesis that the signal has a Breit-Wigner line shape and performs the analysis in a restricted search window around the hypothetical mass of the $Z^\prime$-boson. This approach is theoretically motivated by the benchmark models, all predicting a narrow width $Z^\prime$-boson, and by the will to perform an analysis as much as possible model independent. One should stress that the CMS analysis makes use of a dedicated cut on the invariant mass of the di-lepton pairs: $|M_{l\bar l}-M_{Z^\prime}|\le 0.05\times E_{\rm LHC}$ where $E_{\rm LHC}$ is the collider energy. This cut was designed so that the error in neglecting the (model-dependent) Finite Width (FW) and interference effects (between $\gamma , Z, Z'$) are kept below $O(10\% )$ for all models and for the full range of allowed $Z^\prime$ masses under study, thus following the recommendations of \cite{Accomando:2013sfa}.
This procedure thus continues to allow for a straightforward interpretation of the extracted mass bounds in the context of any theory predicting a narrow $Z'$-boson. At this stage  of our own analysis, we work under the very same setup, for validation purposes, with the
notable exception that we allow for the aforementioned FW and interference effects, unlike the experimental results which assume the so-called Narrow Width Approximation (NWA), wherein the (narrow) $Z^\prime$ is actually produced on-shell. Table \ref{tab:events_8tev} summarizes the bounds we obtain. They reproduce  the CMS limits very well in general, within the accuracy of  $1-2\%$. The only slight exception is the $Q$-model in the GSM class where our limit, 
based on the present analysis accounting for full width and interference, is different from the CMS one by about $5\%$.
Yet, this is well in line with expectations, as this model is the one yielding the largest width. 

\begin{figure}[t]
\centering
\subfigure[]{
\includegraphics[width=0.8\textwidth]{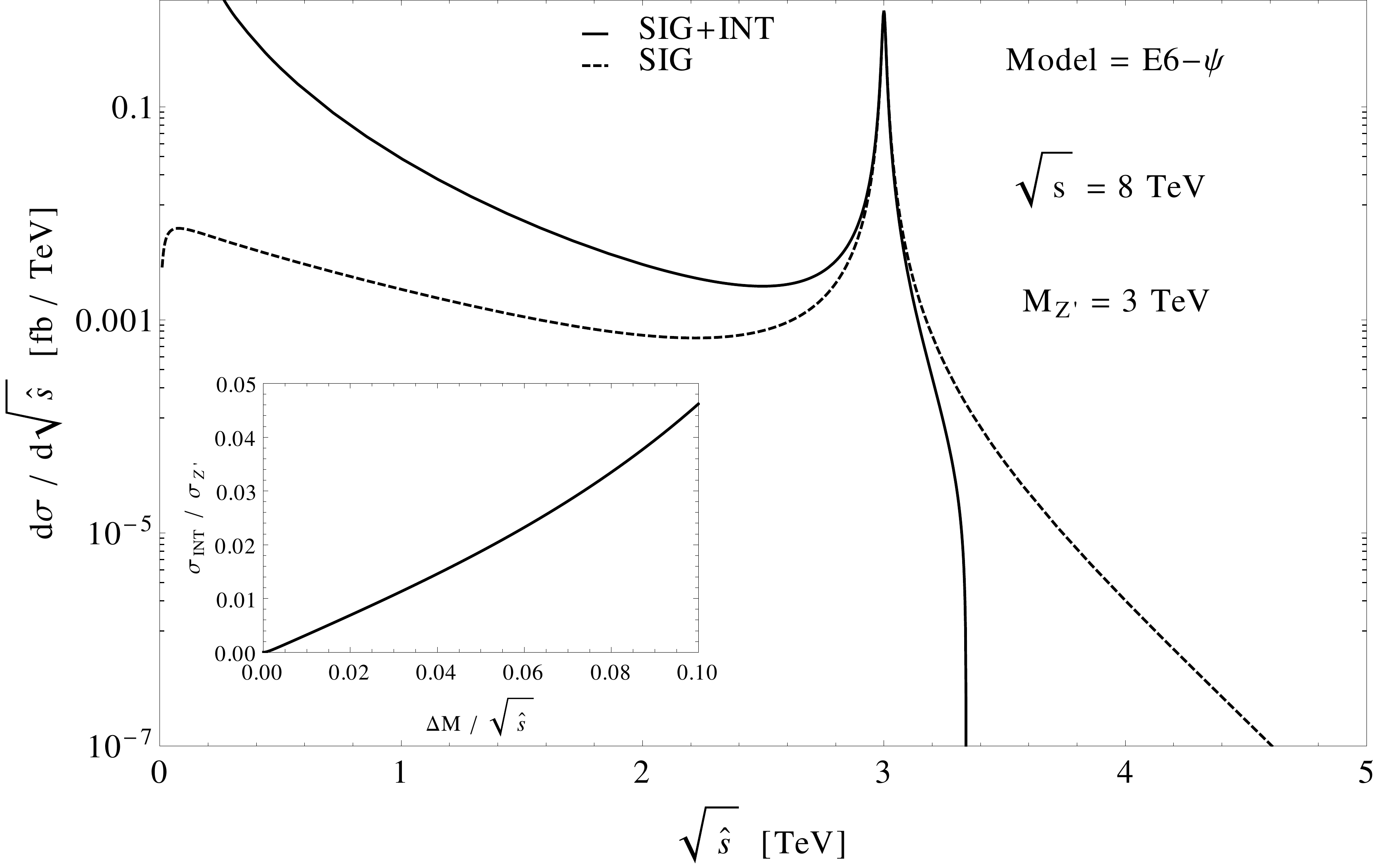}
\label{fig:signala}
}%
\\
\subfigure[]{
\includegraphics[width=0.8\textwidth]{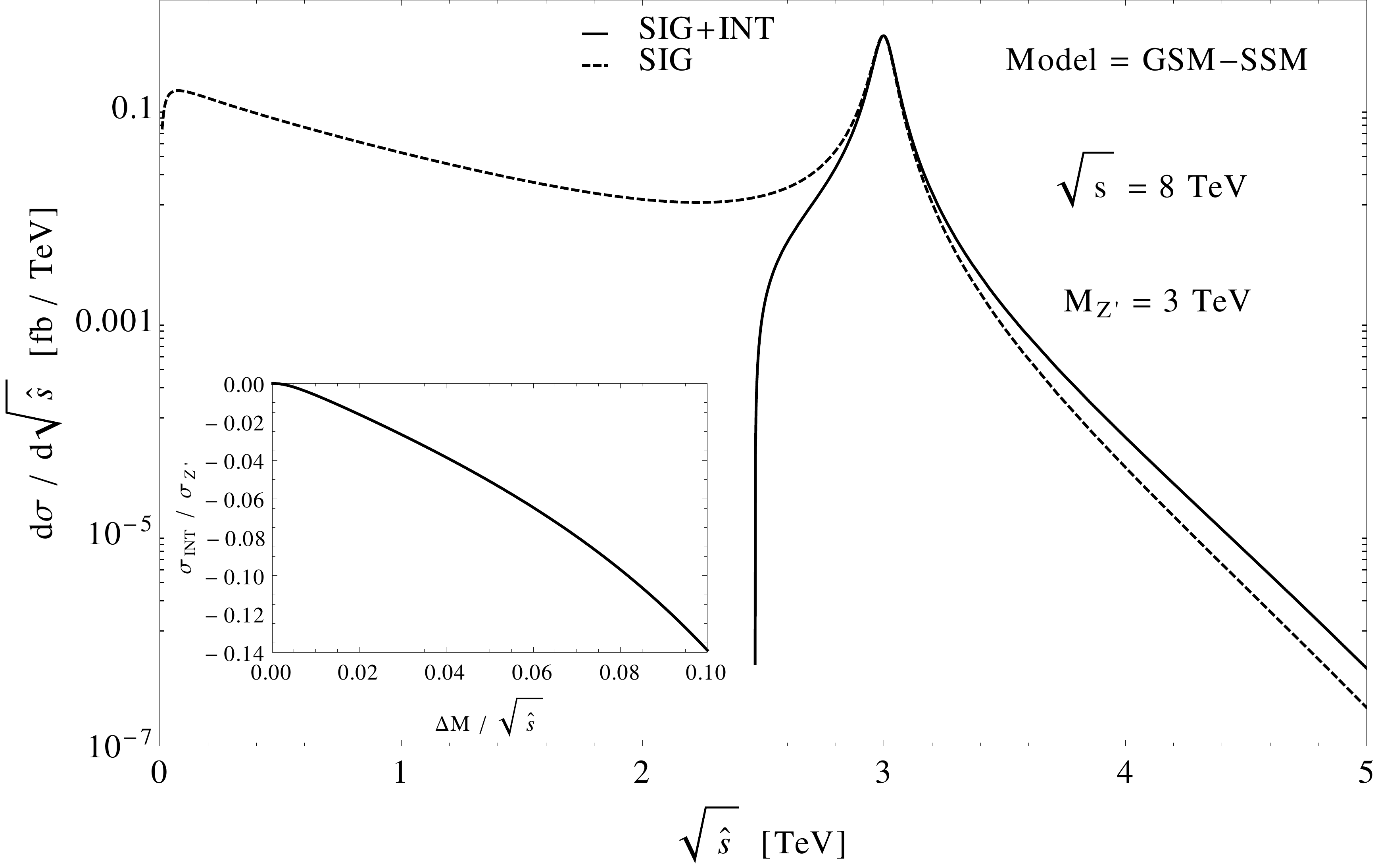}
\label{fig:signalb}
}
\caption{
\subref{fig:signala} Differential cross section in the invariant mass of the di-lepton system coming from the $Z^\prime$-boson production and decay in the Drell-Yan channel: $pp\rightarrow  e^+e^-$. We consider the $E_\psi$ model. The solid line shows the complete new physics contribution to the invariant mass distribution, that is the $Z^\prime$ signal plus the interference with the SM background. The dashed line represents instead the pure $Z^\prime$ signal, neglecting the intereference. The plot has been produced for the 8 TeV LHC, no cuts are applied.
\subref{fig:signalb} Same for the SSM benchmark scenario.
}
\label{fig:signal}
\end{figure}

But let us set the stage in some more detail. In Fig. 2 we show the behaviour of the new physics signal for two representative scenarios: the $E_\psi$ model (Fig. 2a) and the SSM benchmark scenario (Fig. 2b). 
The solid line represents the full new physics signal, that is, $Z^\prime$-boson production and decay including the interference with the SM background. The dashed line gives instead the pure $Z^\prime$-boson signal, neglecting the interference. 
(As evident from the plots, in both cases we allow for FW effects of the $Z'$-boson.)
As one can see, the shape of the distribution in the invariant mass of the di-lepton system is quite model dependent off peak. This is due to the fact that it can  be heavily affected by model dependent interference effects there.
The sign of the interference is not defined a priori. It can be either positive, like in the $E_6$ models, or negative, like in the two other classes of GLR and GSM models. In addition, its magnitude can be quite sizeable.
The off peak tail of the signal distribution in the di-lepton invariant mass is thus highly model dependent in the low mass region, leading to either an excess or a depletion of the total number of expected events as compared to the SM background, according to the sign of the interference.

Furthermore, it is clear that the fully integrated cross section for the complete new physics signal, that is, $Z^\prime$-boson production and decay including the interference with the SM background, is not always a uniquely defined variable. In the SSM, and more generally in all GSM and GLR models, the signal can manifest itself as a negative correction to the
differential cross section (solid line) at low masses. Similarly, the fully integrated cross section for the pure $Z^\prime$-boson production and decay, neglecting the interefences, can also give quite a wrong picture. Taking into account the low mass tail of the invariant mass distribution can overestimate the $Z^\prime$-boson signal by a large factor. For a SSM $Z^\prime$ with mass $M_{Z^\prime} = 3$ TeV, we have that the fully integrated cross section for the pure signal is $\sigma_{Z^\prime} = 0.17 fb$ while the complete signal cross section, integrated in the mass window where it is positive definite, is equal to $\sigma_{Z^\prime}$ + Interference = 0.06 $fb$. In this case, taking into account the unphysical tail (which in the SSM is in reality washed out by the destructive interference) leads to overestimate the $Z^\prime$ signal by a factor of 3 and consequently to extract more stringent limits that are erroneous. In essence,
 a shape analysis of the signal over the full invariant mass region is very challenging. Thus, the definition of the observable to be used to interpret the data and extract the mass bounds on the hypothetical $Z^\prime$-boson must indeed be appropriately chosen.

\begin{figure}[t]
\centering
\subfigure[]{
\includegraphics[width=0.5\textwidth]{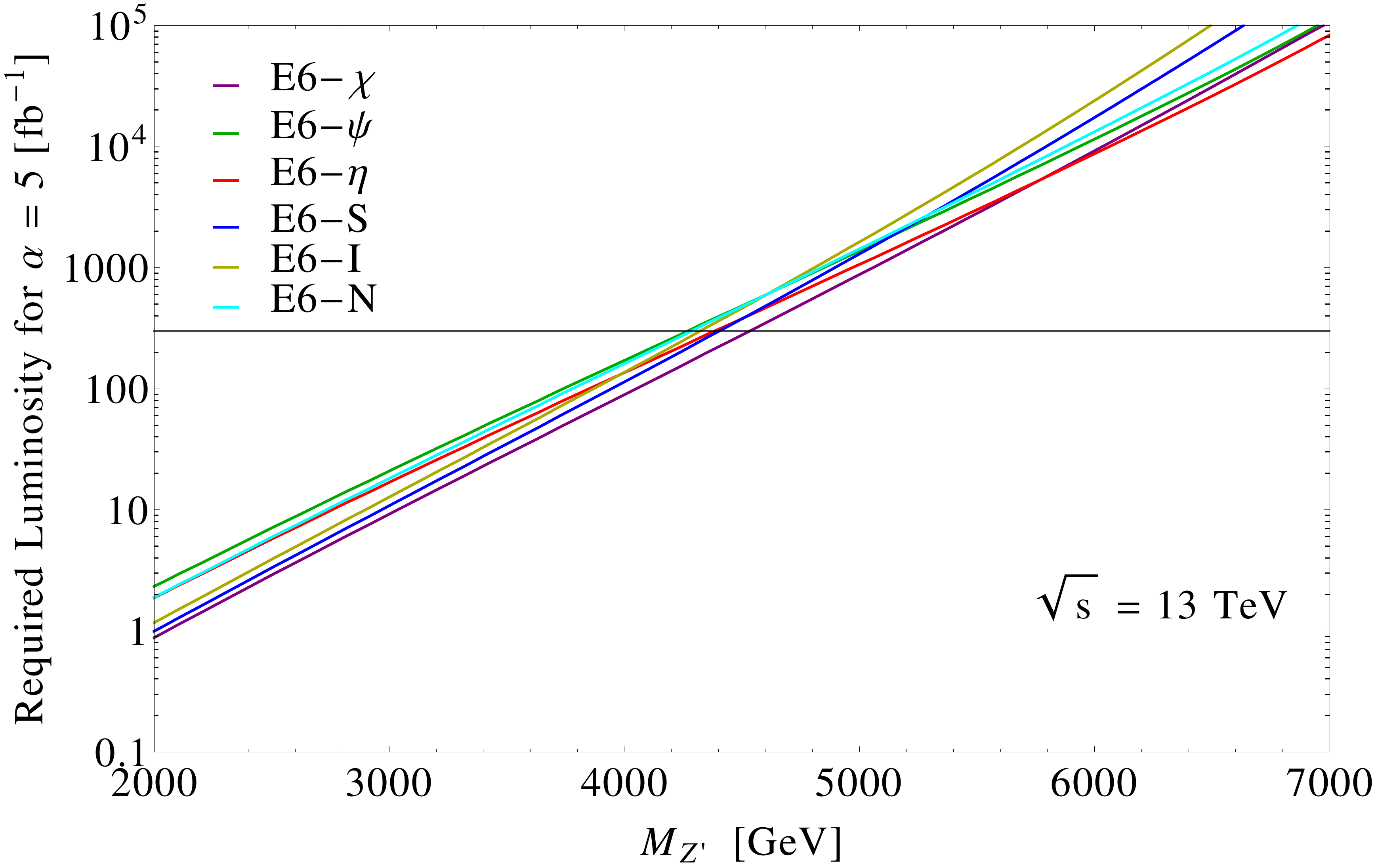}
\label{fig:discoverya}
}%
\subfigure[]{
\includegraphics[width=0.5\textwidth]{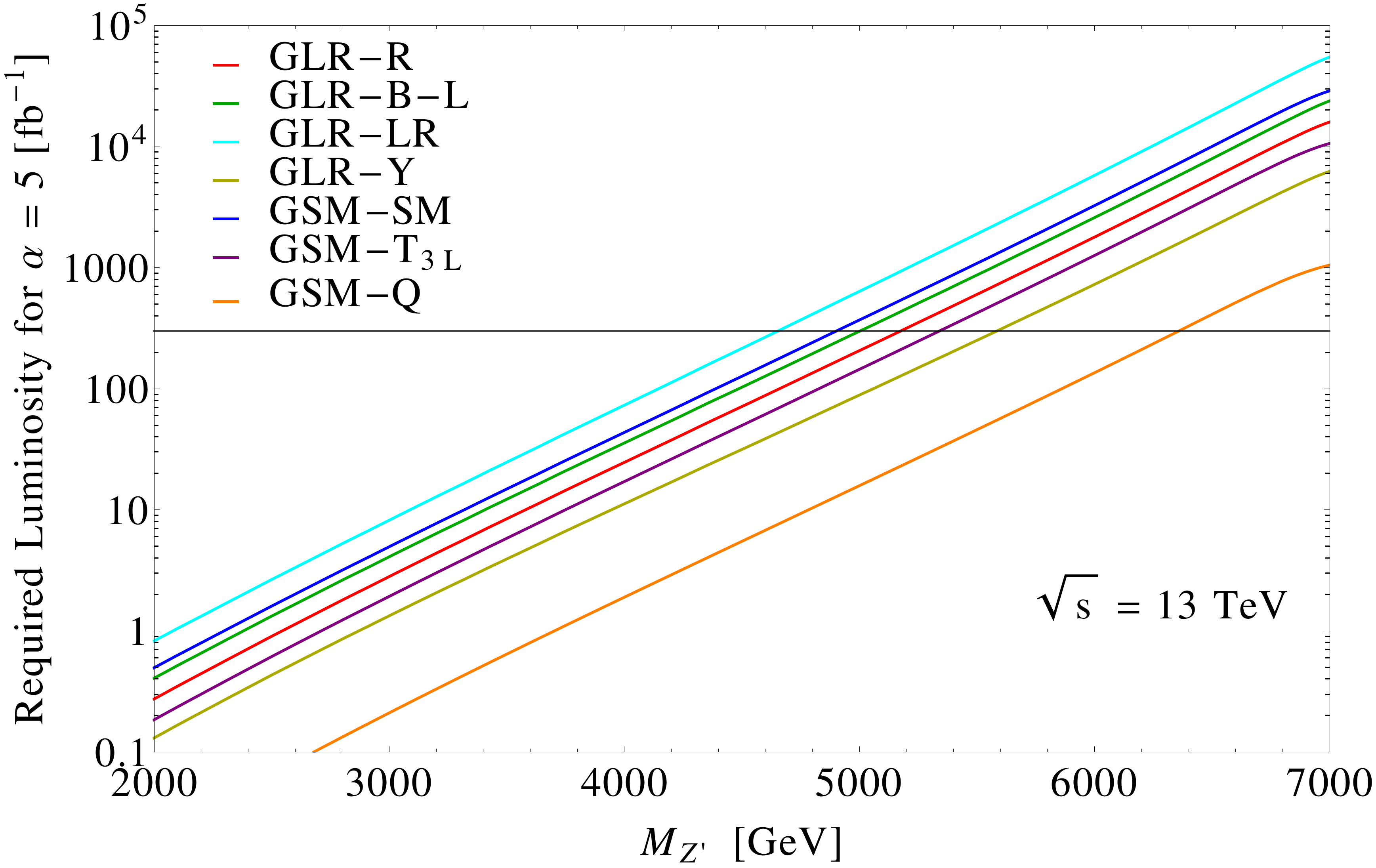}
\label{fig:discoveryb}
}
\caption{(Colour online)
\subref{fig:discoverya} Discovery potential of the 13 TeV LHC for the $E_6$ class of models. We plot the
$5\sigma$ contours as a function of $Z^\prime$-boson mass and luminosity. We perform a combined analysis over $e^+e^-$ and $\mu^+\mu^-$ pairs and assume the $A\times \epsilon$ factor given by CMS at the 8 TeV LHC.
 \subref{fig:discoveryb} Same for the GSM and GLR classes of models.
}
\label{fig:discovery}
\end{figure}


However, thanks to the approach recommended in  \cite{Accomando:2013sfa}, all such extreme effects are avoided and, in the instance,
we can conclude that our code for the simulation of Drell-Yan processes which might receive a contribution from a narrow  $Z'$-boson exchange, $pp\rightarrow\gamma , Z, Z'\rightarrow l^+l^- (l=e, \mu )$, has been validated against the CMS results. In short,
by finally taking into account the published acceptance $\times$  efficiency corrections ($A\times \epsilon$), we can indeed reproduce the above mentioned extracted bounds on the $Z^\prime$-boson mass by self-consistently evaluating signal and background. In doing so, we have applied the Poisson statistics for computing the significance and hence the exclusion limits on $M_{Z^\prime}$.

By using such a code, in Figs. 3 and 4, we now project discovery and exclusion potential of the upgraded LHC, which will run at 13 TeV. The first two plots show the LHC discovery potential for the $E_6$ models (Fig. 3a) and for the remaining other two classes GLR and GSM (Fig. 3b).  Figs. 4a and 4b display instead the LHC exclusion potential for $E_6$, GSM and GLR models, respectively.
In deriving these results, we stress  that we have included the $A\times \epsilon$ factor extracted by the analyses performed by CMS at the 8 TeV LHC, thereby implicitly assuming that no significant departures in this respect are assumed at the upgraded CERN machine.
These projections are valid only for narrow width $Z^\prime$-bosons for which the optimal observable is the invariant mass of the di-lepton system, used in the standard bump search performed by the experimental collaborations. From the above plots, one can conclude that the 13 TeV run of the LHC should be able to discover a $Z'$-boson with mass up to about 4500 GeV and 5600 GeV within the $E_6$ and GSM/GLR class of models, respectively. If nothing is found, the exclusion limits will be pushed up to 5300 GeV and 6400 GeV for an $E_6$ $Z'$-boson and a GSM/GLR $Z'$-boson, respectively. Table \ref{tab:events_13tev} summarizes the LHC Run II potential for discovering or excluding a $Z'$-boson of a certain mass within all considered thirteen models. These  projections  have been obtained for the design value of the integrated luminosity: $L=300 fb^{-1}$. 

\begin{table}[t]
\begin{center}

\begin{tabular}{|c||c|c|c|c|c|c|c|c|c|c|c|c|c|}
\hline
Class 
& \multicolumn{6}{|c|}{$E_6$}
& \multicolumn{4}{|c|}{GLR}
& \multicolumn{3}{|c|}{GSM}
\\ \hline
$U^\prime (1)$ Models 
& $\chi$
& $\psi$
& $\eta$
& $S$
& $I$
& $N$
& $R$
& ${B-L}$
& ${LR}$
& $Y$
& ${SM}$
& ${T_{3L}}$
& $Q$
\\ \hline
$M_{Z^\prime}$ [GeV]
& 4535 
& 4270 
& 4385  
& 4405  
& 4325  
& 4290  
& 5175 
& 5005 
& 4655  
& 5585  
& 4905  
& 5340  
& 6360
 \\  \hline
$M_{Z^\prime}$ [GeV]
& 5330 
& 5150 
& 5275  
& 5150  
& 5055  
& 5125  
& 6020 
& 5855 
& 5495  
& 6435  
& 5750  
& 6180  
& 8835
 \\  \hline

\end{tabular}
\end{center}
\caption{
Projection of discovery limits (first row) and exclusion limits (second row) on the $Z^\prime$-boson mass from direct searches at the forthcoming Run II of the LHC at 13 TeV. We assume the original design value for the integrated luminosited: $L=300 fb^{-1}$. We consider thirteen different models with an extra $U^\prime (1)$ gauge group predicting a new heavy neutral boson characterized by a narrow width. From left to right, the columns indicate the $M_{Z^\prime}$ limit in $GeV$ within the $E_6$, GLR and GSM class of models.
}
\label{tab:events_13tev}
\end{table}

This concludes the section on the state-of-the-art of narrow width $Z^\prime$-bosons and their search using traditional methods.

\begin{figure}[t]
\centering
\subfigure[]{
\includegraphics[width=0.5\textwidth]{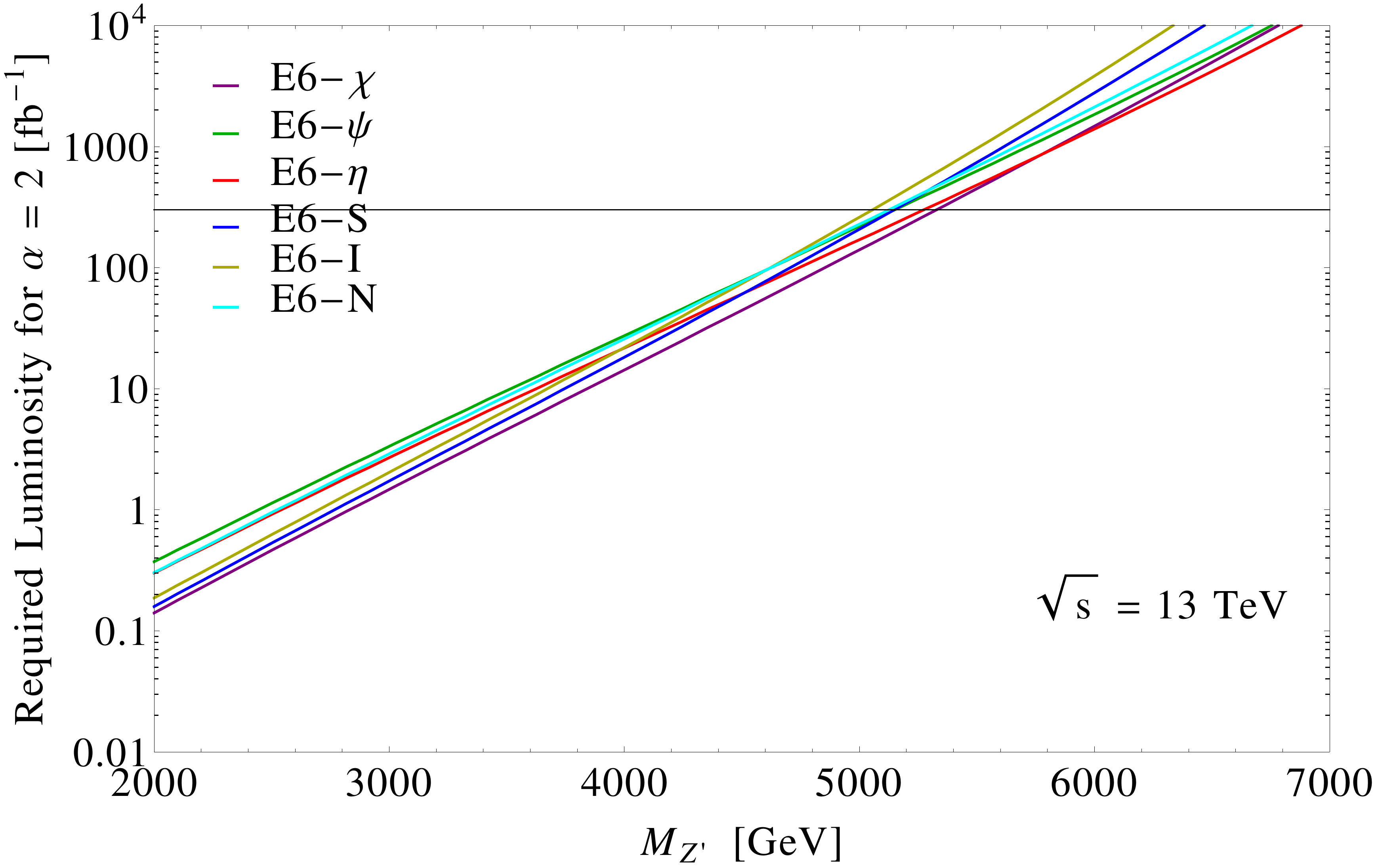}
\label{fig:exclusiona}
}%
\subfigure[]{
\includegraphics[width=0.5\textwidth]{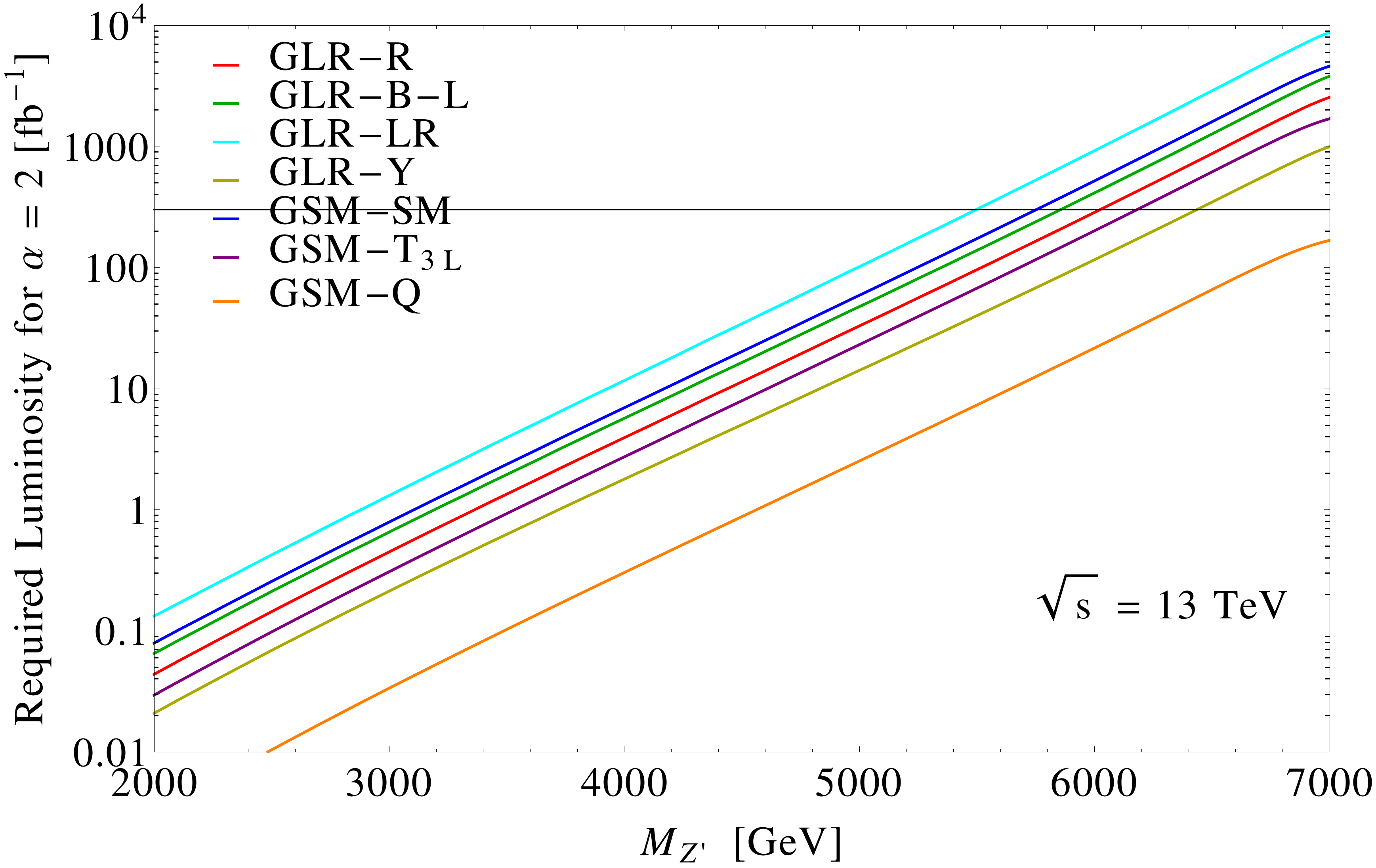}
\label{fig:exclusionb}
}
\caption{(Colour online)
\subref{fig:exclusiona} Exclusion potential of the 13 TeV LHC for the $E_6$ class of models. We plot the $5\sigma$ contours as a function of $Z^\prime$-boson mass and luminosity. We perform a combined analysis over $e^+e^-$ and $\mu^+\mu^-$ pairs and assume the $A\times \epsilon$ factor given by CMS at the 8 TeV LHC.
 \subref{fig:exclusionb} Same for the GSM and GLR classes of models.
}
\label{fig:exclusion}
\end{figure}


\section{The Forward-Backward  Asymmetry}
\label{sec:AFB}

In this section, we define the forward-backward charge asymmetry (AFB) and we discuss its role in $Z^\prime$-boson searches other than data interpretation. In the literature, the AFB has been long exploited  to help disentangling the various theories predicting an extra heavy neutral boson and tracing back the Lagrangian parameters (see, for example,  \cite{Carena:2004xs} and \cite{Petriello:2008zr} and references therein).
 This is not an easy task and the sensitivity of AFB measurements to new physics like additional $Z^\prime$-bosons has therefore received a lot of attention in the past years. For  Drell-Yan processes, AFB is defined from the angular distribution 

\begin{equation}
\label{eq:AFB}
{d\sigma\over {d\cos\theta_l^*}}\propto \sum_{spin,col}\left|\sum_i\mathcal{M}_i\right|^2=\frac{\hat{s}^2}{3}\sum_{i,j}|P^*_iP_j|[(1+\cos^2\theta_l^*)C^{ij}_S+2\cos\theta_l^* C^{ij}_A] 
\end{equation}
where $\theta_l^*$ is the lepton angle with respect to the quark direction in the di-lepton center-of-mass frame (CM), which can be derived from the measured four-momenta of the di-lepton system in the laboratory frame. The AFB is indeed given by the coefficient of the contribution to the angular distribution linear in $\cos\theta_l^*$. In Eq.  \ref{eq:AFB}, $\sqrt{\hat{s}}$ is the invariant mass of the di-lepton system, and $P_i$ and $P_j$ are the propagators of the gauge bosons involved in the process. At the tree-level,  the Drell-Yan production of charged lepton pairs is mediated by three gauge bosons: the SM photon and $Z$-boson and the hypothetical $Z^\prime$-boson. These three vector boson exchanges all participate in the matrix element squared. We thus have:
\begin{equation}
P_iP_j = \frac{(\hat{s}-M_i^2)(\hat{s}-M_j^2)+M_i\Gamma_i M_j\Gamma_j} 
        {\left((\hat{s}-M_i^2)^2+M_i^2\Gamma_i^2\right)
        \left((\hat{s}-M_j^2)^2+M_j^2\Gamma_j^2\right)}
\end{equation}
where $M_i$ and $\Gamma_i$ are the mass and the width of the gauge bosons involved and $i,j = \{\gamma, Z, Z^\prime\}$. Finally, the factors $C_{S}^{ij}$ and $C_{A}^{ij}$ in the angular distribution given in Eq. \ref{eq:AFB} are the parity symmetric and anti-symmetric coefficients which are functions of the chiral quark and lepton couplings, $q_{L/R}^i$ and $e_{L/R}^i$, to the $i$-boson with $i=\{\gamma, Z, Z^\prime\}$:
\begin{align}
\label{eqn:def_C}
C_{S}^{ij}&=(q_L^i q_L^j+ q_R^i q_R^j)(e_L^i e_L^j+ e_R^i e_R^j), \\           
C_{A}^{ij}&=(q_L^i q_L^j- q_R^i q_R^j)(e_L^i e_L^j- e_R^i e_R^j).   
\end{align}
\noindent
One can conveniently compute the forward (F) and backward (B) contributions to the total cross section integrating over opposite halves of the angular phase space:
\begin{align}
\label{eqn:xs_FB}
d\hat{\sigma}_{F}&=\int_0^1\frac{d\hat{\sigma}}{d\cos\theta_l^*}d\cos\theta_l^*=\frac{\hat{s}}{128\pi}\sum_{i,j}\frac{P_iP_j}{1+\delta_{ij}}(3\pi C_S^{ij}+8C_{A}^{ij}),\\             d\hat{\sigma}_{B}&=\int_{-1}^0\frac{d\hat{\sigma}}{d\cos\theta_l^*}d\cos\theta_l^*=\frac{\hat{s}}{128\pi}\sum_{i,j}\frac{P_iP_j}{1+\delta_{ij}}(3\pi C_S^{ij}-8C_{A}^{ij}),
\end{align}
where $i$ and $j$ sum over the mediating resonances, $\{\gamma,Z,Z^\prime\}$.
\par\noindent
From the above expressions one can immediately see that the total cross section, $\sigma = \sigma_F + \sigma_B$, depends uniquely on the parity symmetric coefficient $C_S$. Conversely, the difference between forward and backward cross sections, $\sigma_F - \sigma_B$, preserves only the contribution proportional to the parity antisymmetric coefficient $C_A$. This is the term which is related to the AFB. One can thus define the AFB as the difference between forward and backward cross sections normalized to the total cross section:
    \begin{align}
        \begin{split}\label{eqn:xs_afb}
        d\hat{\sigma}&=d\hat{\sigma}_F+d\hat{\sigma}_B=\frac{3\hat{s}}{128}\sum_{i,j}\frac{P_iP_j}{1+\delta_{ij}}C_S^{ij},\\
        A_{FB}&=\frac{d\hat{\sigma}_F-d\hat{\sigma}_B}{d\hat{\sigma}_F+d\hat{\sigma}_B}
        =\frac{8\hat{s}}{3\pi\hat{\sigma}}\sum_{i,j}\frac{P_iP_j}{1+\delta_{ij}}C_{A}^{ij}.
        \end{split}
    \end{align} 
    with the SM background corresponding to $ij=\gamma\gamma,ZZ,\gamma Z$ and the new physics given by $ij=\gamma Z^\prime,Z Z^\prime,Z^\prime Z^\prime$. In the light of the above discussion, total cross section and AFB depend on different combinations of $Z^\prime$-boson couplings to ordinary matter. For that reason, the AFB can give complementary information about the structure of such couplings when compared to the total cross section. This feature has motivated several authors to study the potential of the AFB observable in interpreting a possible $Z^\prime$-boson discovery obtained in the usual resonance hunt as in Ref. \cite{Carena:2004xs, Petriello:2008zr, Rizzo:2009pu}. Our point is that AFB can also be a powerful tool to search for new physics.

\subsection{The reconstructed AFB}

The AFB is obtained by integrating the lepton angular distribution forward and backward with respect to the quark direction.
As in $ pp$ collisions the original quark direction is not known, one has to extract it from the kinematics of the di-lepton system. In this analysis, we follow the criteria of Ref. \cite{Dittmar:1996my} and simulate the quark direction from the boost of the di-lepton system with respect to the beam axis ($z$-axis). This strategy is motivated by the fact that at the $pp$ LHC the di-lepton events at high invariant mass come from the annihilation of either valence quarks with sea antiquarks or sea quarks with sea antiquarks. As the valence quarks carry away, on average,  a much larger fraction of the proton momentum than the sea antiquarks, the boost direction of the di-lepton system should give a good approximation of the quark direction. A leptonic forward-backward asymmetry can thus be expected with respect to the boost direction. In contrast, the subleading number of di-lepton events which originate from the annihilation of quark-antiquark pairs from the sea must be symmetric.

As a measure of the boost, we define the di-lepton rapidity
\begin{equation}
y_{l\bar{l}}={1\over 2} \rm{ln}\left [{{E+P_z}\over {E-P_z}}\right ]
\label{eq:yll}
\end{equation}
\noindent
where $E$ and $P_z$ are the energy and the longitudinal momentum of the di-lepton system, respectively. We identify the quark direction through the sign of $y_{l\bar{l}}$. In this way, one can define the reconstructed forward-backward asymmetry, from now on called $A_{FB}^\ast$. 
Namely, we have defined $A_{FB}^\ast$ using the $\theta_l^\ast$ reconstructed angle, which is the angle between the final state lepton and the incoming quark direction in the center-of-mass of the di-lepton system.
As the AFB reconstruction  procedure relies on the correlation between the boost variable, $y_{l\bar{l}}$, and the direction of the incoming valence quark, it is therefore more likely to pick up the true direction of the quark for higher values of $y_{l\bar{l}}$. Increasing the probability of identifying the direction of the quark would lead to an observed value of $A_{FB}^\ast$ that is closer to the `true' value of $A_{FB}$ if one were able to access the partonic CM frame. The tradeoff occurs in the reduction of statistics which impacts the significances the other way. The general definition of significance $S$ between predictions of an observable $O$ with uncertainty $\delta O$ from two hypotheses is
$$
S=\frac{|O_1-O_2|}{\sqrt{\delta O^2_1+\delta O^2_2}}.
$$
The statistical uncertainty on the AFB is given by
$$
\delta A_{FB} =\sqrt{\frac{4}{\mathcal{L}}\frac{\sigma_{F}\sigma_{B}}{(\sigma_{F}+\sigma_{B})^3}} = 
\sqrt{\frac{(1-A_{FB}^2)}{\sigma\mathcal{L}}} = \sqrt{\frac{(1-A_{FB}^2)}{N}},
$$
where $\mathcal{L}$ is the integrated luminosity and $N$ the total number of events. One can thus see that the significance is proportional to the root of the total number of events. Imposing a stringent cut on the boost variable, $y_{l\bar{l}}$, would then improve the reconstructed AFB guiding it  towards its true line shape, but it will decrease the statistics. In the next subsection, the subtle balance between line shape gain and statistics loss in maximizing the significance via the di-lepton rapidity cut will be discussed in detail. 

\subsection{On the di-lepton rapidity cut}

\begin{figure}[t]
\centering
\subfigure[]{
\includegraphics[width=7.7cm]{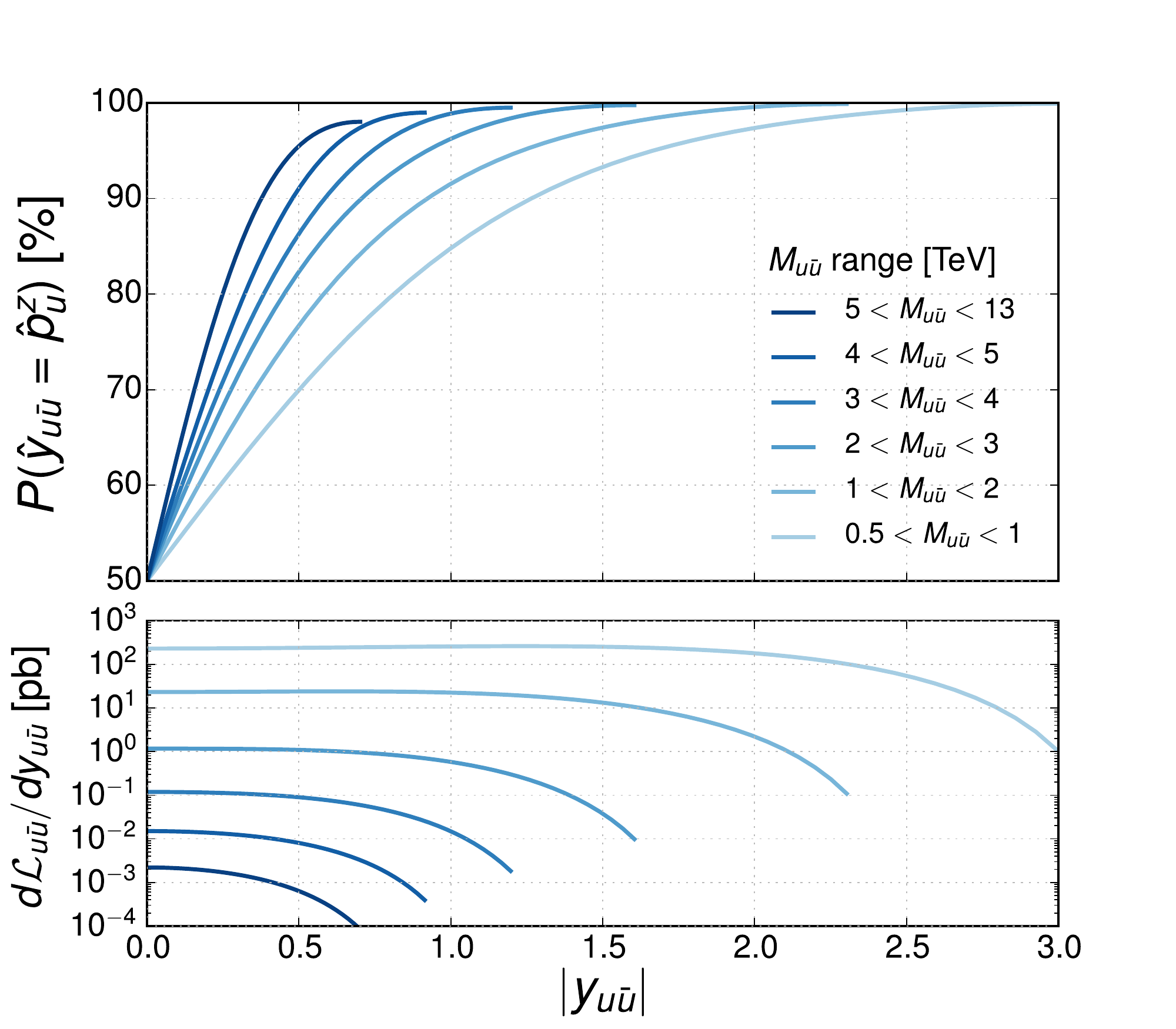}
\label{fig:fractiona}
}
\subfigure[]{
\includegraphics[width=7.7cm]{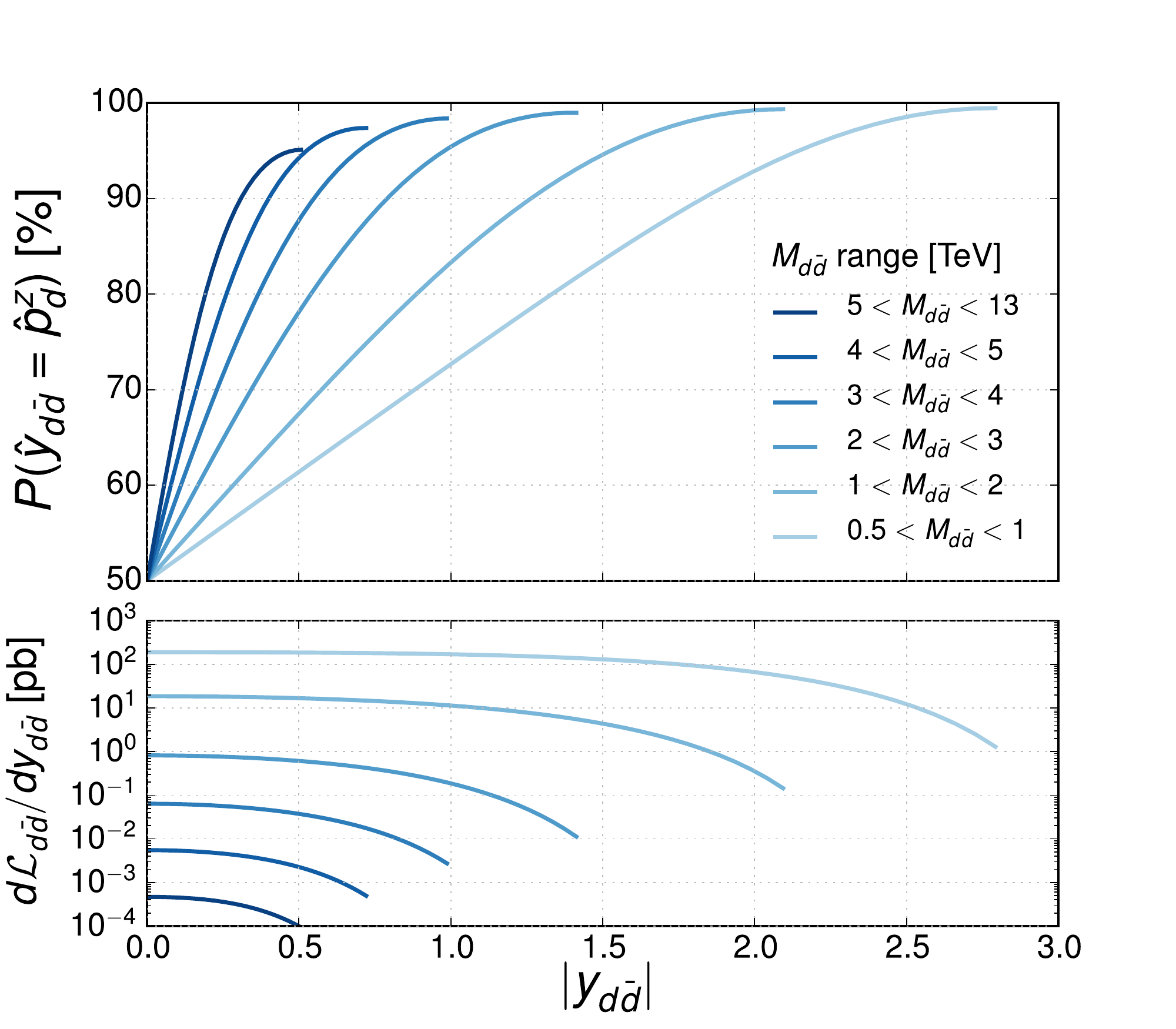}
\label{fig:fractionb}
}
\caption{(Colour online)
\subref{fig:fractiona}Upper plot: Probability of getting the correct direction of the valence up-quark at the 13 TeV LHC via the boost direction of the di-lepton system, given by the sign of the di-lepton rapidity $y_{l\bar{l}}$, as a function of the modulus $|y_{l\bar{l}}|=|y_{u\bar{u}}|$ for six different invariant mass windows scanning from 500 GeV to 5000 GeV and beyond. Lower plot: Differential luminosity as a function of $|y_{l\bar{l}}|$ for the correctly assigned quark pair (dashed-line) and for the full sample (solid line).
 \subref{fig:fractionb} Same as (a) for valence down-quarks.
 }
\label{fig:fraction}
\end{figure}

As discussed in the previous section, since the true quark direction is not known in $pp$ collisions, at the LHC one has to extract it from the kinematics of the di-lepton system. In this paper, the valence quark direction is approximated by the boost direction of the $l^+l^-$ pairs with respect to the beam axis, that is given by the sign of the di-lepton rapidity $y_{l\bar{l}}$ defined in Eq. \ref{eq:yll}. The correctness of this assignment as a function of $y_{l\bar{l}}$ has been studied in Ref. \cite{Dittmar:1996my} for di-lepton events with invariant masses above 400 GeV. In this section, we further analyse this issue by investigating the energy scale dependence of the probability of getting the true quark direction via the sign of $y_{l\bar{l}}$. In Figs. 5a and 5b, we plot the fraction of events with the correctly assigned direction for up-quarks and down-quarks, respectively, as a function of $|y_{l\bar{l}}|$ for different invariant mass windows of the di-lepton system. The fraction of correctly assigned events increases with the rapidity, confirming the results presented in literature \cite{Dittmar:1996my}. 
\begin{figure}[t]
\centering
\includegraphics[width=7.7cm]{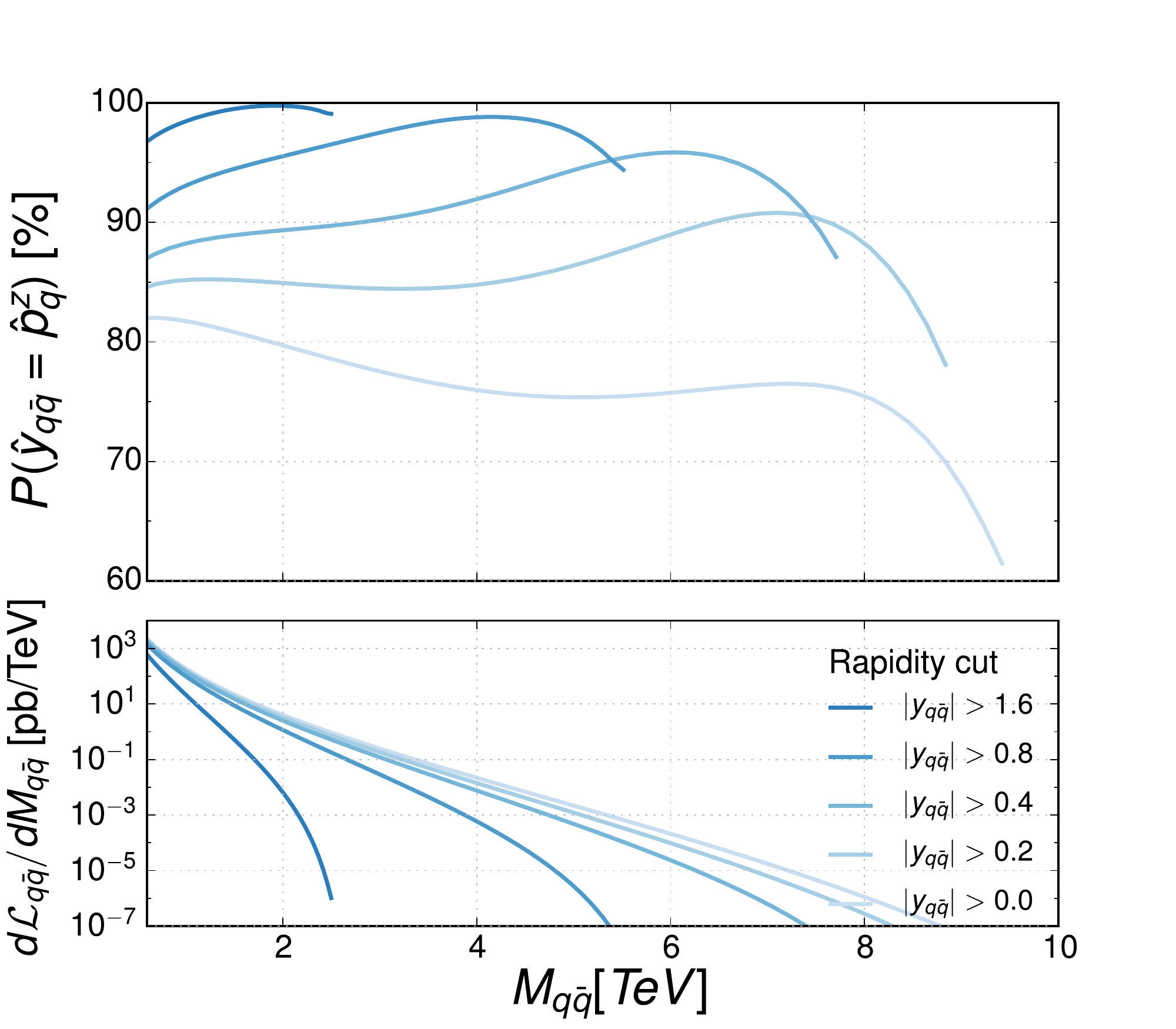}
\caption{(Colour online)
Upper plot: Probability of getting the correct direction of the valence quarks at the 13 TeV LHC via the boost direction of the di-lepton system, given by the sign of the di-lepton rapidity $y_{l\bar{l}}$, as a function of the di-quark (or di-lepton) invariant mass for five different cuts on the di-lepton rapidity. Lower plot: Differential luminosity as a function of the di-lepton invariant mass for the correctly assigned quark pair (dashed-line) and for the full sample (solid line).
Here we average on both up and down-quarks.
}
\label{fig:fraction_M}
\end{figure}
The additional information contained in Fig. 5 is that such an increase depends on the energy scale. For di-lepton invariant masses of TeV order the probability of getting the true quark direction becomes more than 80\% for a rapidity cut $|y_{l\bar{l}}|\ge 0.8$. For higher invariant masses, beyond the present $Z^\prime$-boson limits of $O$(3 TeV), the same probability can be obtained by imposing a lower rapidity cut: $|y_{l\bar{l}}|\ge 0.35$. Moreover, up-quarks and down-quarks respond differently to the $|y_{l\bar{l}}|$ cut. The probability of getting the correct direction is higher for up-quarks than for down-quarks, at fixed $|y_{l\bar{l}}|$ value. In Fig. 6, the fraction of correctly assigned events is shown as a function of the invariant mass for five different cuts on the magnitude of the di-lepton rapidity, $|y_{l\bar{l}}|$. This time, we consider the average over up and down-quarks. From here, one can see that, in searching for extra $Z^\prime$-bosons with masses larger than $O$(3 TeV) the $|y_{l\bar{l}}|$ cut is not mandatory. The true direction of the quark is indeed correctly guessed more than 70\% of the times, even if no cut is applied on the di-lepton rapidity.
This means that, at high di-lepton invariant masses, we should be able to observe a lepton asymmetry with a well approximated shape even without imposing ad hoc cuts. The advantage would be twofold: preserving a small statistical error on that shape, owing to the much larger acceptance one should have in absence of the $|y_{l\bar{l}}|$ cut, and working with an event sample flavour independent up to a large extent. This latter feature would guarantee a more model independent procedure, as the different $Z'$ models have obviously different couplings of the extra gauge boson to up and down-quarks.

\begin{figure}[t]
\centering
\includegraphics[width=7.7cm]{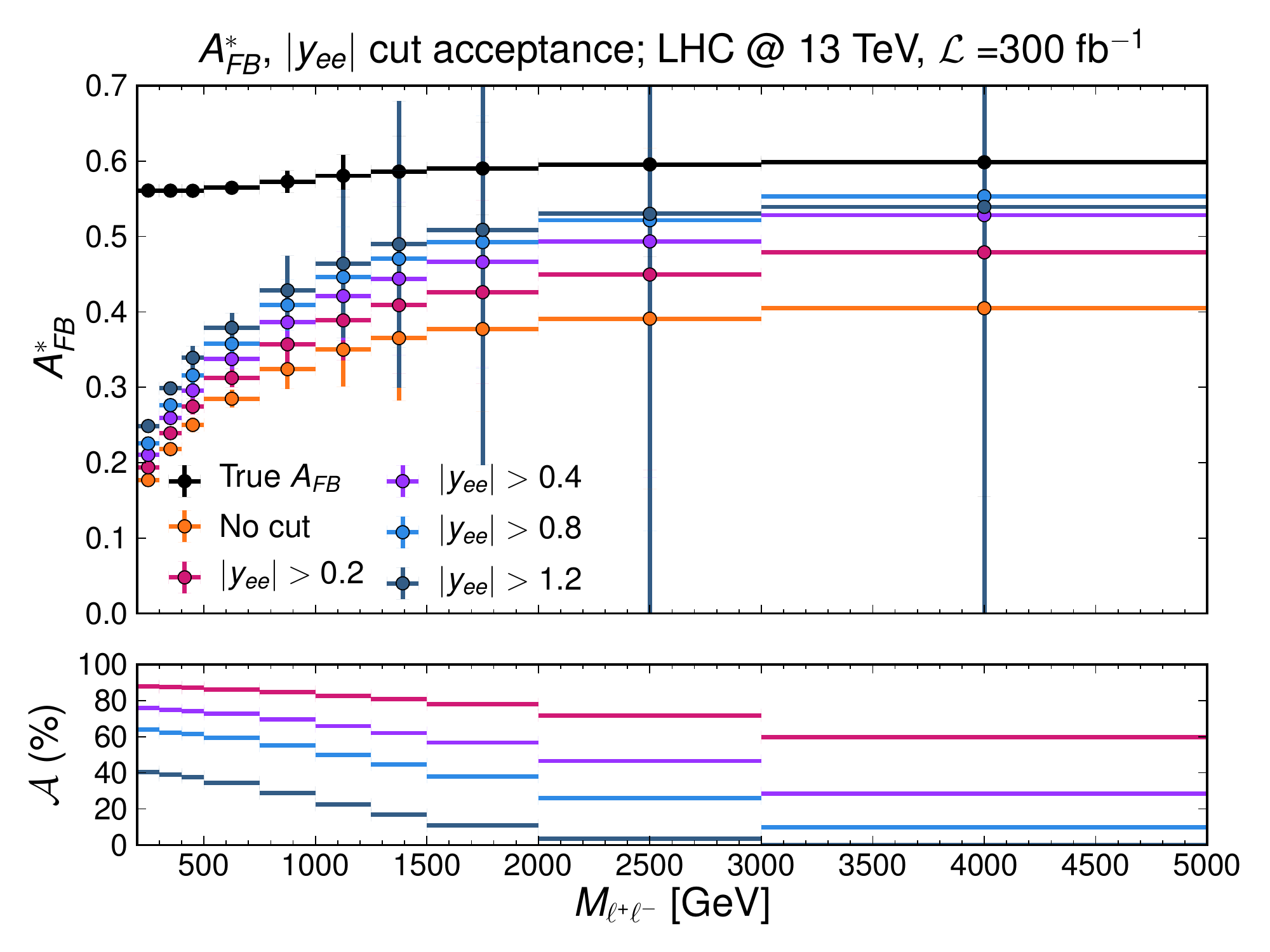}
\caption{(Colour online)
 Upper plot: Reconstructed forward-backward asymmetry as a function of the di-lepton invariant mass within the SM at the 13 TeV LHC with a total integrated luminosity $\mathcal{L}=300 fb^{-1}$ for a set of different rapidity cuts on the di-lepton system. In the legend, $|y_{ee}|$ corresponds to $|y_{l\bar{l}}|$ defined in the text. The black line represents the true AFB for comparison. Lower plot: Acceptance as a function of the di-lepton invariant mass for the same set of di-lepton rapidity cuts as above. }
\label{fig:shape_acceptance}
\end{figure}

In order to quantify the delicate balance between AFB line shape and statistical error, in the upper plot of Fig. 7 we show the shape of the reconstructed lepton asymmetry, $A_{FB}^*$, within the SM as a function of the di-lepton invariant mass for a set of different cuts on $|y_{l\bar{l}}|$. We compare the results to the true AFB, where the direction of the valence quark is taken directly from the Monte Carlo (MC) event generator. In the lower plot of Fig. 7, we display the acceptance as a function of the same variable $M_{l\bar l}$ for the same set of $|y_{l\bar{l}}|$ cuts. Comparing the two plots, one can see that $A_{FB}^*$ tends to the true AFB with increasing the $|y_{l\bar{l}}|$ cut, but at the same time the acceptance heavily decreases. In particular, for masses above 2.5 TeV, if we apply the stringent cut $|y_{l\bar{l}}|\ge 0.8$ used in literature, the number of events goes down by a factor of 3 while the gain in shape is only about 20\% of the true AFB value. With increasing mass, the acceptance decreases indeed more rapidly with the $|y_{l\bar{l}}|$ cut.

\begin{figure}[t]
\centering
\subfigure[]{
\includegraphics[width=7.cm]{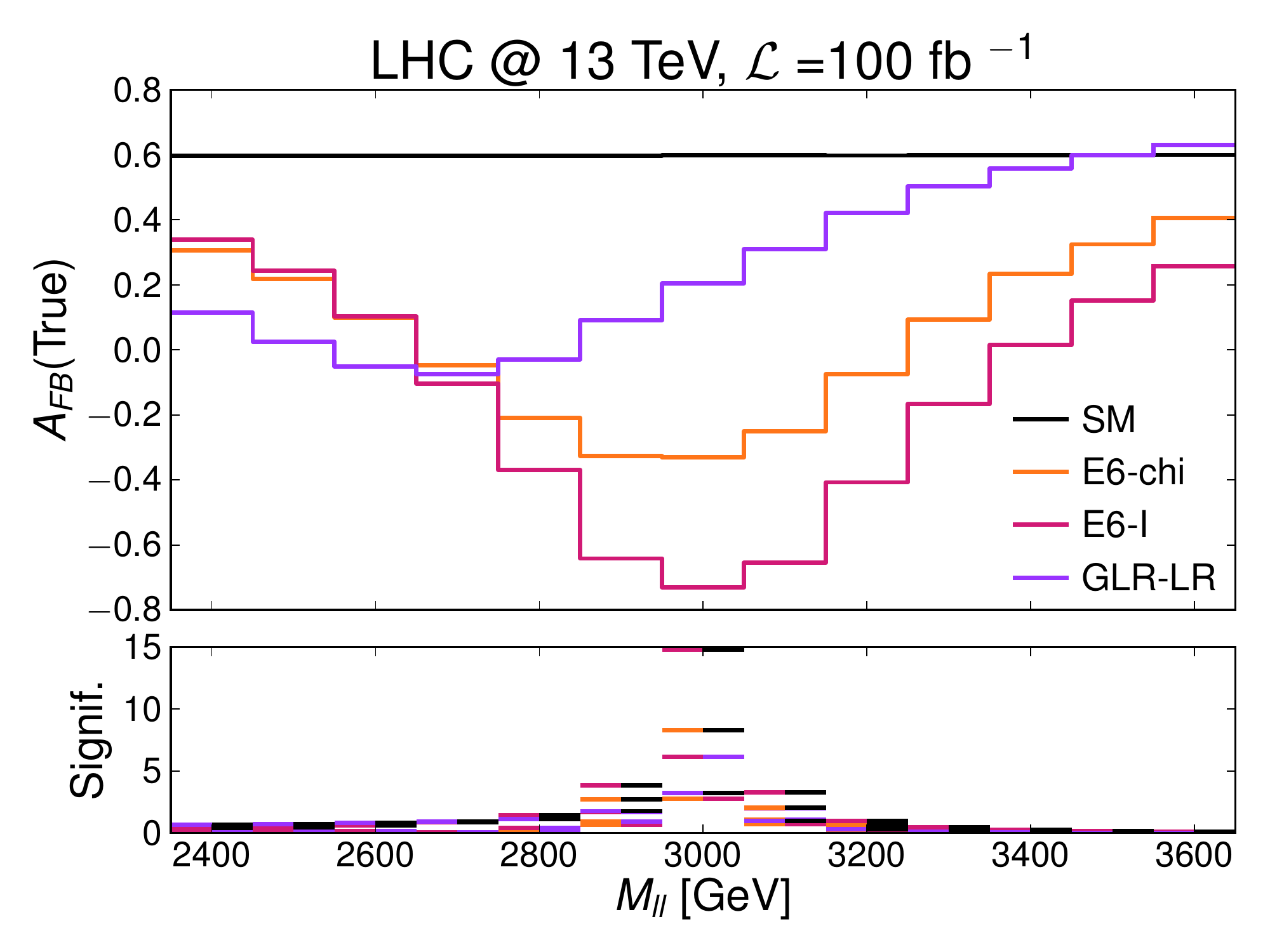}
\label{fig:true_afb}
}
\subfigure[]{
\includegraphics[width=7.cm]{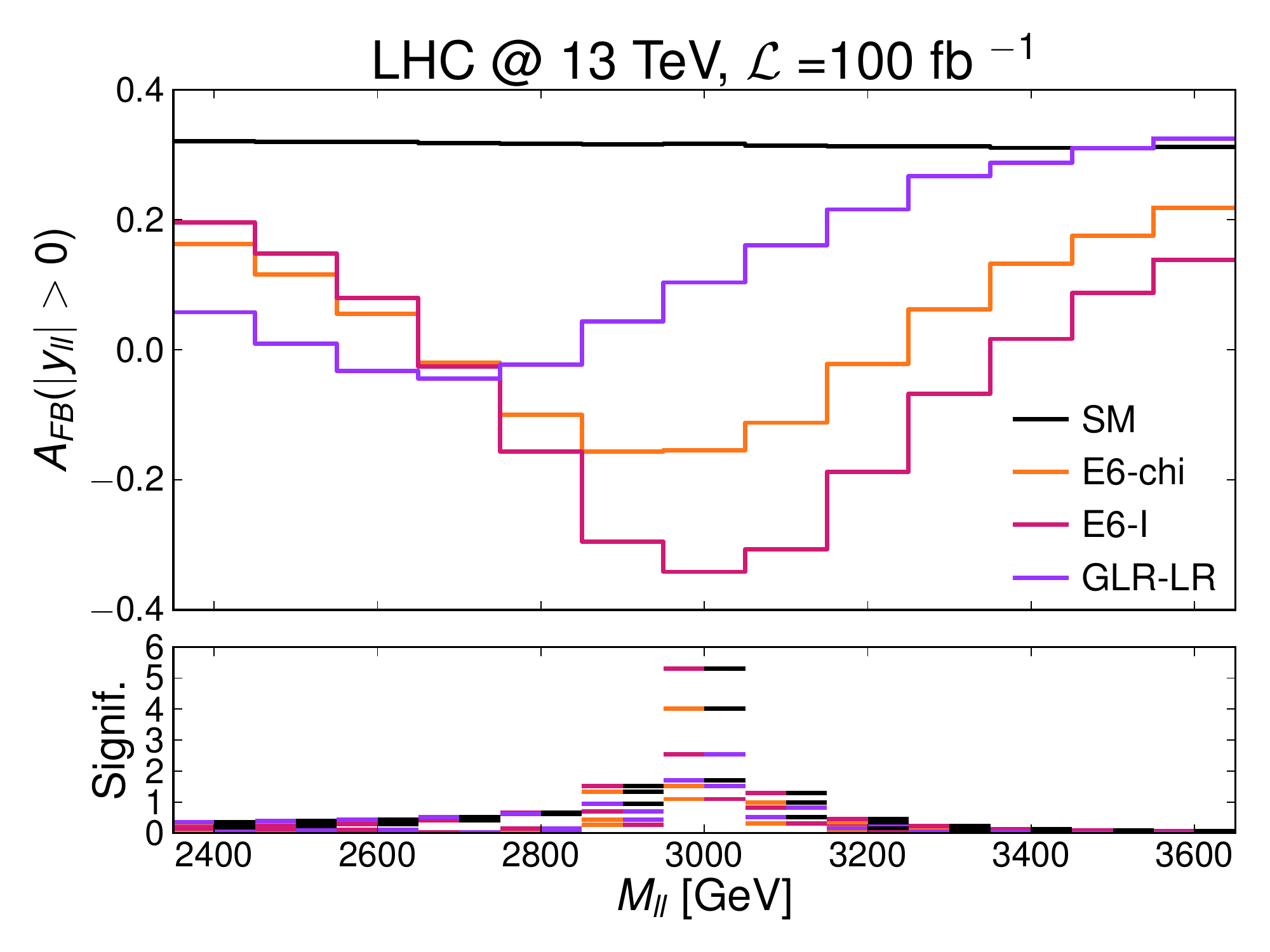}
\label{fig:yll0}
}
\subfigure[]{
\includegraphics[width=7.cm]{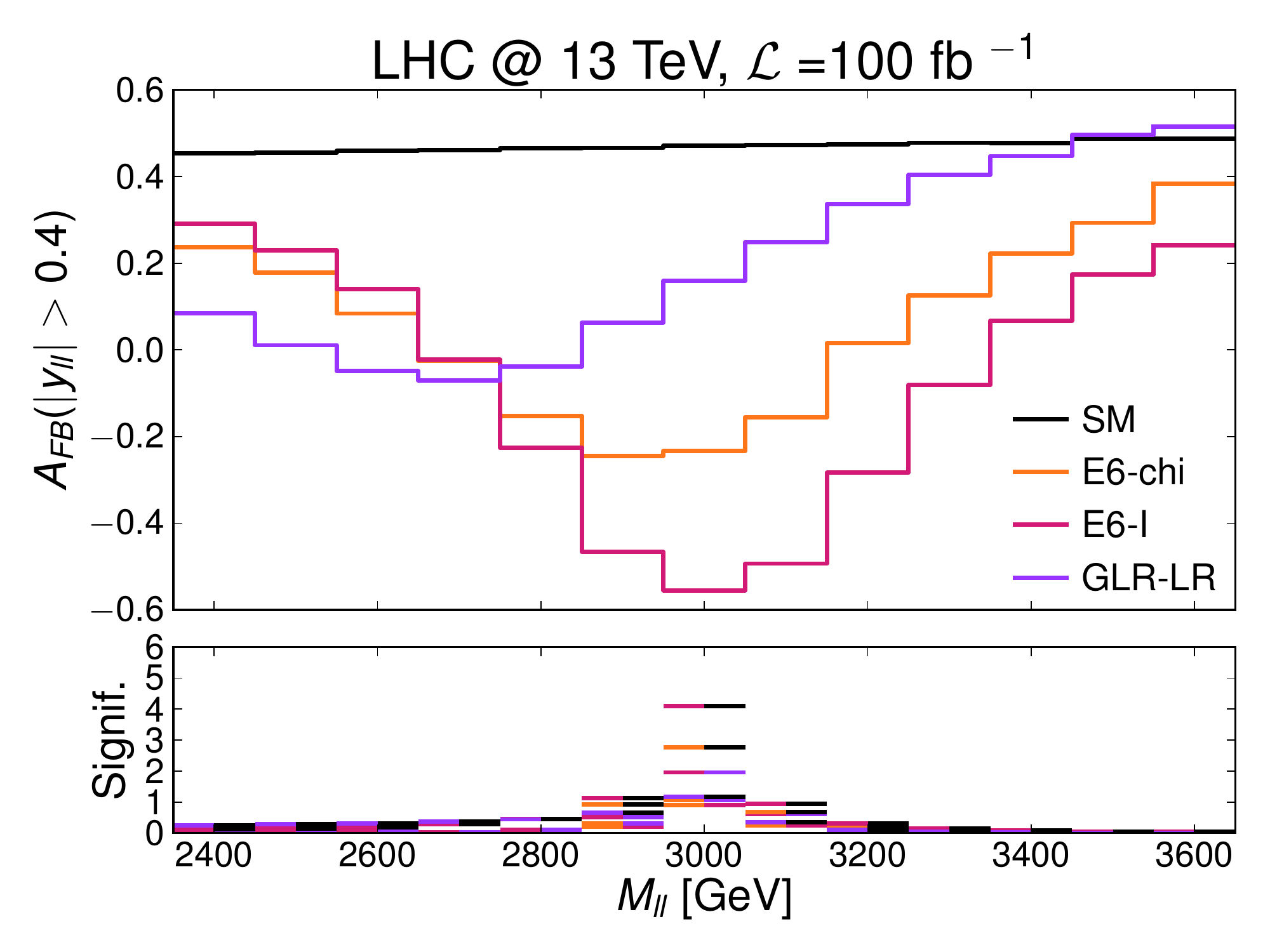}
\label{fig:yll04}
}
\subfigure[]{
\includegraphics[width=7.cm]{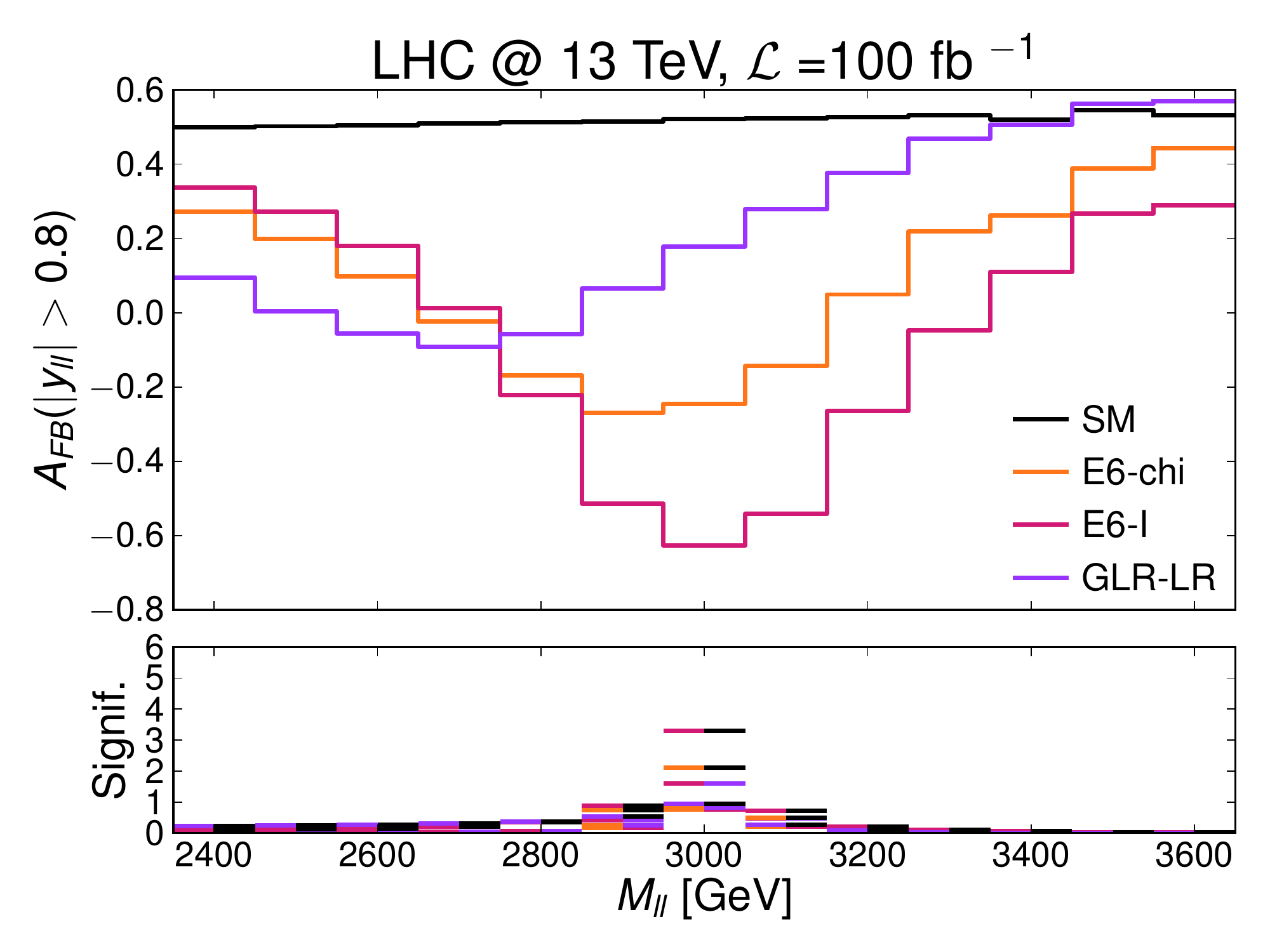}
\label{fig:yll08}
}
\caption{(Colour online)
\subref{fig:true_afb} True forward-backward asymmetry as a function of the di-lepton invariant mass as predicted by the SM (black), the $E_\chi$ (orange), the $E_I$ (magenta) and the $GLR-LR$ (purple) models for a $Z^\prime$-boson with mass $M_ {Z^\prime}$ = 3 TeV. The results are for the LHC at $\sqrt{s}=13$ TeV and $\mathcal{L}=100 fb^{-1}$. Lower plot: the significance in distinguishing models is displayed. The double colour in each bin visualizes the two compared models.
\subref{fig:yll0} Reconstructed forward-backward asymmetry as a function of the di-lepton invariant mass as predicted by the SM (black), the $E_\chi$ (orange), the $E_I$ (magenta) and the $GLR-LR$ (purple) models for a $Z^\prime$-boson with mass $M_ {Z^\prime}$ = 3 TeV. The results are for the LHC at $\sqrt{s}=13$ TeV and $\mathcal{L}=100 fb^{-1}$. No cut on the di-lepton rapidity is imposed: $|y_{l\bar{l}}|\ge 0$. Lower plot: the significance in distinguishing models is displayed.
\subref{fig:yll04} Same as plot (b) with $|y_{l\bar{l}}|\ge 0.4$.
\subref{fig:yll08} Same as plot (b) with $|y_{l\bar{l}}|\ge 0.8$.
}
\label{fig:modelsandcut}
\end{figure}

To visualize how the above features impact the AFB sensitivity to new physics, in Fig. 8 we compare the reconstructed $A_{FB}^*$ observable predicted by three representative $Z^\prime$-models ($E_\chi$, $E_I$ and $GLR-LR$) with the SM expectation at the 13 TeV LHC with total integrated luminosity $\mathcal{L}= 100 fb^{-1}$. As a new physics signal, we consider a hypothetical $Z^\prime$-boson with mass $M_ {Z^\prime}$ = 3 TeV. To quantify the effect of the di-lepton rapidity cut on the significance either in searching for new physics via AFB or in distinguishing between different $Z'$ models, we show results for the commonly used $|y_{l\bar{l}}|\ge 0.8$ setup (Fig. \ref{fig:yll08}) versus the $|y_{l\bar{l}}|\ge 0.4$ (Fig. \ref{fig:yll04}) and no cut (Fig. \ref{fig:yll0}) scenarios. We further display, in Fig. \ref{fig:true_afb}, the ideal situation represented by the true forward-backward asymmetry, $A_{FB}$.

As one can see, imposing a strong di-lepton rapidity cut helps in recovering the true shape and magnitude of the forward-backward asymmetry. However, the consequent decrease of the number of events is so substantial that the significance diminishes drastically with 
increasing the $|y_{l\bar{l}}|$ cut. Moreover, the implementation of the $|y_{l\bar{l}}|$ cut accentuates the flavour dependence of the results or, in other words, the model dependence of the analysis. As the probability of guessing the correct direction of the quark in the reconstruction procedure of the AFB as a function of the $|y_{l\bar{l}}|$ cut depends on the type of quark (up and down-quarks react differently to the cut), the reconstructed AFB shows an increased model dependence in its response to the $|y_{l\bar{l}}|$ cut. To exemplify this concept, let us take the third bin from the left in Fig. \ref{fig:modelsandcut}. There (Fig. \ref{fig:true_afb}), the $E_\chi$ and $E_I$ models are degenerate as far as the true asymmetry is considered. When we compare the reconstructed asymmetry, we see that the two models are not degenerate any more in that bin. The splitting increases with the $|y_{l\bar{l}}|$ cut, as the two models react differently to such a cut, having different couplings of the corresponding $Z'$-boson to up-  and down-quarks. In order to minimize the presence of model dependent elements in the analysis, it is thus advisable not to include the di-lepton rapidity cut. Hence, in the following. we will be working in a setup where we do not impose such a restriction.

\section{The role of AFB in Z' searches: narrow heavy resonances}
\label{sec:narrowZ}
 
The AFB is the observable where the effects of the interference between new physics and SM background are maximal. In the  Drell-Yan processes, these effects are of course present also in the total cross section. They are readily seen  in both cases in the di-lepton invariant mass. 
As mentioned repeatedly, constraining the search window for new physics within the interval $|M_{l\bar l}-M_{Z^\prime}|\le 0.05\times E_{\rm{LHC}}$ guarantees that finite width and interference effects are below the $O(10\% )$ level when compared to the complete new physics signal. Such effects are instead an intrinsic part of the AFB and dominate its dynamics. For such a reason, the AFB is an intrinsically model dependent variable and in literature has therefore been traditionally considered for disentangling different models predicting a spin-1 heavy neutral particle. Its role has therefore been cornered so far to the interpretation of a possible $Z^\prime$-boson discovery obtained via the default bump search. 

In this paper, we aim to show that AFB can also be used for searches, directly, as a primary variable alongside the cross section itself.
In this section we focus on $Z^\prime$-bosons characterized by a narrow width. This is the most common kind of particle predicted by theories with an extra $U^\prime (1)$ gauge group.
This is also the scenario mostly studied in literature. 
The experimental searches for such an object
are tailored on this expectation and the corresponding results coming from the data collected at the 8 TeV LHC have been summarized in Section II. With respect to the `AFB search', the $Z^\prime$ models can be divided into two 
categories: $Z^\prime$ models with AFB centred on the $Z^\prime$-boson mass 
and $Z^\prime$ models with shifted AFB. In the next two subsections, we discuss their properties in turn.

\subsection{Z' models with AFB centred on peak}

In this subsection, we discuss models where the AFB is peaked on the $Z'$-boson mass. These models belong to the $E_6$ class of theories which predict new narrow width spin-1 resonances. In the literature, it is known that such models contain one extra neutral gauge boson whose width cannot exceed a few percent of its mass: $\Gamma_{Z'}/M_{Z'}\le 5\%$. Even the inclusion of new $Z'$-boson decay channels into exotic states would not change this estimate. 

\begin{figure}[t]
\centering
\subfigure[]{
\includegraphics[width=7.cm]{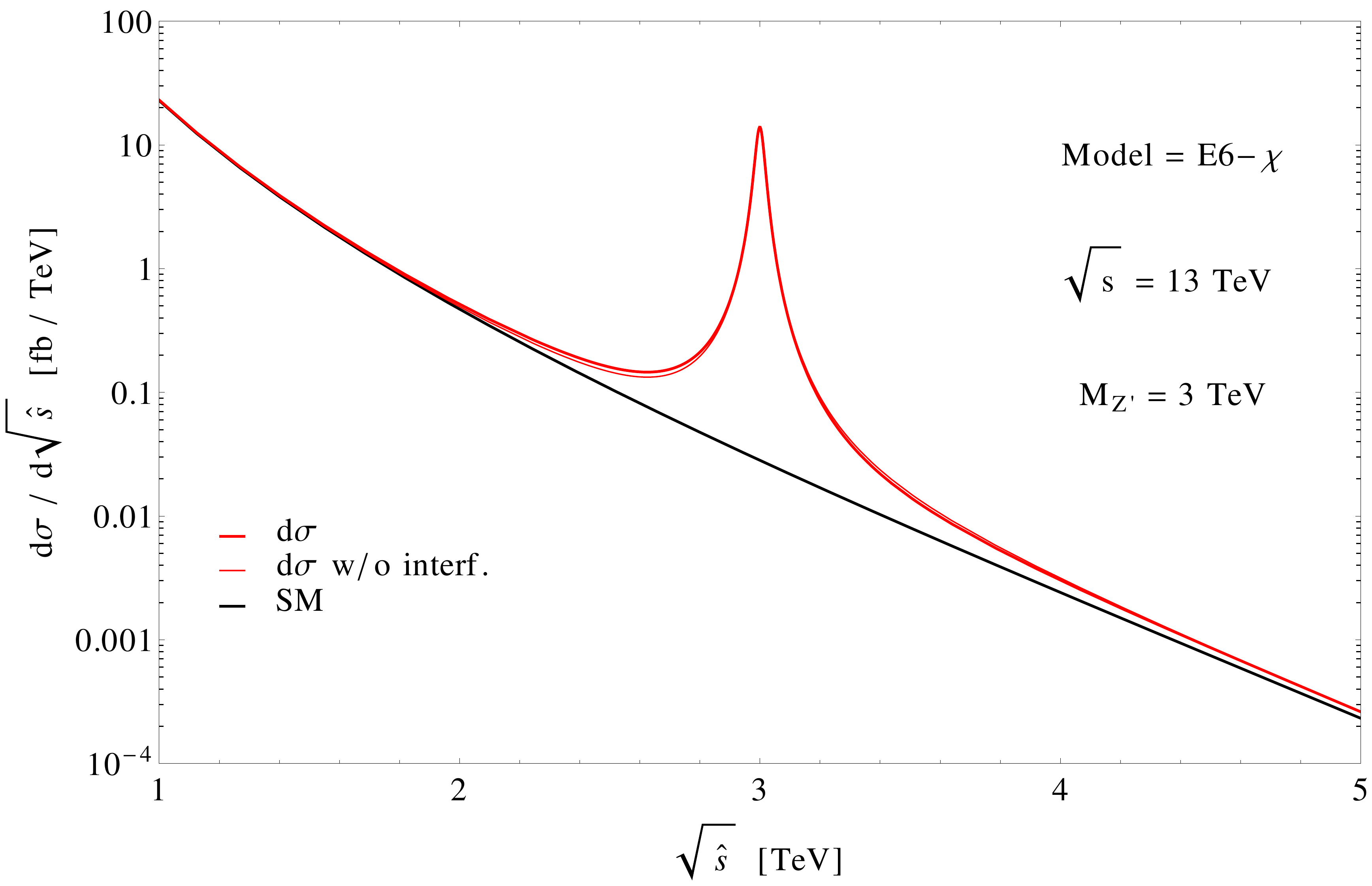}
\label{fig:sigma_chi_analytic}
}
\subfigure[]{
\includegraphics[width=7.cm]{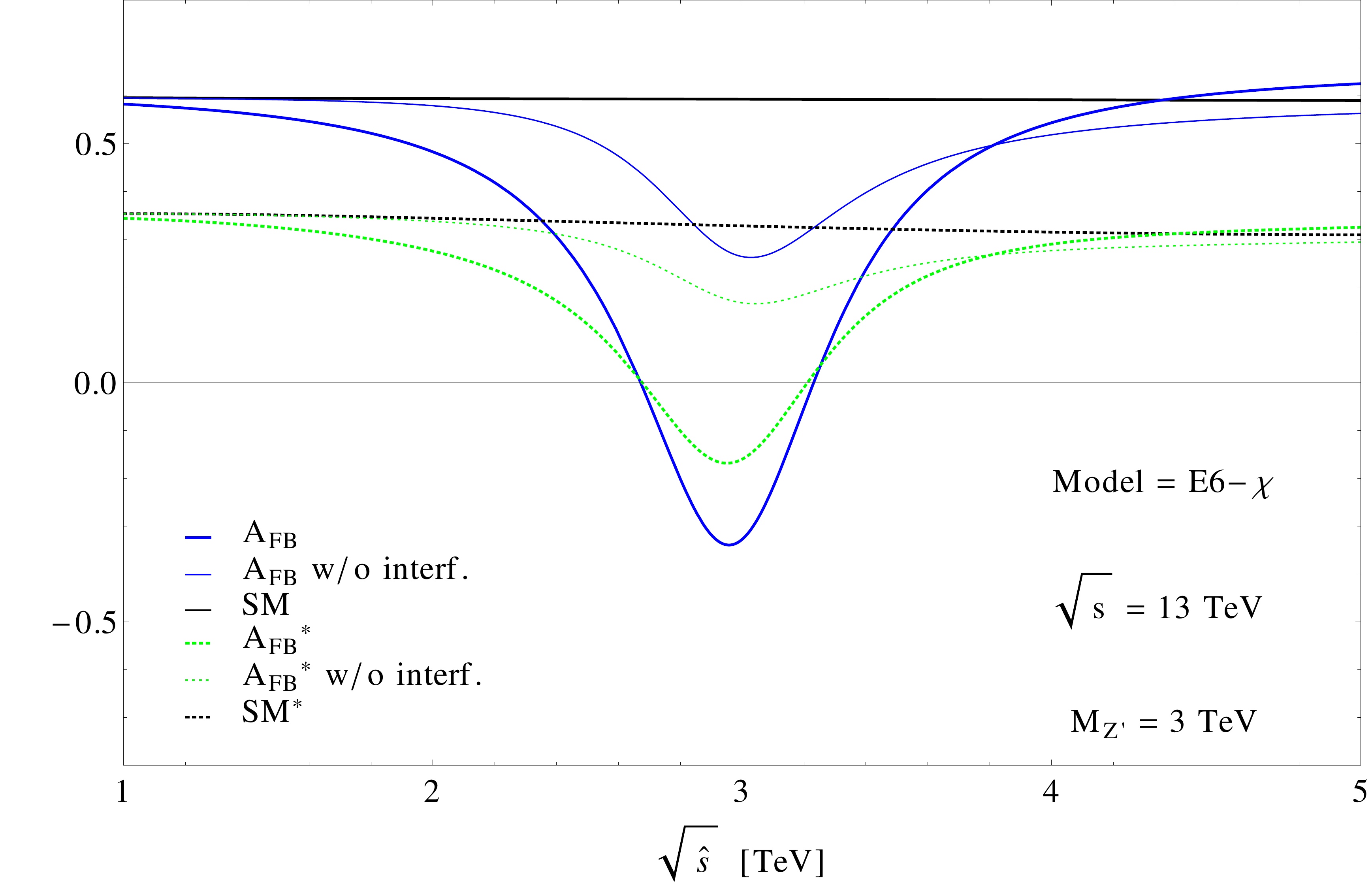}
\label{fig:afb_chi_analytic}
}
\subfigure[]{
\includegraphics[width=7.cm]{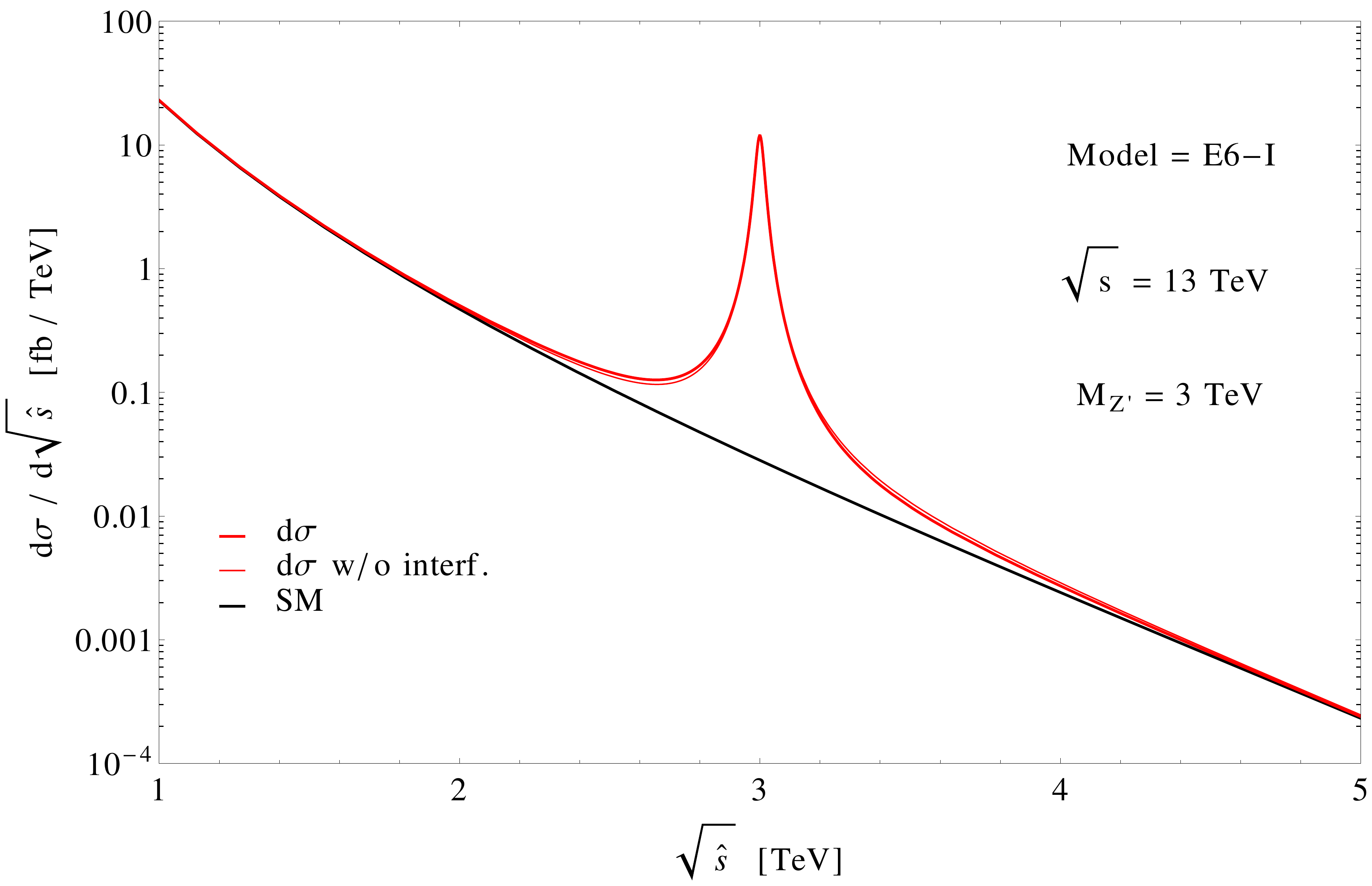}
\label{fig:sigma_I_analytic}
}
\subfigure[]{
\includegraphics[width=7.cm]{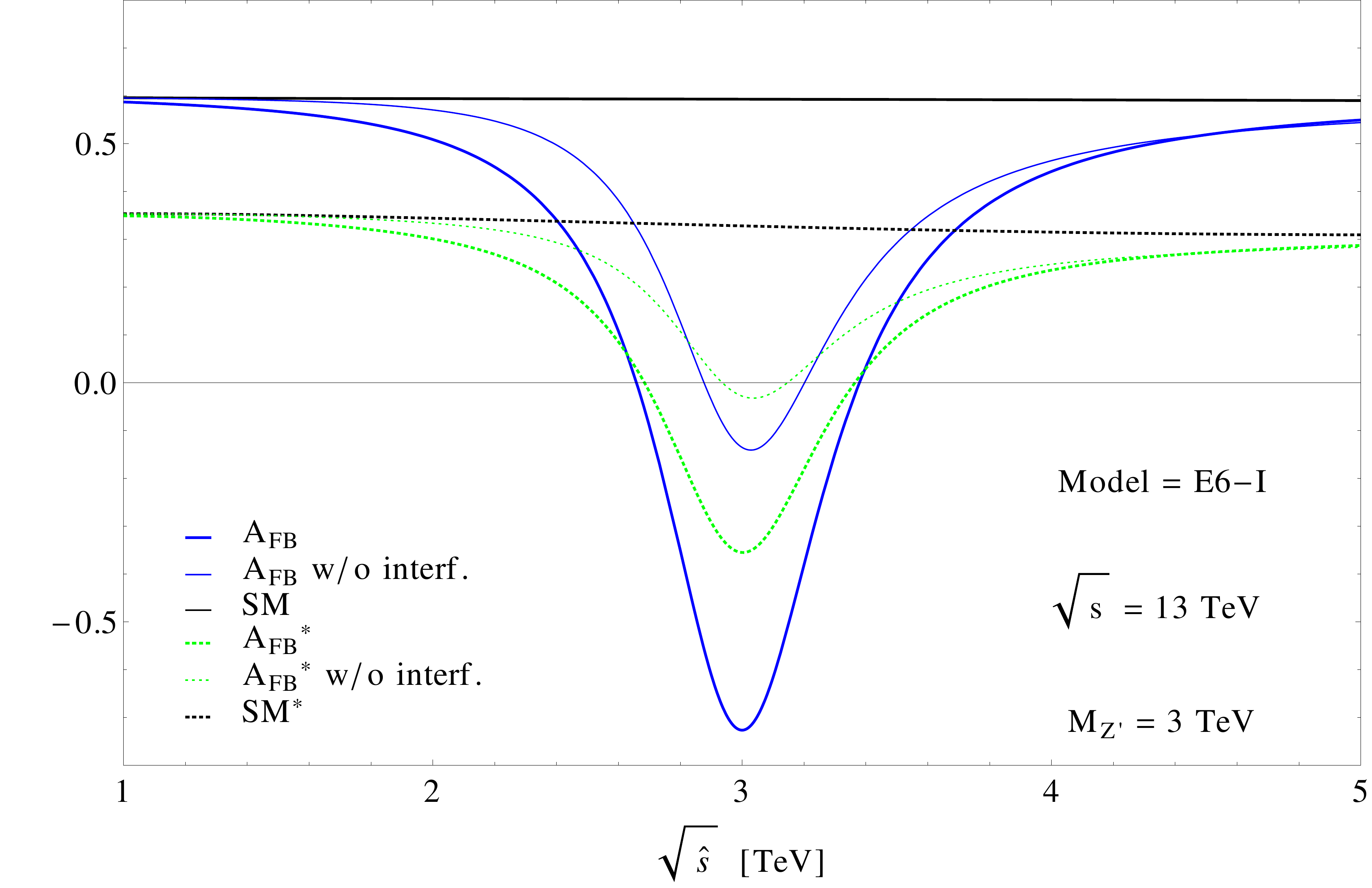}
\label{fig:afb_I_analytic}
}
\caption{(Colour online)
\subref{fig:sigma_chi_analytic} Differential cross section as a function of the di-lepton invariant mass as predicted by the $E_\chi$ model for a $Z^\prime$-boson with mass $M_ {Z^\prime}$ = 3 TeV. The results are for the LHC at $\sqrt{s}$=13 TeV. 
\subref{fig:afb_chi_analytic} Reconstructed forward-backward asymmetry as a function of the di-lepton invariant mass as predicted by the $E_\chi$ model for a $Z^\prime$-boson with mass $M_ {Z^\prime}$ = 3 TeV. The results are for the LHC at $\sqrt{s}$=13 TeV. No cut on the di-lepton rapidity is imposed: $|y_{l\bar{l}}|\ge 0$. 
\subref{fig:sigma_I_analytic} Same as plot (a) within the $E_I$ model.
\subref{fig:afb_I_analytic} Same as plot (b) within the $E_I$ model.
}
\label{fig:E6models}
\end{figure}

\begin{figure}[t]
\centering
\subfigure[]{
\includegraphics[width=7.7cm]{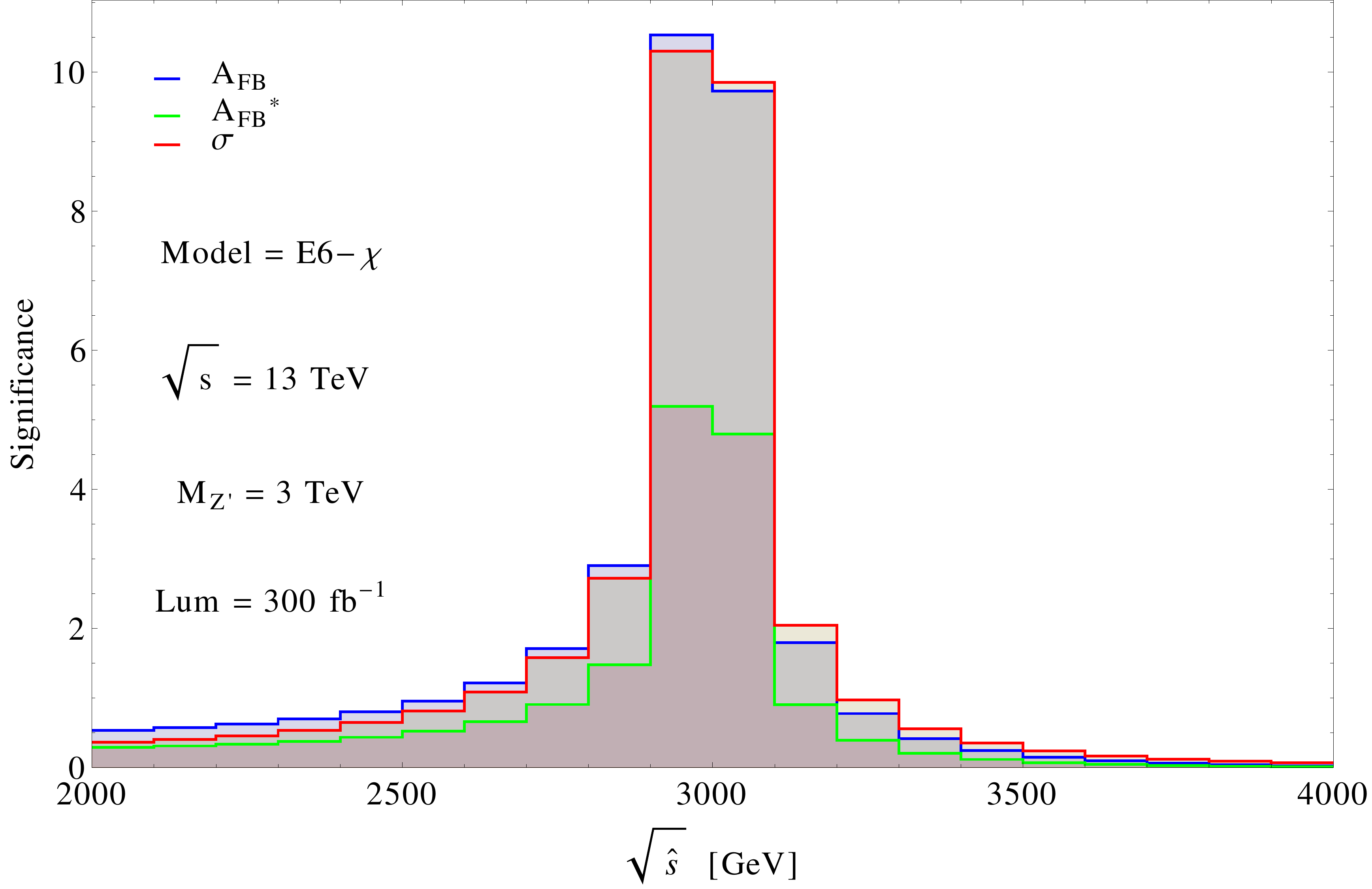}
\label{fig:significance_chi}
}
\subfigure[]{
\includegraphics[width=7.7cm]{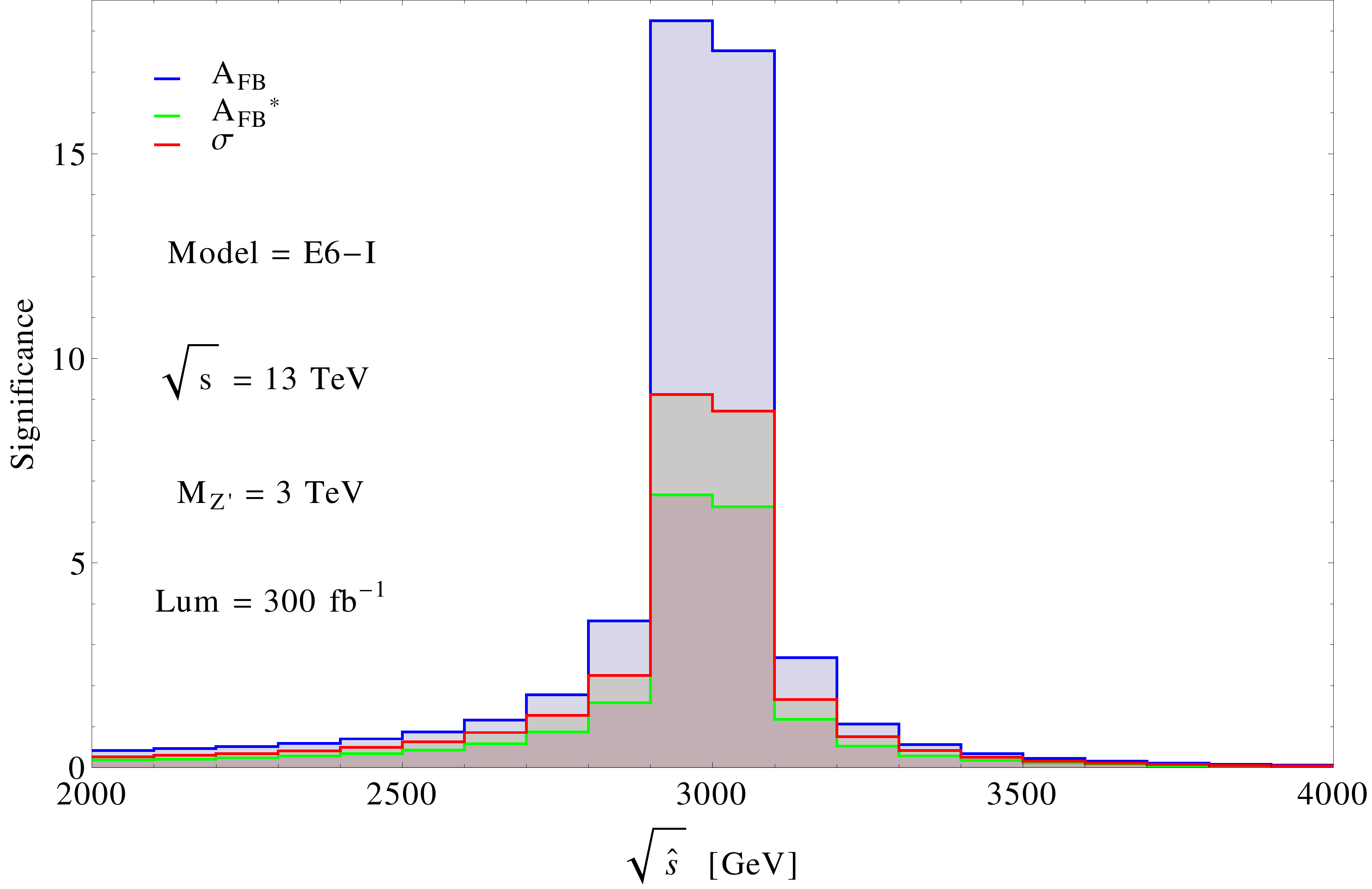}
\label{fig:significance_I}
}
\caption{(Colour online)
\subref{fig:significance_chi} Binned significance as a function of the di-lepton invariant mass as predicted by the $E_\chi$ model for a $Z^\prime$-boson with mass $M_ {Z^\prime}$ = 3 TeV.
The red line represents the significance corresponding to the invariant mass distribution. The blue and green lines show the significance extracted by an ideal measurement of true and reconstructed AFB, respectively. The results are for the LHC at $\sqrt{s}=13$ TeV and $\mathcal{L}=300 fb^{-1}$. 
\subref{fig:significance_I} Same as (a) for the $E_I$ model.
}
\label{fig:E6models_analyticsignificance}
\end{figure}

We first compare the shape of the AFB distribution as a function of the di-lepton invariant mass, $M_{l\bar l}$, with the differential cross section in the same variable. In Fig. 9, we show results for two representative $E_6$ models: $E_\chi$ and $E_I$. We consider a hypothetical $Z'$-boson with mass $M_{Z'}=3$ TeV. In Figs. \ref{fig:afb_chi_analytic} and \ref{fig:afb_I_analytic}, we display both the true and the reconstructed AFB within the two chosen $E_6$ models with and without taking into account the interference between the extra $Z'$-boson and the SM background. As one can see, the role played by the interference is extremely important. The AFB shape is drastically modified by getting its peak heavily accentuated. In contrast, the invariant mass distribution is almost interference free if the $|M_{l\bar l}-M_{Z^\prime}|\le 0.05\times E_{\rm{LHC}}$ cut is imposed, as shown in Figs. \ref{fig:sigma_chi_analytic} and \ref{fig:sigma_I_analytic} for the two representative $E_6$ models. In interpreting the experimental data coming from AFB measurements it is then mandatory to include the interference, no matter what kinematical cut is applied.

\begin{figure}[t]
\centering
\subfigure[]{
\includegraphics[width=7.2cm]{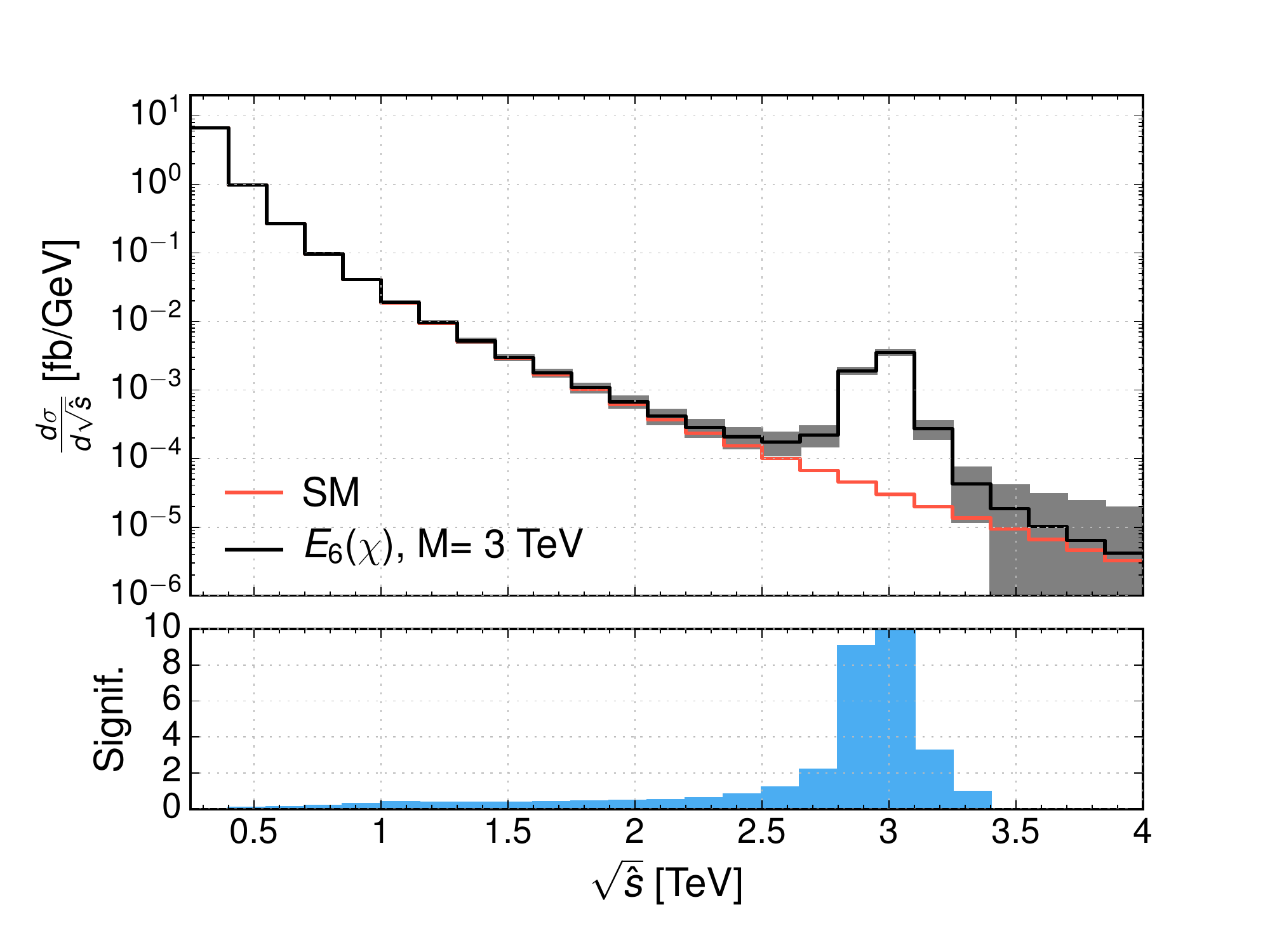}
\label{fig:sigma_chi_realistic300}
}
\subfigure[]{
\includegraphics[width=7.2cm]{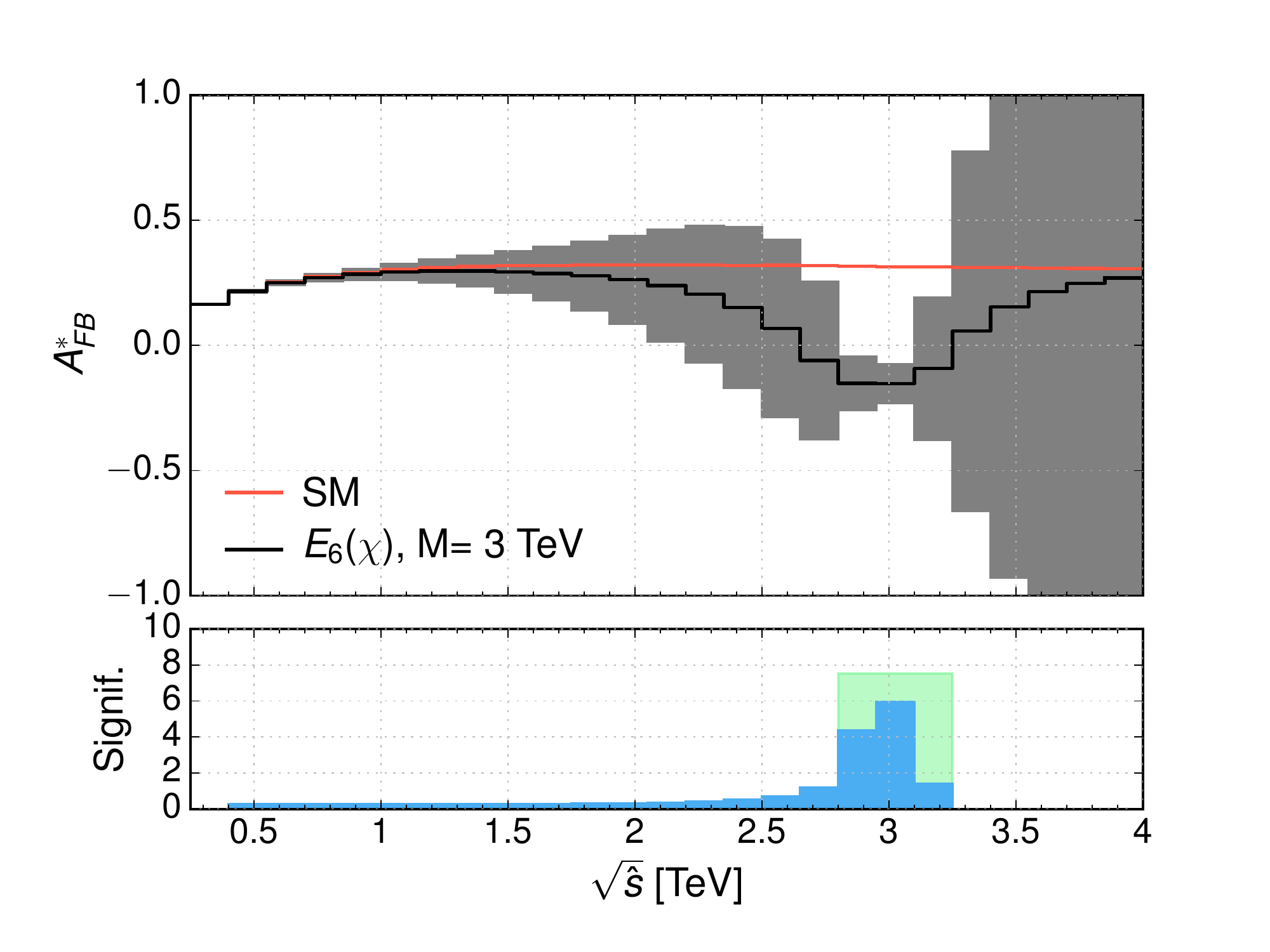}
\label{fig:AFB_chi_realistic300}
}
\subfigure[]{
\includegraphics[width=7.2cm]{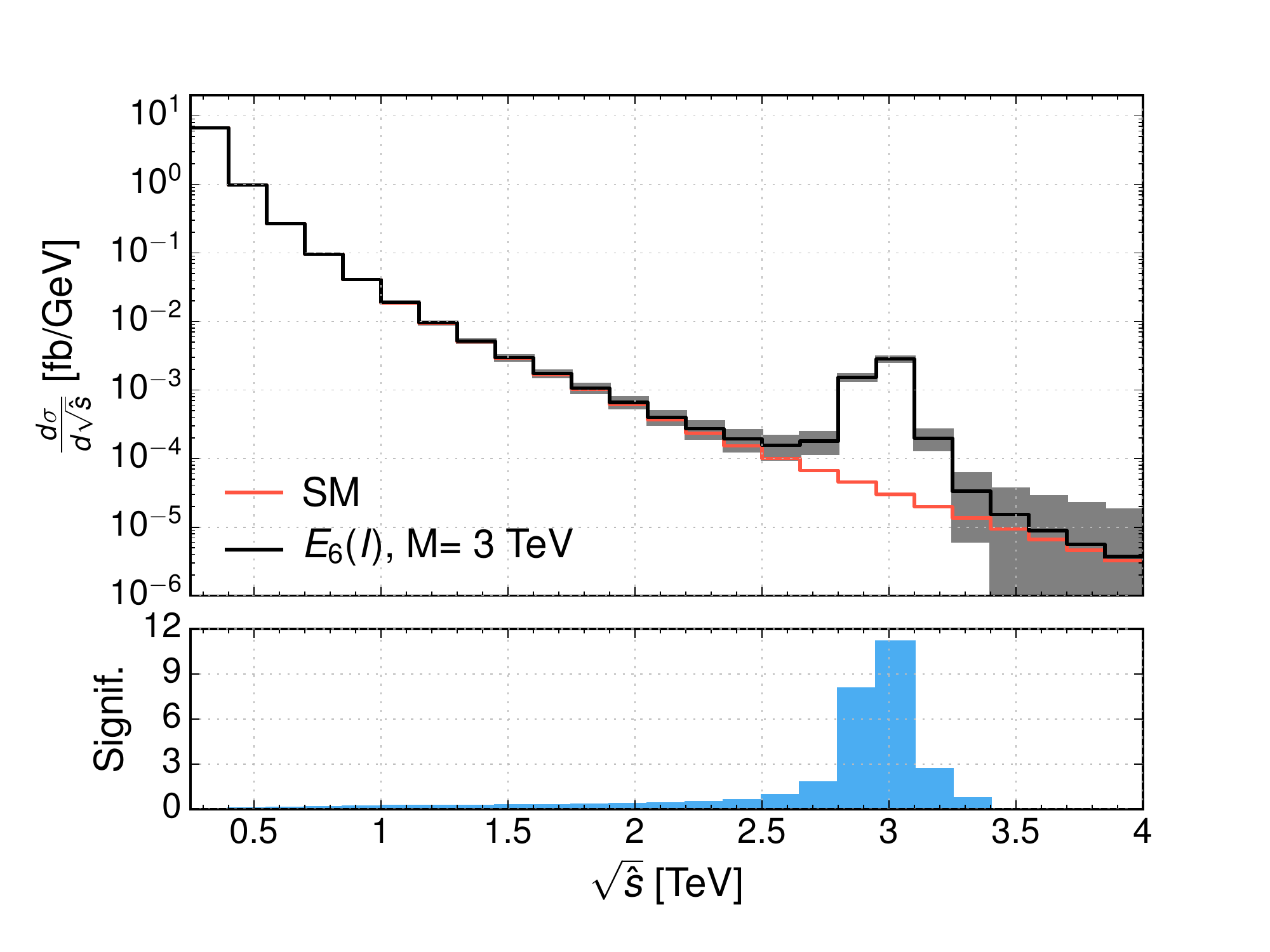}
\label{fig:sigma_I_realistic300}
}
\subfigure[]{
\includegraphics[width=7.2cm]{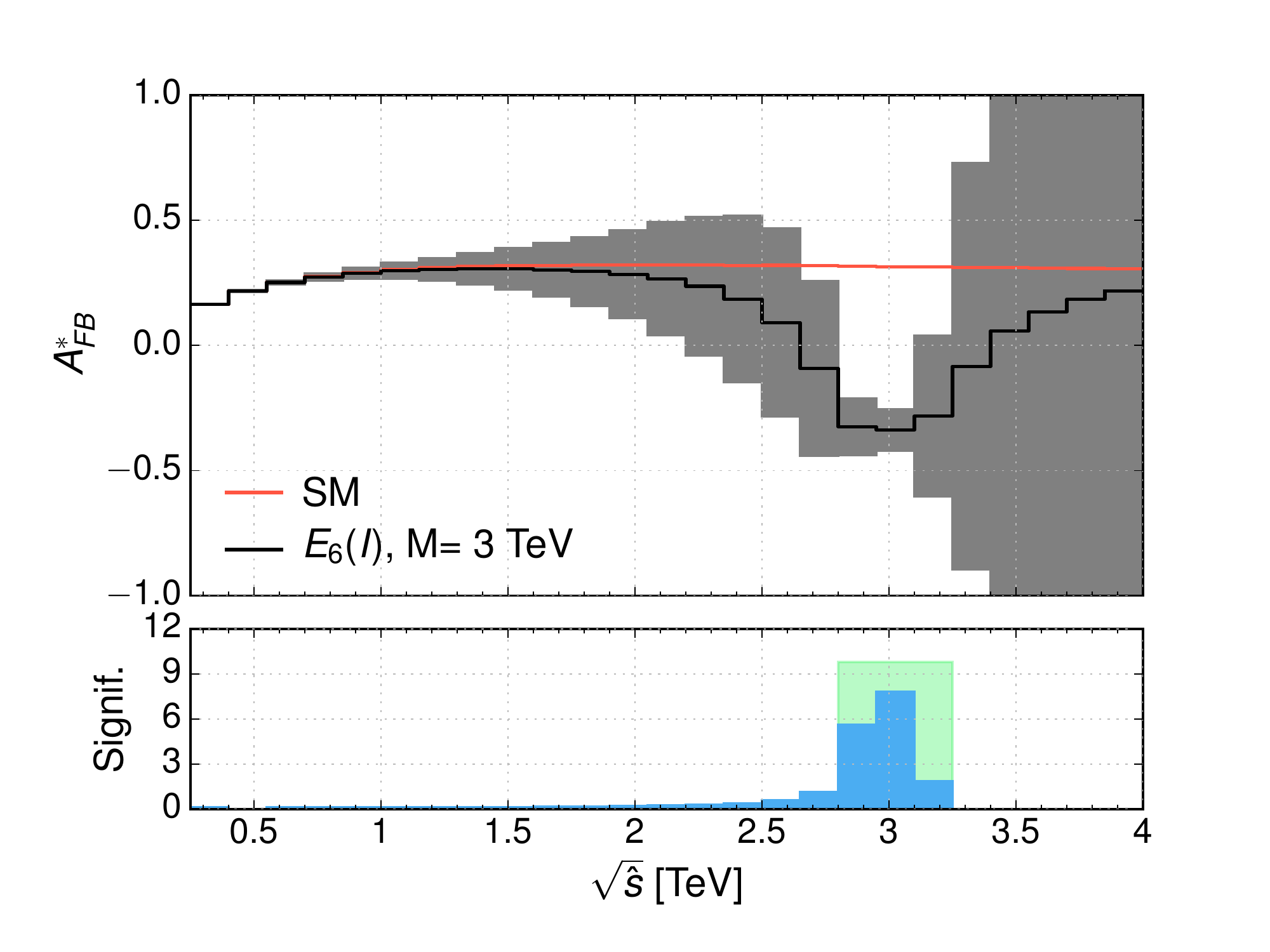}
\label{fig:AFB_I_realistic300}
}
\caption{(Colour online)
\subref{fig:sigma_chi_realistic300} Binned differential cross section as a function of the di-lepton invariant mass within the $E_\chi$ model for a $Z^\prime$-boson with mass $M_ {Z^\prime}$ = 3 TeV. Error bars are included. The results are for the LHC at $\sqrt{s}$=13 TeV and $\mathcal{L}=300 fb^{-1}$. Acceptance cuts are imposed (see text). The lower plot shows the  significance.
\subref{fig:AFB_chi_realistic300} Binned $A_{FB}^*$ as a function of the di-lepton invariant mass within the $E_\chi$ model for a $Z^\prime$-boson with mass $M_ {Z^\prime}$ = 3 TeV.
Error bars are included. The results are for the LHC at $\sqrt{s}=13$ TeV and $\mathcal{L}=300 fb^{-1}$. Acceptance cuts are imposed (see text). In the lower plot, the blue histogram shows the binned  significance while the green area indicates the total significance integrated over that invariant mass region.
\subref{fig:sigma_I_realistic300} Same as (a) for the $E_I$ model.
\subref{fig:AFB_I_realistic300} Same as (b) for the $E_I$ model.
}
\label{fig:E6models_realistic}
\end{figure}

In terms of significance, the search for narrow width $Z'$ models with AFB centred on the $Z'$ mass is summarized in Fig. 10 for the two representative models $E_\chi$ and $E_I$. Within the $E_\chi$ model, the true AFB would give rise to a significance slightly lower than that one coming from the usual bump search, as shown in Fig. 10a. The reconstruction procedure of the AFB depletes this result but still the two significances from cross section and $A_{FB}^*$ are comparable over the full di-lepton invariant mass range . Fig. 10b shows that the $E_I$ model is more accessible through the AFB than the cross section. There, indeed, the significance from the true AFB is a factor two bigger than the significance coming from the bump search. Once again, the AFB reconstruction pollutes the ideal result. The significance from the reconstructed AFB gets reduced, but its value remains anyhow only slightly lower than that one coming from the resonance search. The $E_I$ model is not unique in this respect, also the $E_S$ model shares the same property.

\begin{figure}[t]
\centering
\subfigure[]{
\includegraphics[width=7.2cm]{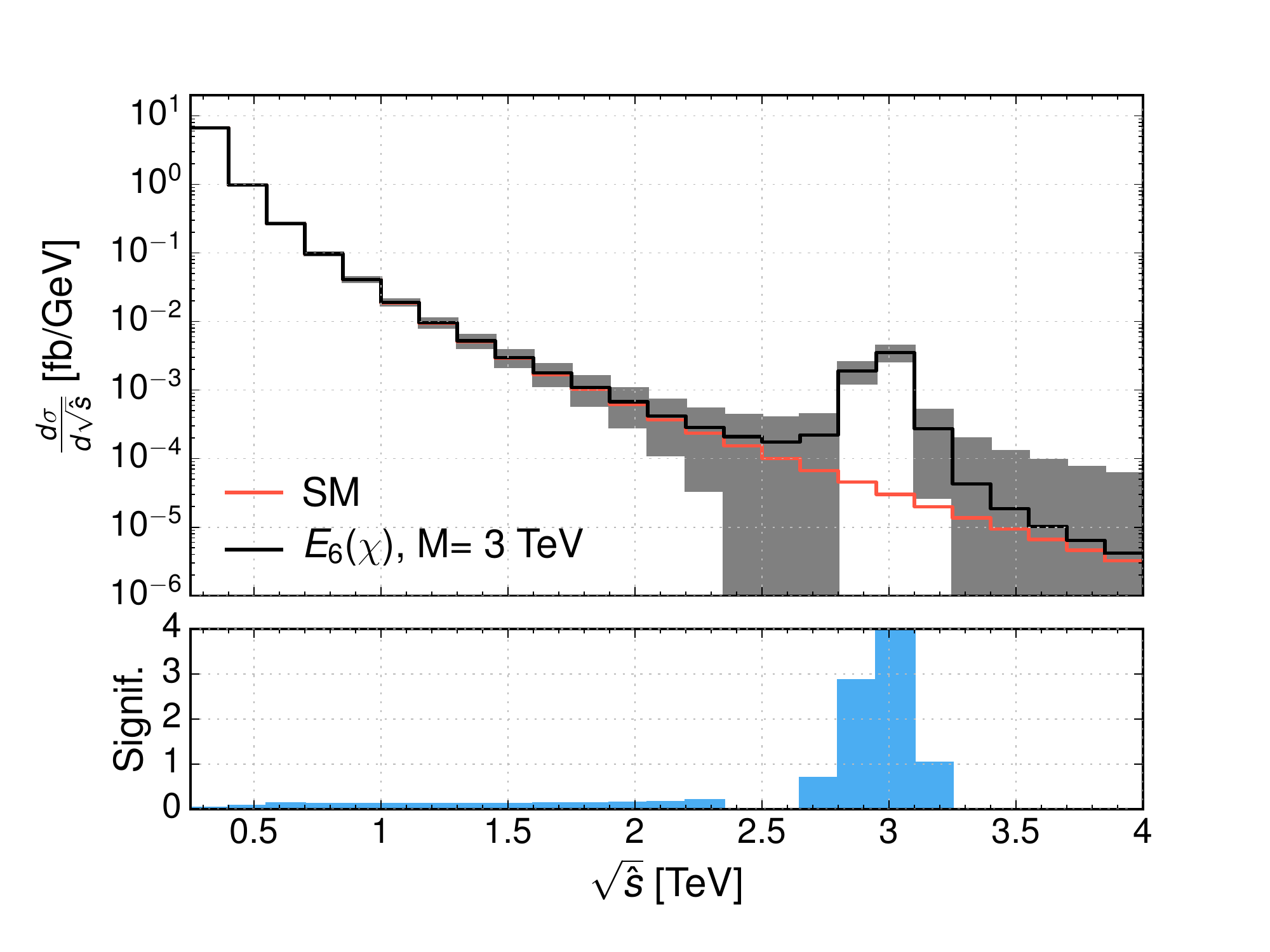}
\label{fig:sigma_chi_realistic30}
}
\subfigure[]{
\includegraphics[width=7.2cm]{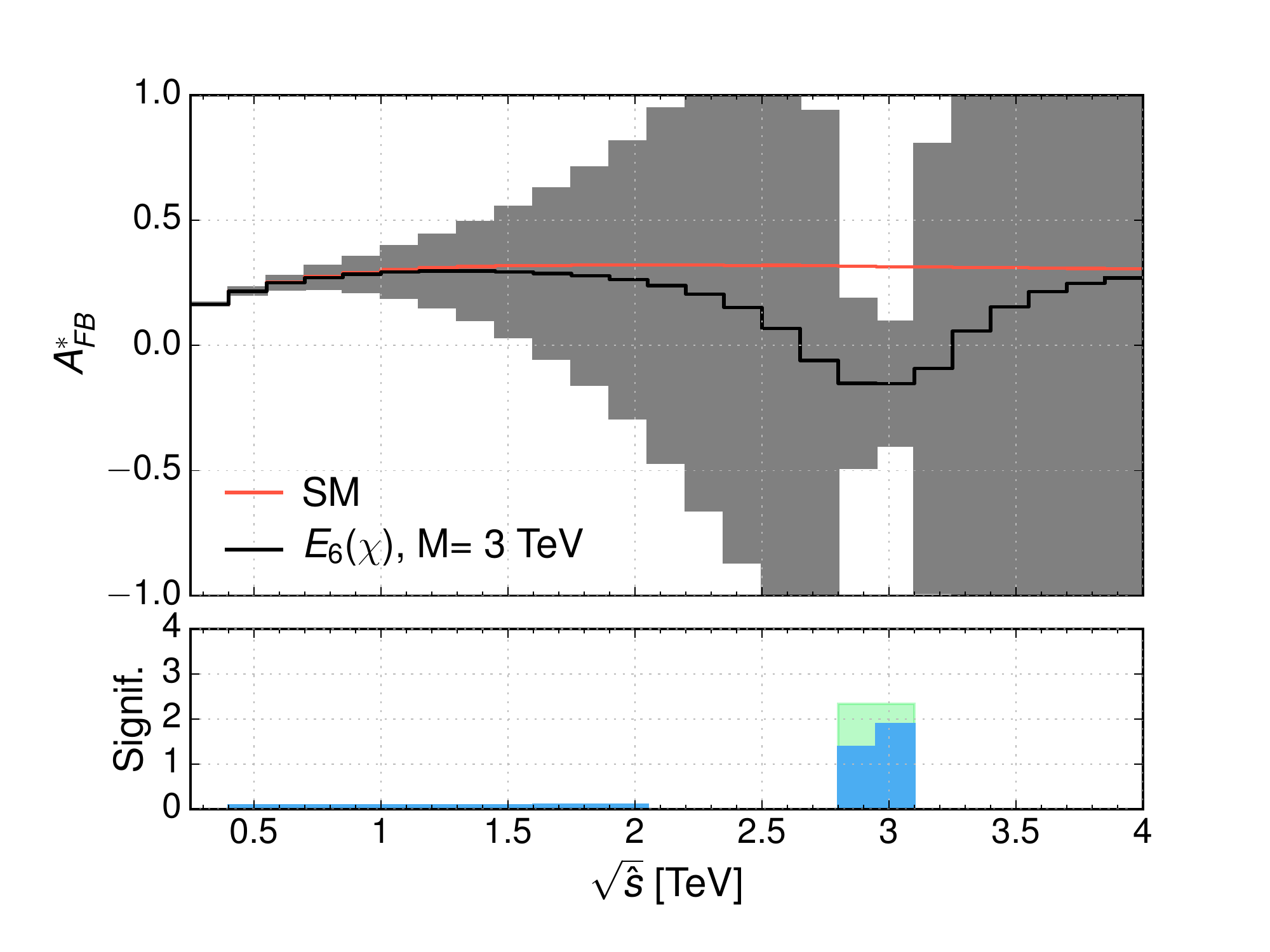}
\label{fig:AFB_chi_realistic30}
}
\subfigure[]{
\includegraphics[width=7.2cm]{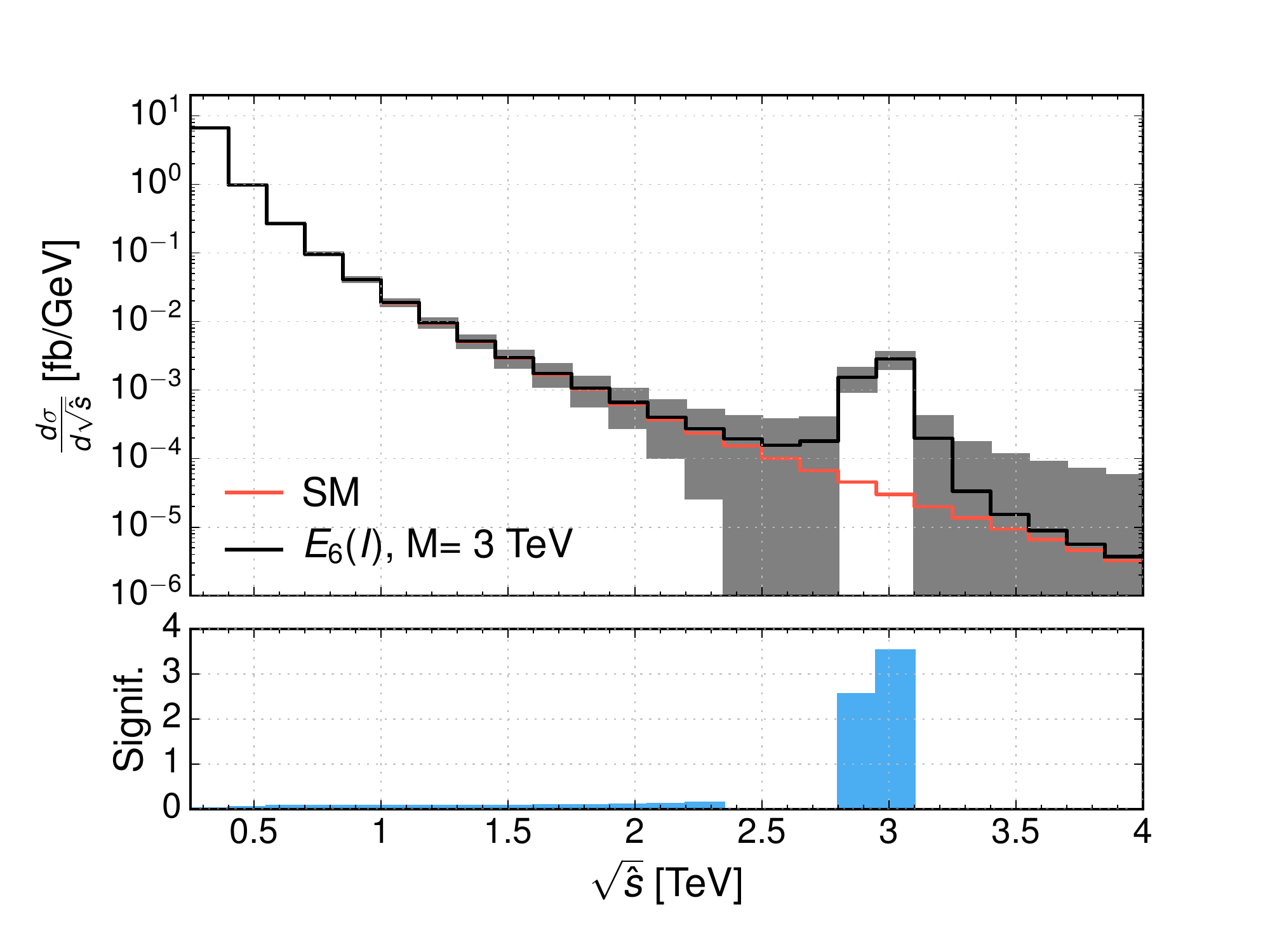}
\label{fig:sigma_I_realistic30}
}
\subfigure[]{
\includegraphics[width=7.2cm]{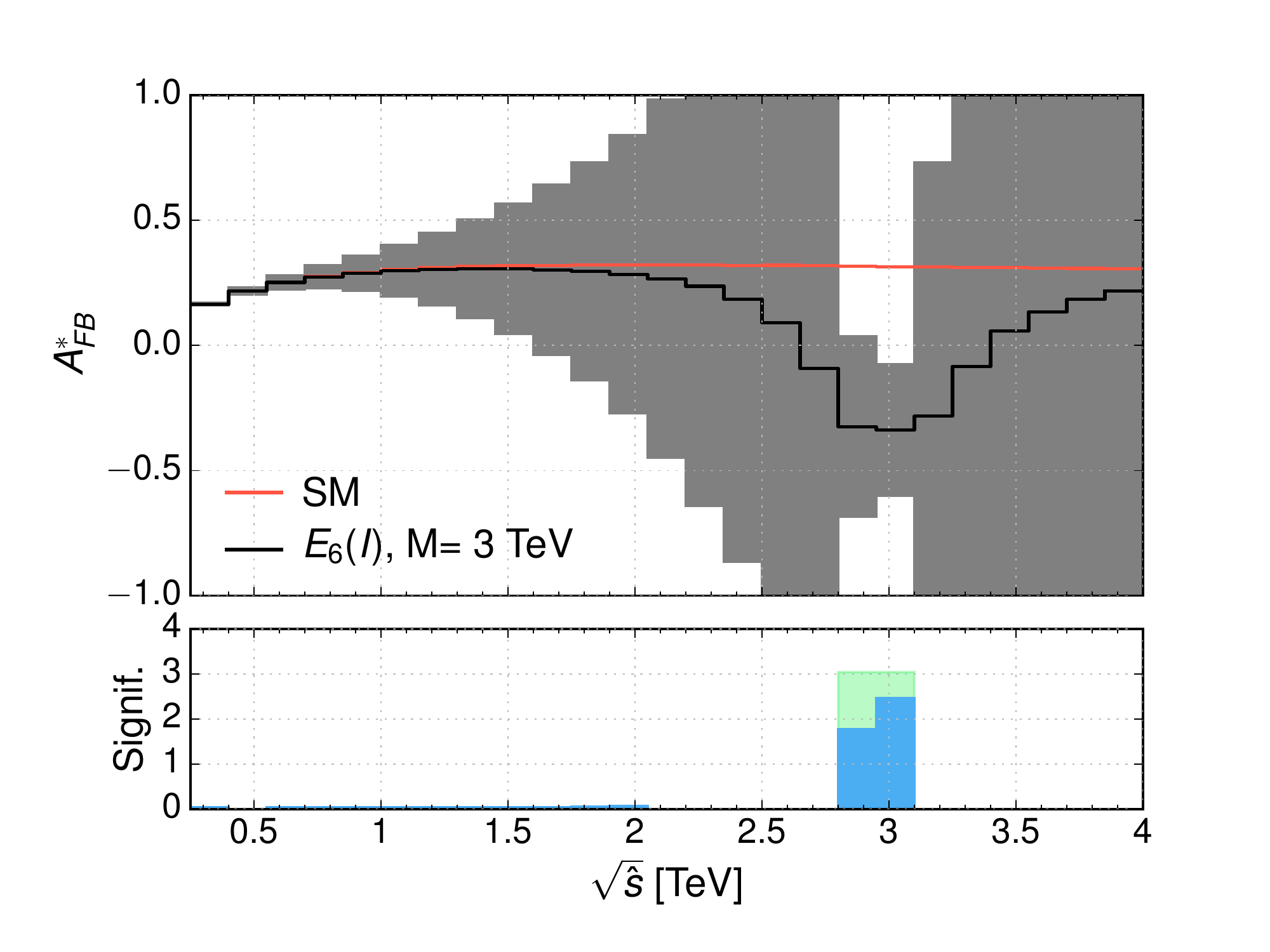}
\label{fig:AFB_I_realistic30}
}
\caption{(Colour online)
\subref{fig:sigma_chi_realistic30} Binned differential cross section as a function of the di-lepton invariant mass as predicted by the $E_\chi$ model for a $Z^\prime$-boson with mass $M_ {Z^\prime}$ = 3 TeV. Error bars are included. The results are for the LHC at $\sqrt{s}$=13 TeV and $\mathcal{L}=30 fb^{-1}$. Acceptance cuts are imposed (see text). The lower plot shows the significance.
\subref{fig:AFB_chi_realistic30} Binned $A_{FB}^*$ as a function of the di-lepton invariant mass as predicted by the $E_\chi$ model for a $Z^\prime$-boson with mass $M_ {Z^\prime}$ = 3 TeV.
Error bars are included. The results are for the LHC at $\sqrt{s}$=13 TeV and $\mathcal{L}=30 fb^{-1}$. Acceptance cuts are imposed (see text). In the lower plot, the blue histogram shows the significance bin by bin while the green area indicate the total significance integrated over that invariant mass region.
\subref{fig:sigma_I_realistic30} Same as (a) for the $E_I$ model.
\subref{fig:AFB_I_realistic30} Same as (b) for the $E_I$ model.
}
\label{fig:E6models_realistic_30invfb}
\end{figure}

Similar trends are shown by all models belonging to the $E_6$ class of theories and do not change when a more realistic setup is considered. Implementing the acceptance cuts extracted by the CMS analysis at the 8 TeV LHC ($P_T(l)\ge 25$ GeV and $\eta (l) \le 2.5$ with $l=e, \mu$), the shape of the reconstructed AFB, $A_{FB}^*$, including error bars would in fact appear as in Fig. 11. The significances coming from the AFB and the  cross section are indeed equivalent in magnitude, if only the statistical error is included. We thus expect that the use of the $A_{FB}^*$ observable, when associated to the default resonance search, could improve the discovery potential of new narrow width $Z'$-bosons. Further, being a ratio of differential cross sections, the reconstructed  $A_{FB}^*$ could help in minimizing the systematical errors thus rendering the measurement much more accurate. 

This is in particular the case when we should be in presence of an evidence for a new 
$Z'$-boson in the resonance search at the 3-4 sigma level. In these conditions, one couldn't claim the discovery of a new gauge boson just looking at the resonant peak in the di-lepton invariant mass distribution. However, if a signal of similar strength were to be discovered in an independent observable, the suggestion of the possible presence of new physics would turn into a robust claim. This is the role that the AFB would play. In Fig. 12, we plot the differential cross section and  $A_{FB}^*$ as a function of   in the di-lepton invariant mass, $M_{l\bar l}=\sqrt{\hat s}$, within the $E_\chi$ and $E_I$ models, at the forthcoming Run II of the LHC at 13 TeV with $\mathcal{L}=30 fb^{-1}$, that is the integrated luminosity which should be collected at the end of 2016. There, a new physics evidence at barely 4 sigma in the bump search could be reinforced by the simultaneous measurement of the reconstructed AFB, showing a signal at the 2-sigma level.
  
\subsection{Z' models with shifted AFB}

\begin{figure}[t]
\centering
\subfigure[]{
\includegraphics[width=7.7cm]{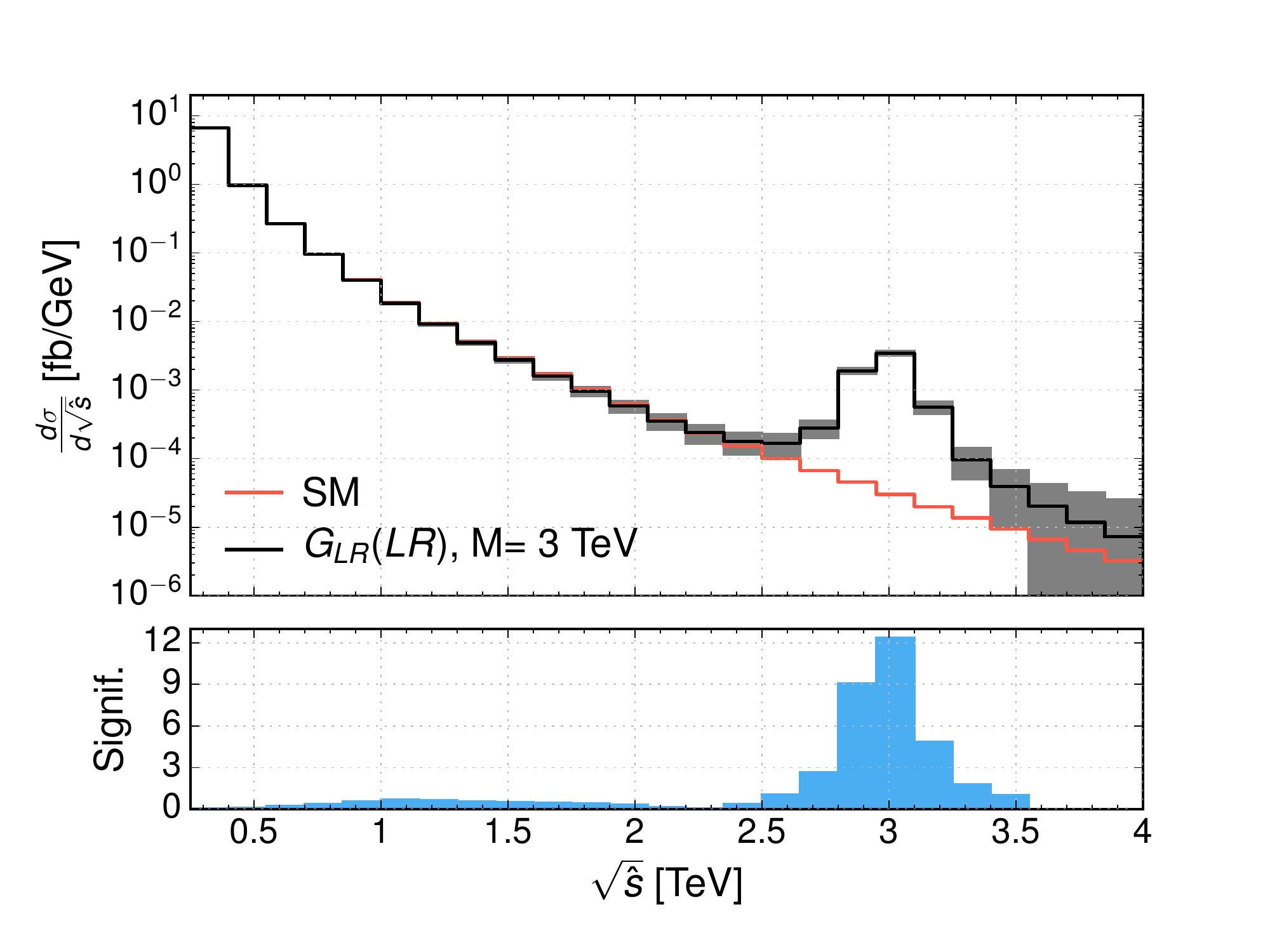}
\label{fig:sigma_glr_realistic}
}
\subfigure[]{
\includegraphics[width=7.7cm]{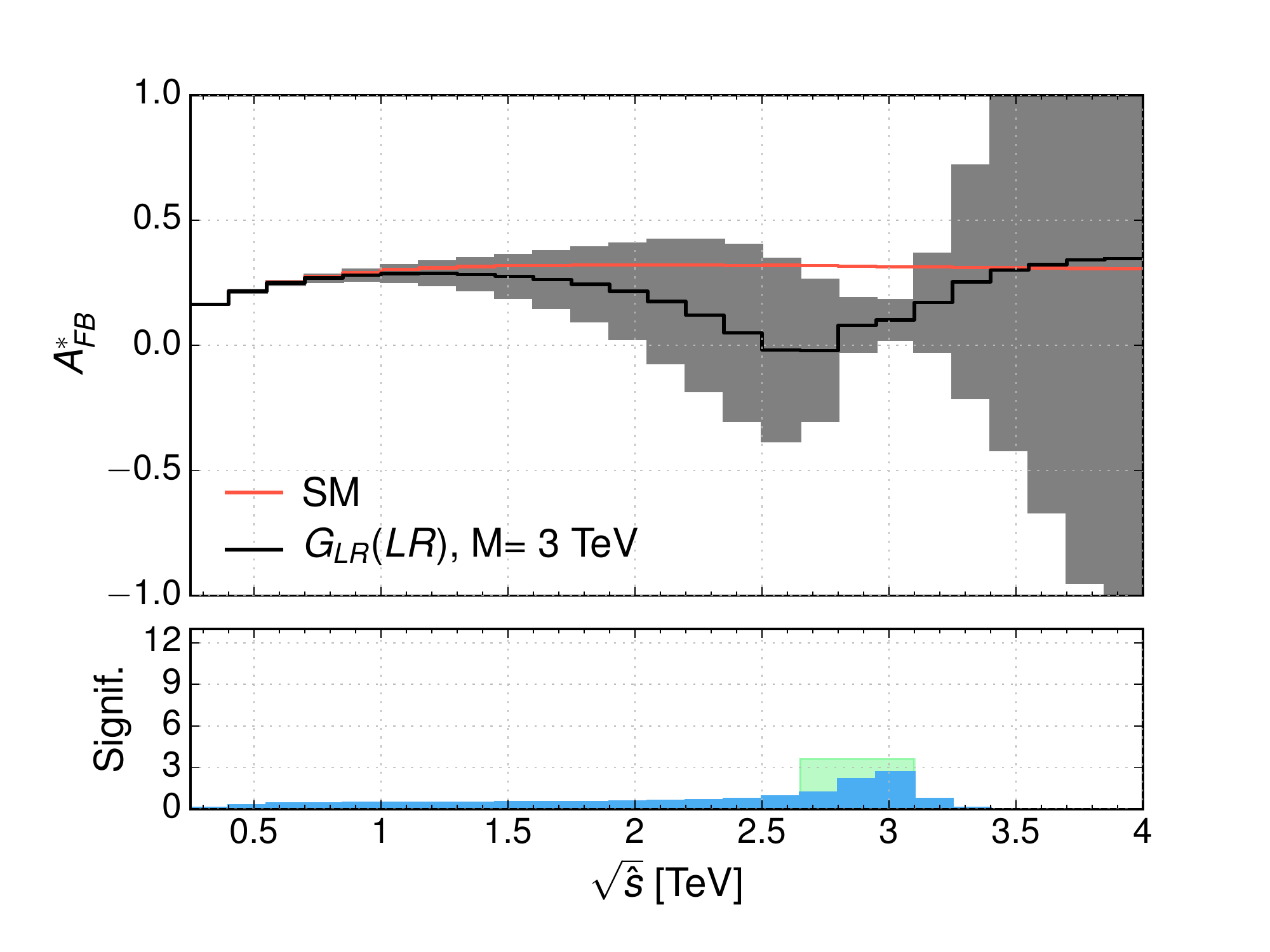}
\label{fig:AFB_glr_realistic}
}
\subfigure[]{
\includegraphics[width=7.7cm]{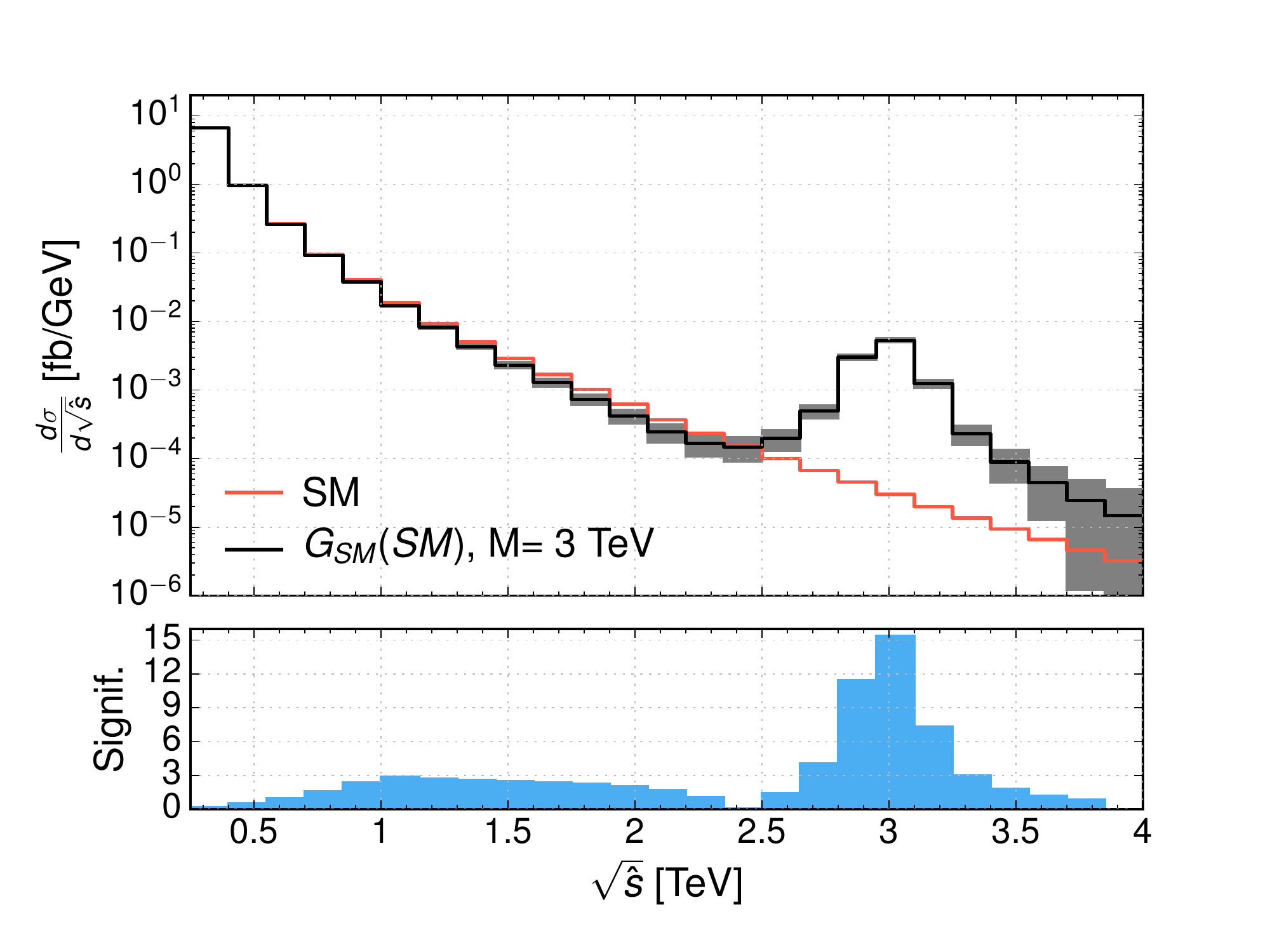}
\label{fig:sigma_ssm_realistic}
}
\subfigure[]{
\includegraphics[width=7.7cm]{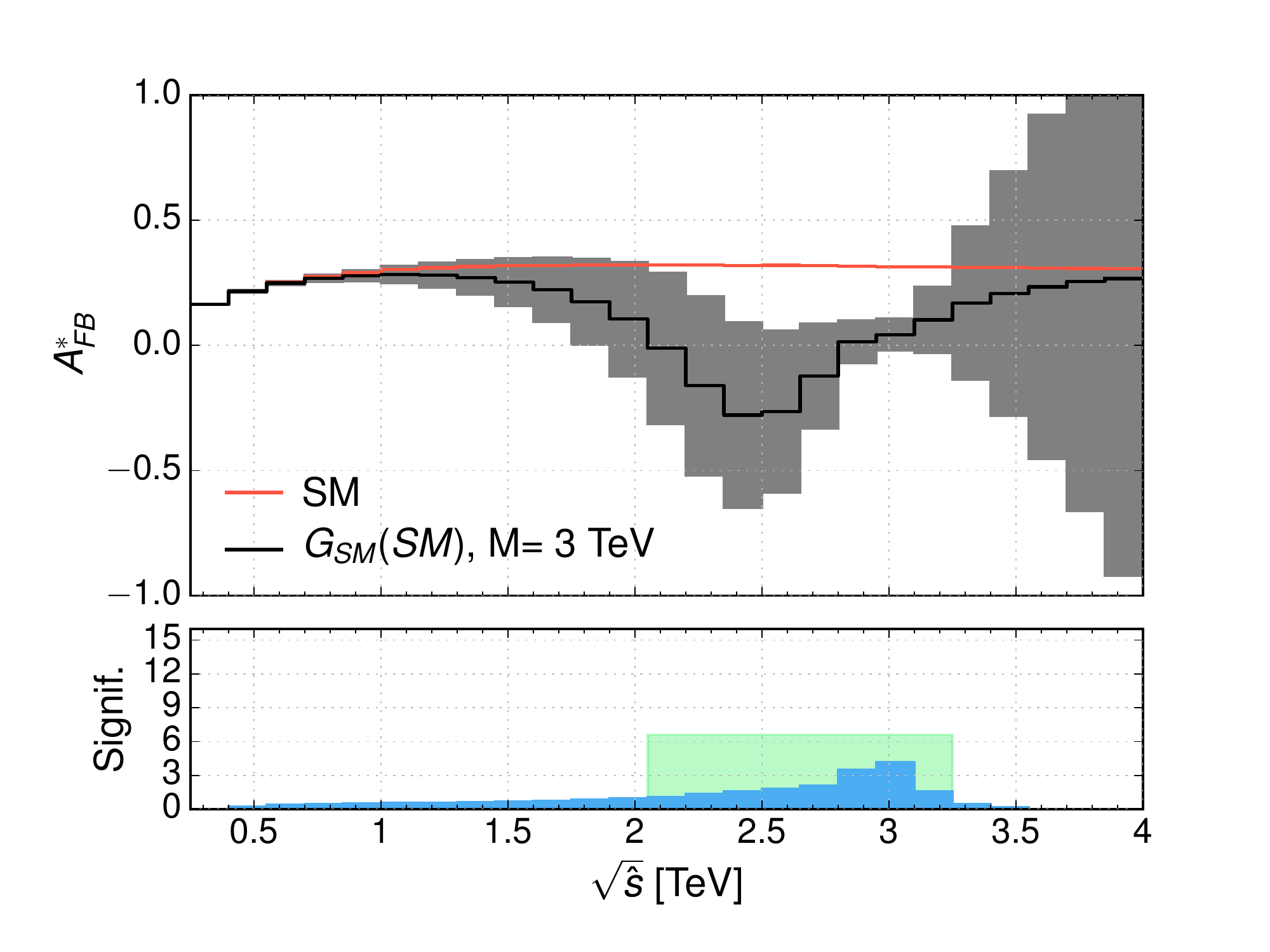}
\label{fig:AFB_ssm_realistic}
}
\caption{(Colour online)
\subref{fig:sigma_glr_realistic} Binned differential cross section as a function of the di-lepton invariant mass as predicted by the $GLR-LR$ model for a $Z^\prime$-boson with mass $M_ {Z^\prime}$ = 3 TeV. Error bars are included. The results are for the LHC at $\sqrt{s}$=13 TeV and $\mathcal{L}=300 fb^{-1}$. Acceptance cuts are included (see text).
\subref{fig:AFB_glr_realistic} Binned $A_{FB}^*$ as a function of the di-lepton invariant mass as predicted by the $GLR-LR$ model for a $Z^\prime$-boson with mass $M_ {Z^\prime}$ = 3 TeV.
Error bars are included. The results are for the LHC at $\sqrt{s}$=13 TeV and $\mathcal{L}=300 fb^{-1}$. Acceptance cuts are included (see text).
\subref{fig:sigma_ssm_realistic} Same as (a) for the Generalized Sequential SM, $GSM-SM$.
\subref{fig:AFB_ssm_realistic} Same as (b) for the $GSM-SM$ model.
}
\label{fig:gLRmodels_realistic}
\end{figure}

In this section, we discuss narrow width $Z'$ models where the AFB  has a shifted peak, that is, not centred on the $Z'$-boson mass. These models belong to the GLR class. The same behaviour is also displayed by the 
SSM scenario taken as benchmark model by the LHC experimental collaborations.

In principle, the reconstructed $A_{FB}^*$ could reveal the presence of a new spin-1 particle at energy scales lower than its mass, as the shape of this observable as a function of the di-lepton invariant mass is accentuated at mass scales smaller than $M_ {Z^\prime}$. This behaviour is shown in Fig. 13b where we plot the reconstructed AFB versus $M_{l\bar l}=\sqrt{\hat s}$ for the representative model GLR-LR. We consider a new $Z'$-boson with mass $M_ {Z^\prime}$ = 3 TeV. As one can see, the peak of $A_{FB}^*$ is shifted on the left-hand side of the physical $Z'$-boson mass and it appears at around 2.6 TeV. This feature is interesting. However, the significance is quite low as shown in Fig. 13b, owing to the poor statistics in that region. For $M_{l\bar l}$ values around the physical mass of the $Z'$-boson, which are statistically relevant, the significance coming from AFB is always much smaller than the significance obtained via the measurement of the differential cross section, displayed in Fig. 13a. For this kind of models, the $A_{FB}^*$ observable is therefore not particularly appropriate for $Z'$ searches. The same conclusion holds for the SSM (see Figs 13c and 13d). Hence, this benchmark model is not an advisable playground for studying the benefits of using the AFB in searching for new $Z'$-bosons. 

\section{The role of AFB in Z' searches: wide heavy resonances}
\label{sec:wideZ}
In this section, we discuss the role of the reconstructed AFB, $A_{FB}^*$, in searches for a new $Z'$-boson characterized by a large width. Such a heavy and wide particle is predicted by different models. A benchmark scenario for experimental analyses is the wide version of the SSM described in Ref. \cite{Altarelli:1989ff}. The proposal is to have a heavy copy of the SM neutral gauge boson, $Z$, with same couplings to ordinary matter and SM gauge bosons. Owing to the $Z'$-boson decay into SM charged gauge bosons, whose rate grows with the third power of the $Z'$-boson mass, the total width of the new heavy particle can be quite large: $\Gamma_{Z'}/M_{Z'}\simeq 50\%$ and above. 

In this case, the invariant mass distribution of the two final state leptons does not show in the cross section
a resonant (or peaking) structure around the physical mass of the $Z'$-boson standing sharply over a smooth background, but just a shoulder spread over the SM background. This result is plotted in Fig. 14a, where we consider a $Z'$-boson with mass $M_ {Z^\prime}$ = 1.5 TeV and width $\Gamma_{Z'}/M_{Z'}= 80 \%$. As the line shape of the resonance is not well defined and these parton level results could be worsened by detector smearing effects giving rise to an even broader spectrum, the $A_{FB}^*$ observable could help to interpret a possible excess of events. The results are shown in Fig. 14b. There, one can see that the $A_{FB}^*$ shape could be visible at the 2$\sigma$ level.

\begin{figure}[t]
\centering
\subfigure[]{
\includegraphics[width=7.7cm]{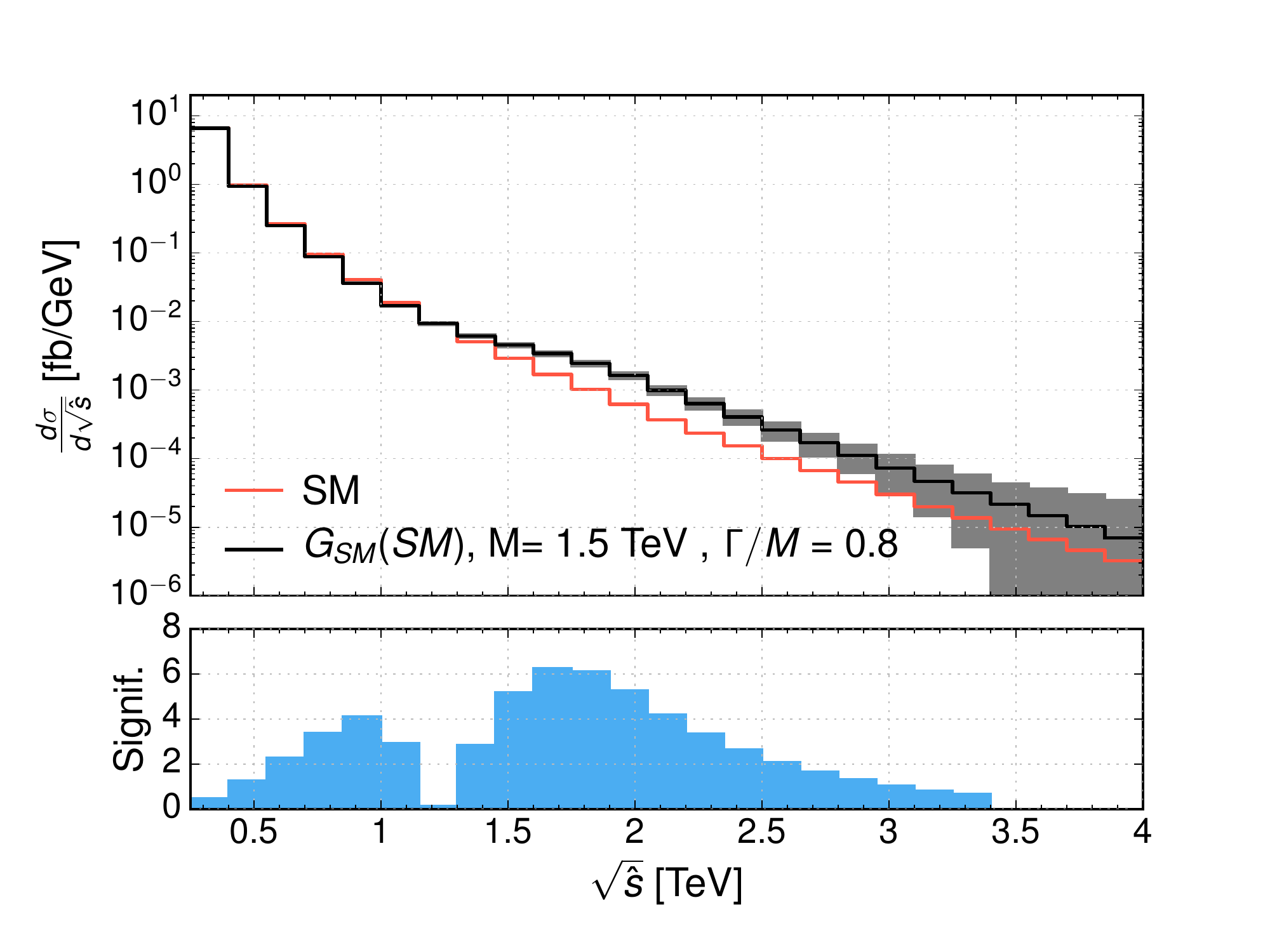}
\label{fig:sigma_gsmwide_realistic}
}
\subfigure[]{
\includegraphics[width=7.7cm]{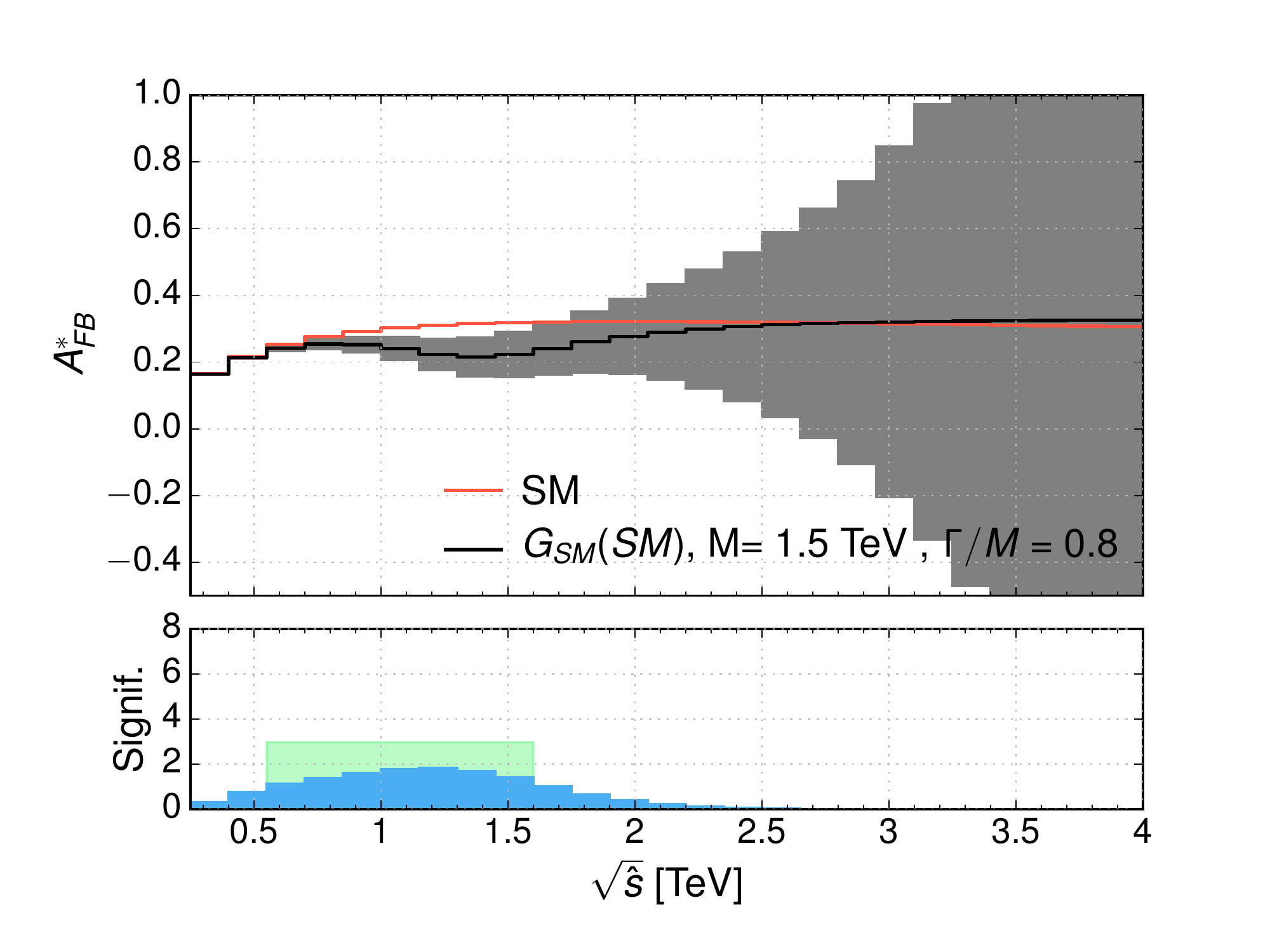}
\label{fig:AFB_gsmwide_realistic}
}
\caption{(Colour online)
\subref{fig:sigma_gsmwide_realistic} Binned differential cross section as a function of the di-lepton invariant mass as predicted by the $GSM-SM$ model for a $Z^\prime$-boson with mass $M_ {Z^\prime}$ = 1.5 TeV and $\Gamma_{Z'}/M_{Z'}= 80 \%$. Error bars are included. The results are for the LHC at $\sqrt{s}$=13 TeV and $\mathcal{L}=300 fb^{-1}$. Acceptance cuts are included (see text).
\subref{fig:AFB_gsmwide_realistic} Binned $A_{FB}^*$ as a function of the di-lepton invariant mass as predicted by the $GSM-SM$ model for a $Z^\prime$-boson with mass $M_ {Z^\prime}$ = 1.5 TeV and $\Gamma_{Z'}/M_{Z'}= 80 \%$.
Error bars are included. The results are for the LHC at $\sqrt{s}$=13 TeV and $\mathcal{L}=300 fb^{-1}$. Acceptance cuts are imposed (see text).
}
\label{fig:gsmmodels_realistic}
\end{figure}

A framework, theoretically more grounded than the wide SSM, which predicts a heavy and broad $Z'$-boson is the so-called non-universal $SU(2)$ model \cite{Kim:2014afa, Malkawi:1999sa}. In this theory, the third generation of fermions is subjected to a new $SU(2)$ dynamics different from the usual weak interaction advocated by the SM. On the contrary, the first two families of fermions only feel the SM weak interaction. As a consequence, a new spectrum of gauge bosons emerges in the model. These new vector bosons can be either narrow or wide. The only constraint comes from the EW Precision Tests (EWPTs) which bound the $Z'$-boson to have a mass $M_ {Z^\prime}\ge$  1.7 TeV. For the analysis of the direct and indirect limits on this model we refer to \cite{Malkawi:1999sa} and references therein.

Within this framework, we consider the wide $Z'$-boson case with  $\Gamma_{Z'}/M_{Z'}=50\%$. In order to fulfil both the limits quoted in Ref.\cite{Malkawi:1999sa} and the direct limits coming from direct searches at the 8 TeV LHC \cite{Khachatryan:2014fba}, we assume $M_ {Z^\prime}$ = 5.5 TeV. The latter analysis performed at the LHC has been optimized for searches of new physics with no resonant peaking structure. The outcome is that there are no events for di-lepton invariant masses above 1.8 TeV. We have taken this limit into account when evaluating the $Z'$-boson mass and width.

This model is a very good playground to test whether the AFB can be used as a primary variable in searches for wide objects. In this case, in fact, the new physics signal appears as an excess of events spread over the SM background. Almost no line shape is present in the di-lepton invariant mass distribution usually measured. Searches are performed relying on a pure counting strategy,  a procedure which does not allow much interpretation of the hypothetical signal. The exploitation of the reconstructed $A_{FB}^*$ could help in this respect.

In Fig. 15, we compare the $Z'$-boson spectrum (15a) and the reconstructed $A_{FB}^*$ (15b) as functions of the di-lepton invariant mass. As one can see in Fig. 15a, the cross section spectrum at parton level is already very broad. Its slope might be lost or mistaken in the SM background normalization. Even if, in the best case, a plateau would be visible over the SM background, its interpretation would be very difficult. Fig. 15b shows that the  $A_{FB}^*$ observable has a sharper line-shape which can reveal the  presence of a spin-1 particle beyond error bars. Such a shape is quite shifted at low energy scales though, compared to the $Z'$-boson mass. Hence, its extraction should enable one to help the discovery of a new vector boson with very high mass. In short, here, the AFB measurement could become particularly useful at the edge of the LHC discovery limits, when new particles can be too heavy and broad to be easily detected via a standard resonant peak search.

\begin{figure}[t]
\centering
\subfigure[]{
\includegraphics[width=7.7cm]{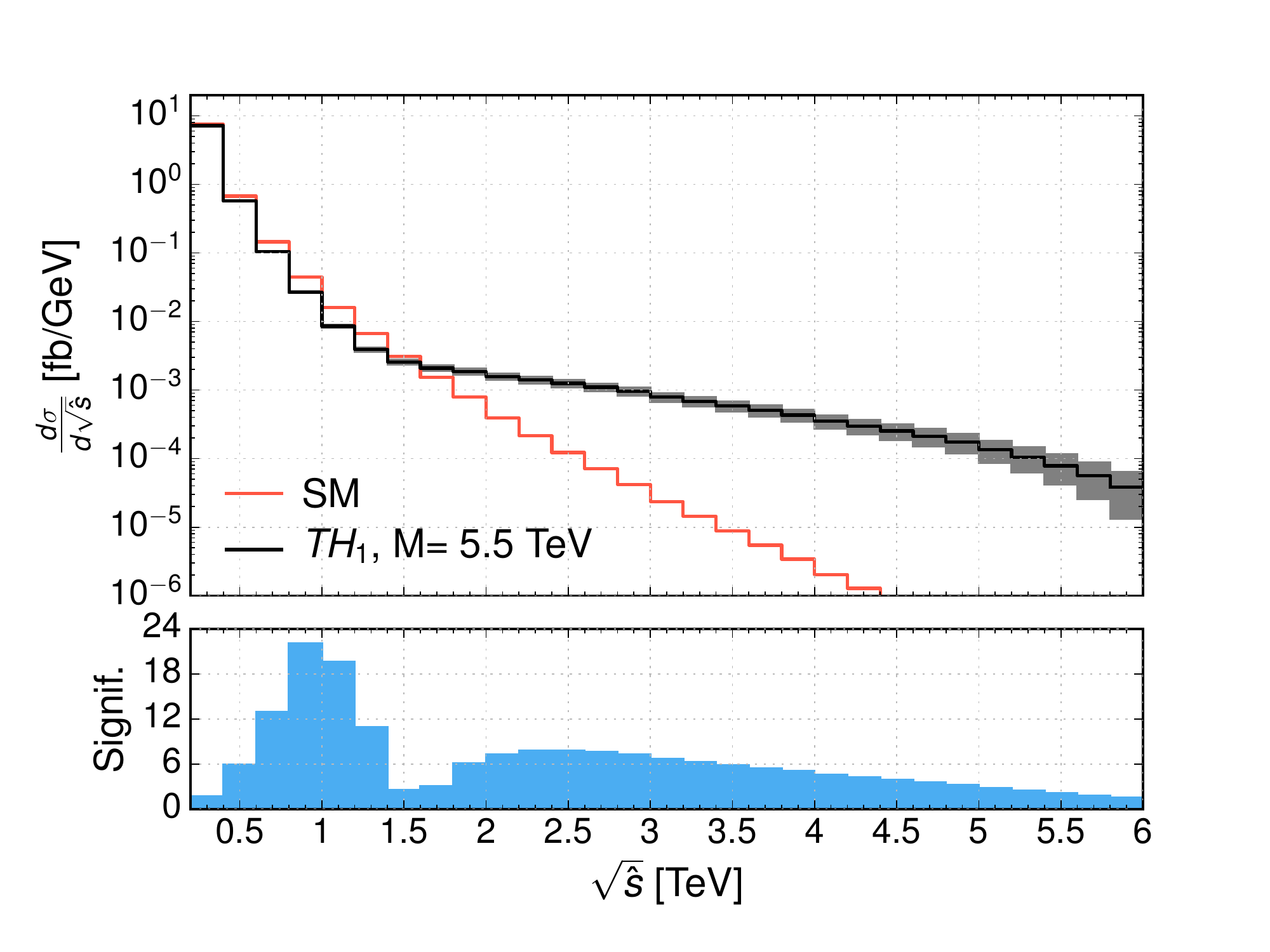}
\label{fig:sigma_thwide_realistic}
}
\subfigure[]{
\includegraphics[width=7.7cm]{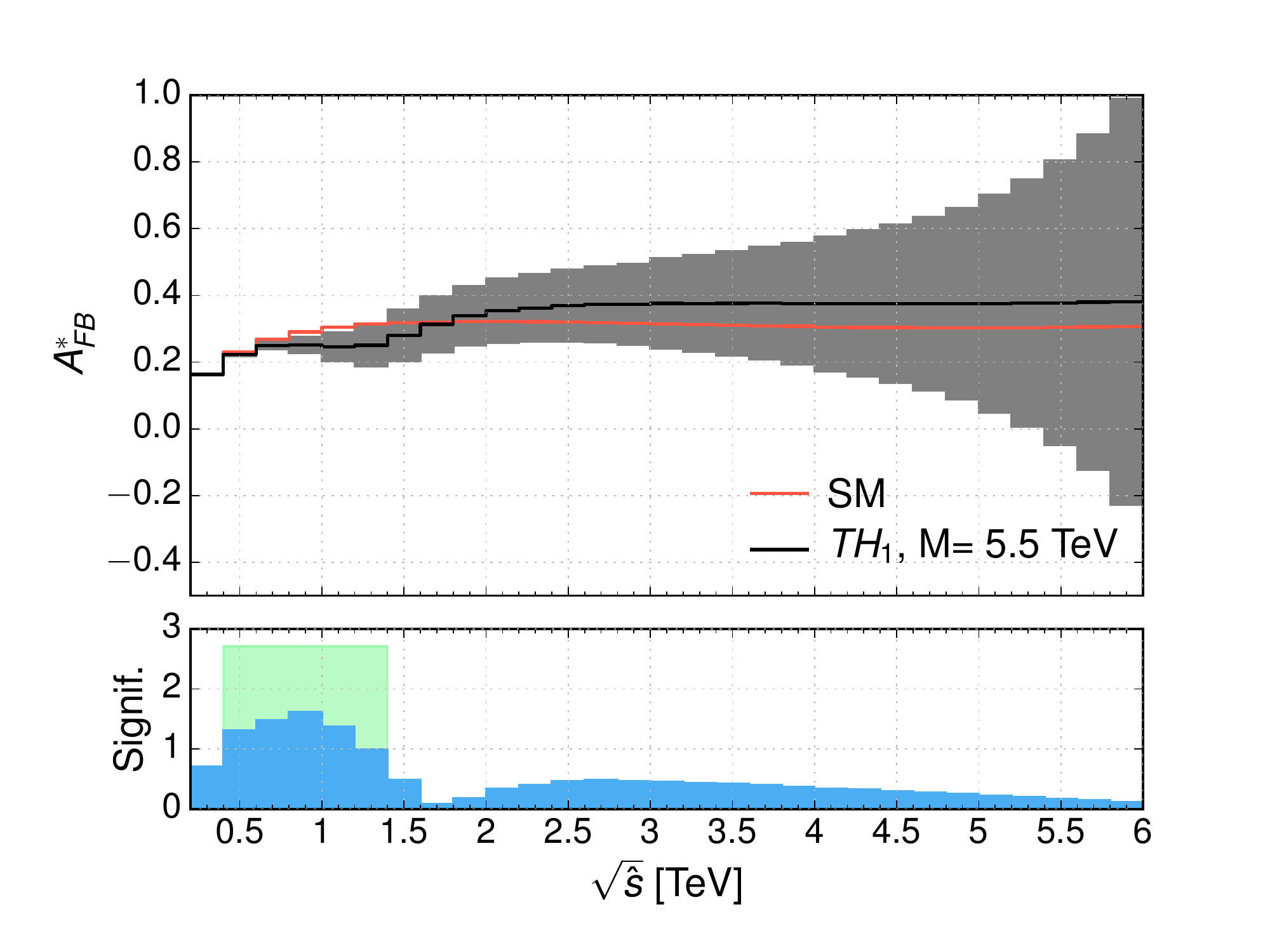}
\label{fig:AFB_thwide_realistic}
}
\caption{(Colour online)
\subref{fig:sigma_thwide_realistic} Binned differential cross section as a function of the di-lepton invariant mass as predicted by the $TH$ model for a $Z^\prime$-boson with mass $M_ {Z^\prime}$ = 5.5 TeV and $\Gamma_{Z'}/M_{Z'}\simeq 50 \%$. Error bars are included. The results are for the LHC at $\sqrt{s}$=13 TeV and $\mathcal{L}=300 fb^{-1}$. Acceptance cuts are imposed (see text). 
\subref{fig:AFB_thwide_realistic} Binned $A_{FB}^*$ as a function of the di-lepton invariant mass as predicted by the $TH$ model for a $Z^\prime$-boson with mass $M_ {Z^\prime}$ = 5.5 TeV and $\Gamma_{Z'}/M_{Z'}\simeq 50 \%$.
Error bars are included. The results are for the LHC at $\sqrt{s}$=13 TeV and $\mathcal{L}=300 fb^{-1}$. Acceptance cuts are imposed (see text).
}
\label{fig:gsmwide_models_realistic}
\end{figure}

The aforementioned scenarios are in fact particularly challenging for experimentalists. The non-resonant analyses of wide objects have been performed by searching for a smooth deviation from the SM background. The number of events above a given lower cut on the di-lepton invariant mass is compared with the total number of expected background events. An optimal minimum mass threshold is chosen to maximize the sensitivity to new physics. Clearly, such an analysis depends quite strongly on the SM background estimate. Usually, the simulated background is normalized to the event number in a mass window of $\pm$ 30 GeV around the $Z$-boson mass. A control region is then selected at higher di-lepton invariant masses in order to perform a data driven modelling of the SM background and recast it in a functional form easy to implement  in the likelihood used for extracting the limit on the $Z'$-boson mass. The method is based on the assumption that the control region is new physics free. But, this is not the case for wide $Z'$-bosons. In these scenarios, the interference between the extra $Z'$-boson and the SM $\gamma , Z$ is so sizeable that it can invade the control region. Being absolutely model-dependent, it can be either constructive or destructive. In any case, it would change accordingly the shape of the di-lepton spectrum. If the interference is negative, it would led to a depletion of events at low mass scales on the left-hand side of the $Z'$-boson resonance. This is exactly the example shown in Figs. 14 and 15 corresponding to the SSM and non-universal $SU(2)$ scenarios, respectively. If not correctly interpreted, these interference effects could induce one to underestimate the SM background with the consequence of overestimating the extracted mass bounds.  
Having all these uncertainties to deal with, the support of a second observable like the AFB is strongly recommended for the non-resonant analyses.

\section{On the robustness of AFB against PDF uncertainties}

In this section, we discuss how robust is the shape of the forward-backward asymmetry against the theoretical uncertainties on the PDFs. We further compare the systematic error induced by the PDFs uncertainty on the differential cross section and on the reconstructed AFB. 

For the determination of the PDF uncertainty we follow  \cite{Alekhin:2011sk} and references therein. Here, we just highlight the key points of the procedure. We compute the Hessian PDF uncertainty for our two observables: di-lepton invariant mass distribution in cross section and reconstructed AFB.
For Hessian PDF sets, both a central set and error sets are given. The number of error sets is twice the number of eigenvectors. For the CTEQ6.6 PDF that we use, the number of error sets is equal to 40. Defining $X_i^\pm$ the value of the variable using the PDF error set corresponding to the "$\pm$" direction for the eigenvector $i$, the symmetric error on the variable $X$ is given by:
\begin{equation}
\Delta X = {1\over 2}\sqrt{\sum_{i=1}^N |X_i^+ - X_i^-|^2}.
\label{PDFuncertainty}
\end{equation}
With this definition, we are now ready to compute the PDF uncertainty of any function $f(X)$. For the differential cross section, we apply Eq. \ref{PDFuncertainty} directly. For the AFB, the computation is slightly more involved. We consider as independent variables the forward and backward (differential) cross sections: $\sigma_F$ and $\sigma_B$. The PDF degrees of freedom of these two observables are correlated, that is the quantity 

\begin{equation}
\cos\phi = {1\over {4\Delta\sigma_F\Delta\sigma_B}}\sum_{i=1}^N ({\sigma_F}_i^+ - {\sigma_F}_i^-) ({\sigma_B}_i^+ - {\sigma_B}_i^-)
\label{PDFcorrelation}
\end{equation}

is equal to 1. Under this condition, by applying the error chain rule, we get
\begin{equation}
\Delta A_{FB}^* = {1\over 2}(1-{A_{FB}^*}^2)|{\Delta\sigma_F\over\sigma_F} - {\Delta\sigma_B\over\sigma_B}|.
\label{AFBuncertainty}
\end{equation}

The sign appearing in the above formula is crucial for the AFB. It indeed clearly shows that there is a partial cancellation of the PDF error on the reconstructed $A_{FB}^*$ due to the fact that this observable is a ratio of (differential) cross sections. Compared to the differential cross section, the AFB is then more robust against PDF uncertainties. This is shown in Fig. \ref{fig:Es_pdf300}  where we compare the effect of PDF and statistical errors on the shape of the di-lepton invariant mass distribution of the cross section and  AFB for two reference models, $E_I$ and $E_\chi$.

\begin{figure}[t]
\centering
\subfigure[]{
\includegraphics[width=7.7cm]{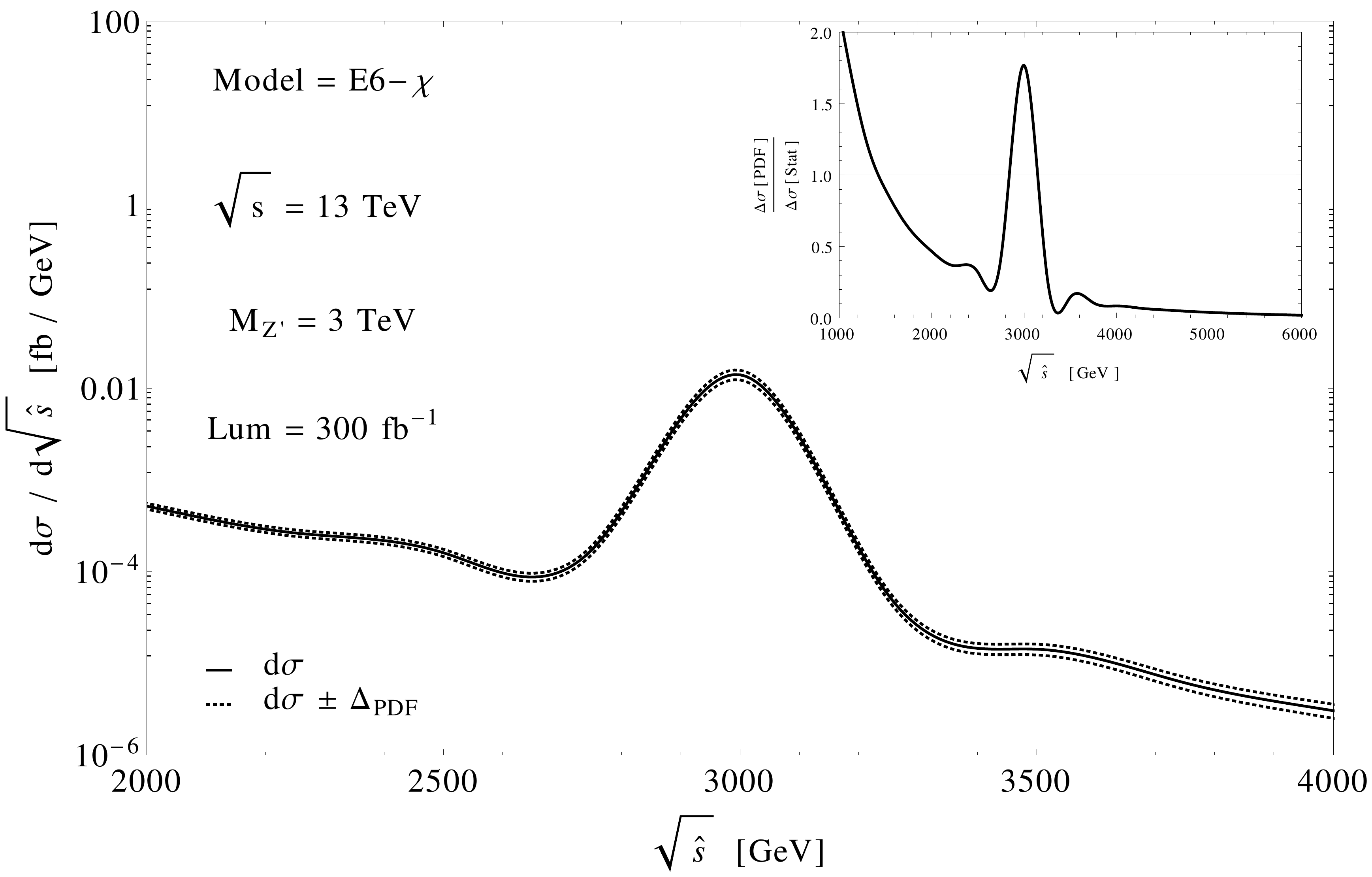}
\label{fig:sigma_chi}
}
\subfigure[]{
\includegraphics[width=7.7cm]{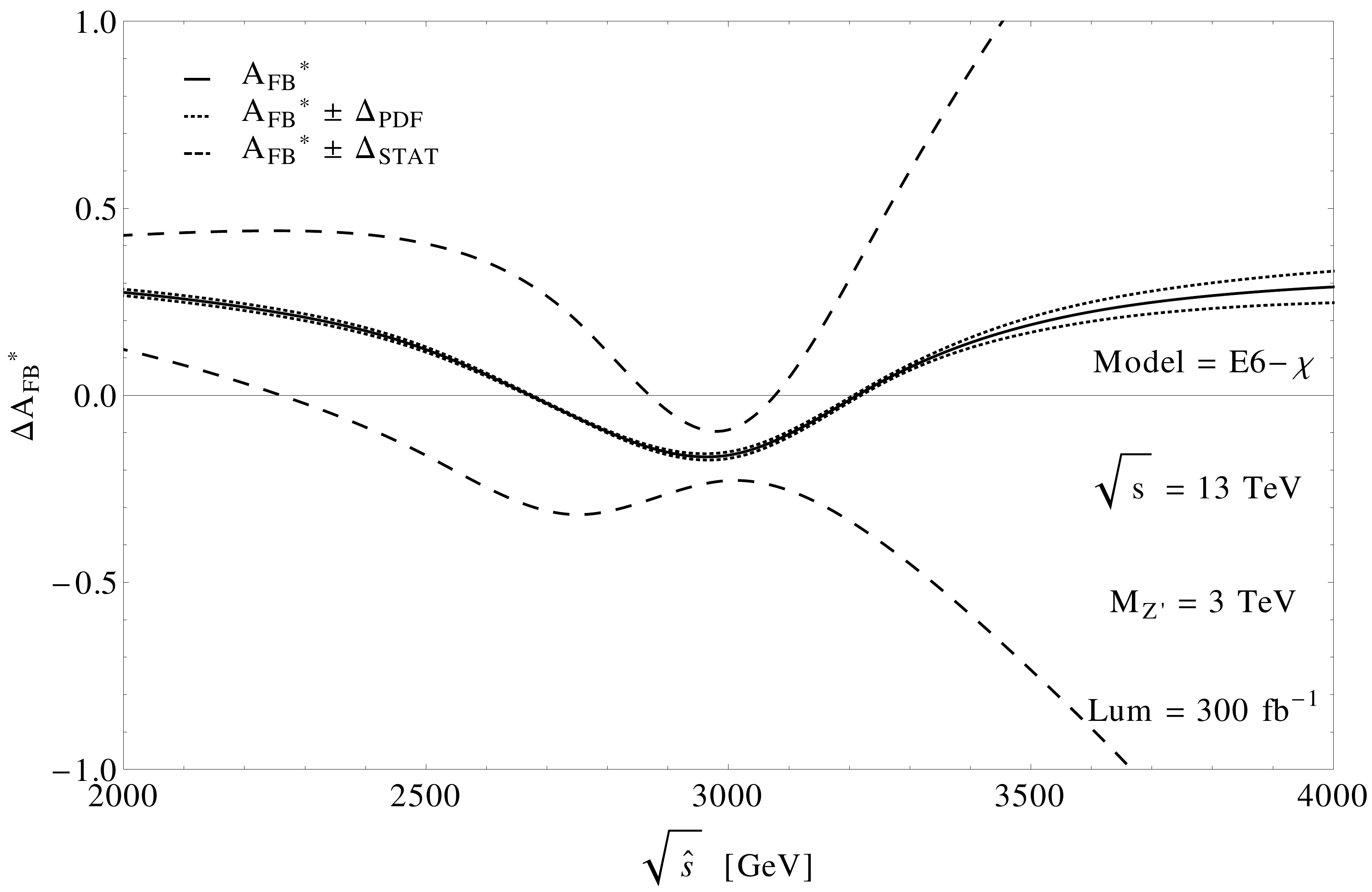}
\label{fig:AFB_chi}
}
\subfigure[]{
\includegraphics[width=7.7cm]{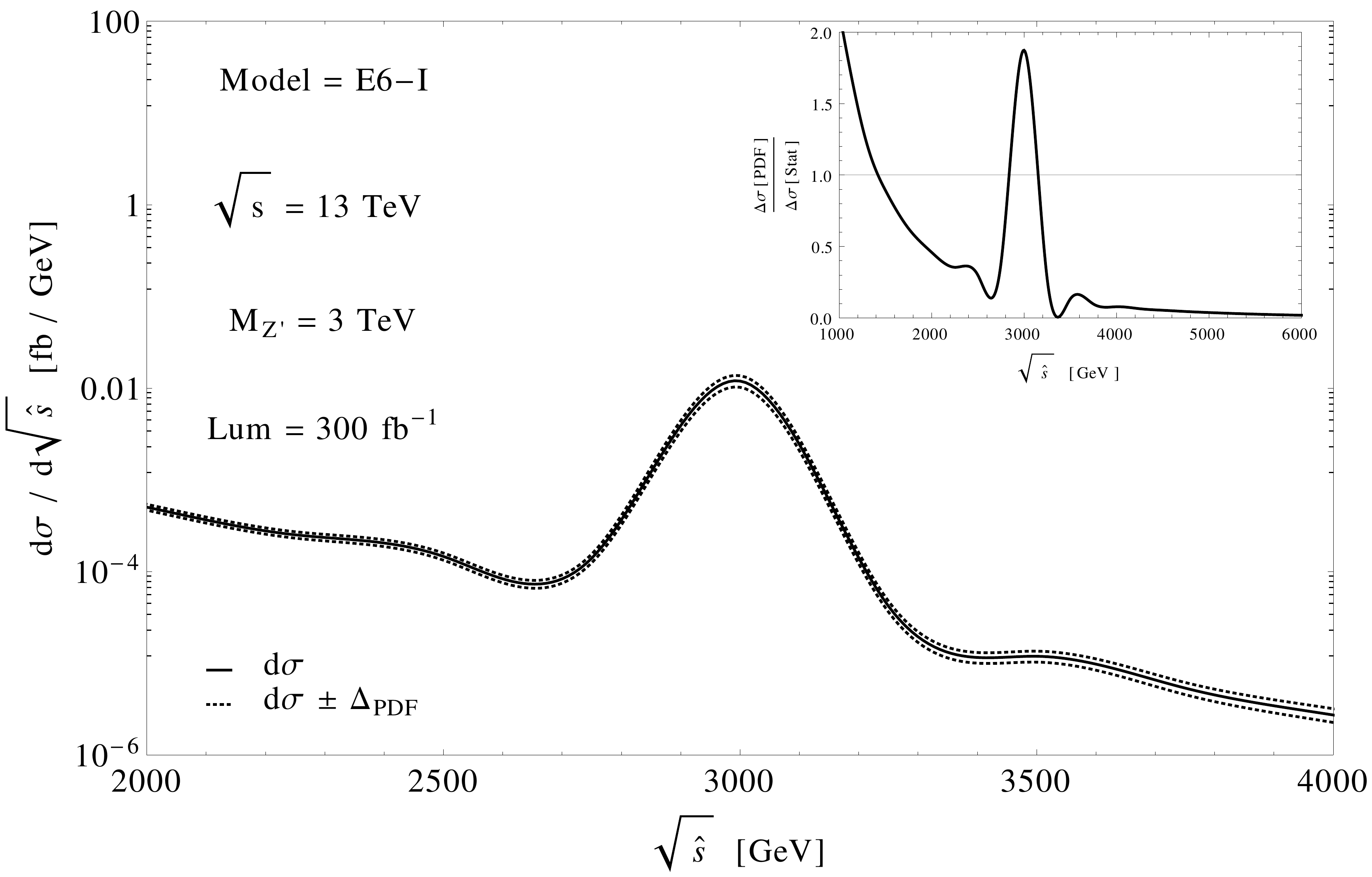}
\label{fig:sigma_I}
}
\subfigure[]{
\includegraphics[width=7.7cm]{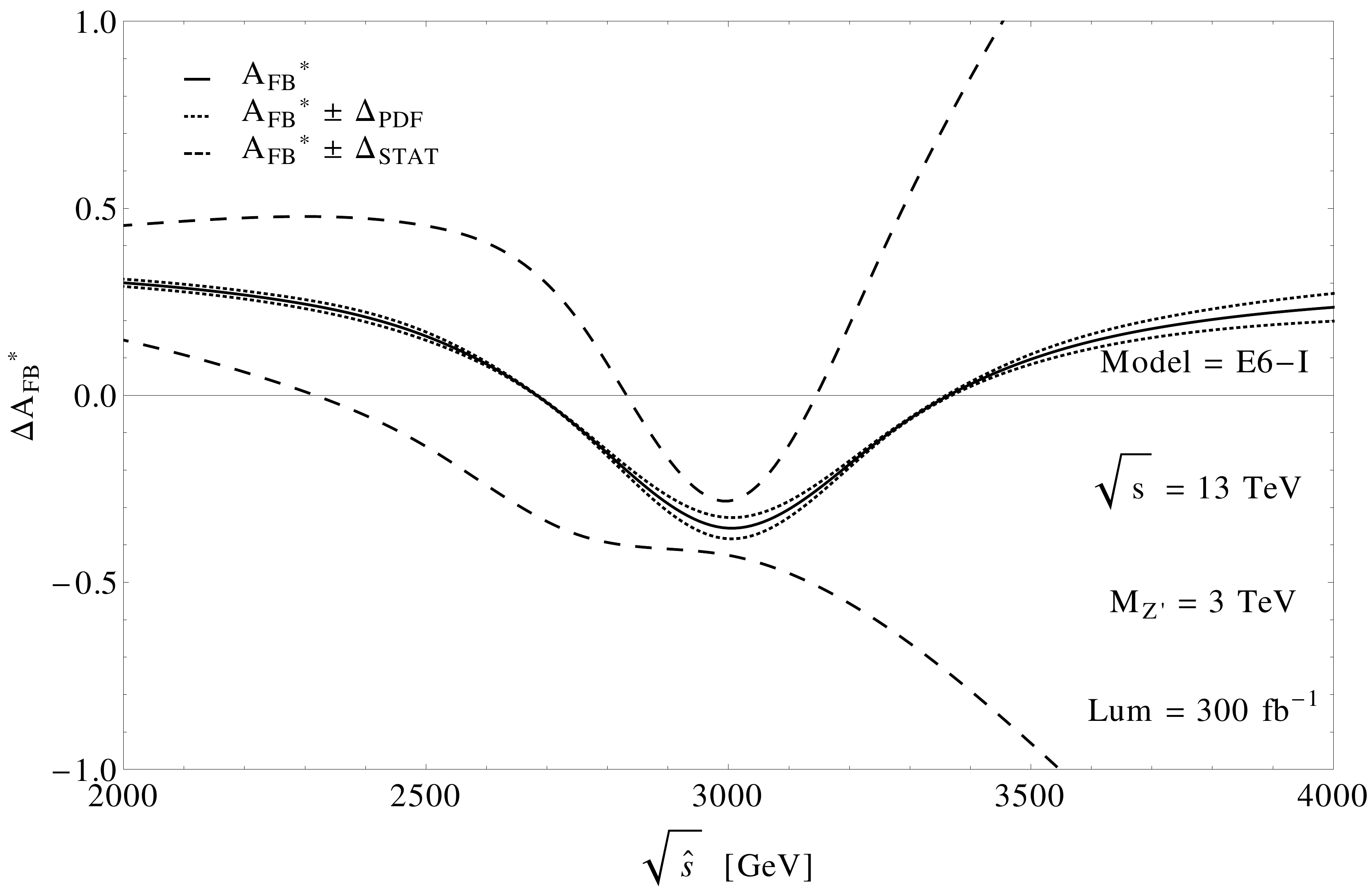}
\label{fig:AFB_I}
}
\caption{
\subref{fig:sigma_chi} Differential cross section as a function of the di-lepton invariant mass as predicted by the $E_\chi$ model for a $Z^\prime$-boson with mass $M_ {Z^\prime}$ = 3 TeV. The results are for the LHC at $\sqrt{s}$=13 TeV and $\mathcal{L}=300 fb^{-1}$. The solid line shows the central value, the dotted line the PDF uncertainty. The inset plot displays the ratio between PDF and statistical errors.
\subref{fig:AFB_chi} $A_{FB}^*$ as a function of the di-lepton invariant mass as predicted by the $E_\chi$ model for a $Z^\prime$-boson with mass $M_ {Z^\prime}$ = 3 TeV. The results are for the LHC at $\sqrt{s}$=13 TeV and $\mathcal{L}=300 fb^{-1}$. The dotted lines show the PDF error band, while the dashed lines define the statistical error band.
\subref{fig:sigma_I} Same as (a) for the $E_I$ model.
\subref{fig:AFB_I} Same as (b) for the $E_I$ model.
}
\label{fig:Es_pdf300}
\end{figure}
As one can see, the behaviours of cross section and AFB are opposite. The differential cross section in the di-lepton invariant mass is dominated by the PDF error on and off-peak. In the region around the peak, the PDF uncertainty is a factor 2 bigger than the statistical error. On the contrary, the AFB is dominated by the statistical error on and off-peak. Moreover, the PDF uncertainty is quite reduced owing to the minus sign in Eq.\ref{AFBuncertainty}. The shape of the AFB is thus not affected by the PDF error, so this observable is theoretically well defined.

In the light of these results, we can revisit Figs 11 and 12. We immediately observe that, if the error from the PDF is a factor 2 bigger than the statistical one for the invariant mass distribution around the $Z'$-boson mass, then the total error becomes 3 times the statistical one (adding linearly the systematic and statistical errors). As a consequence, the significances in Figs 11a and 11c decrease by the same factor. This means that the significances from the corresponding 
AFB distribution are much bigger than those ones from the differential cross section. If no refitting procedure is applied for the PDF in order to minimize their uncertainty, then the AFB observable 
appears to be much more robust than the resonance peak.

Of course, one needs to consider the energy scale dependence of the PDF errors, if no refitting procedure is employed. In Fig. 17a, we plot PDF and statistical errors on the total cross section integrated around the mass of the $Z'$-boson. We integrate in the window $\pm 5\%\ \ E_{\rm{LHC}}$ around the hypothetical $M_{Z'}$ where interference and finite width effects can be neglected. In evaluating the statistical error, we assume the design value for the luminosity: $L=300 fb^{-1}$.
We then vary the value of $M_{Z'}$ to see how statistical and PDF errors change in magnitude.
We consider four theoretical frameworks: $E_\chi$, $E_I$,  GLR-LR and  SSM. The figure shows that, up to roughly a 4 TeV scale, the cross section is dominated by the PDF uncertainty. In contrast, the asymmetry integrated in the same peak region is heavily dominated by the statistics for all possible $Z'$ masses, as shown in Fig. 17b.

\begin{figure}[t]
\centering
\subfigure[]{
\includegraphics[width=7.7cm]{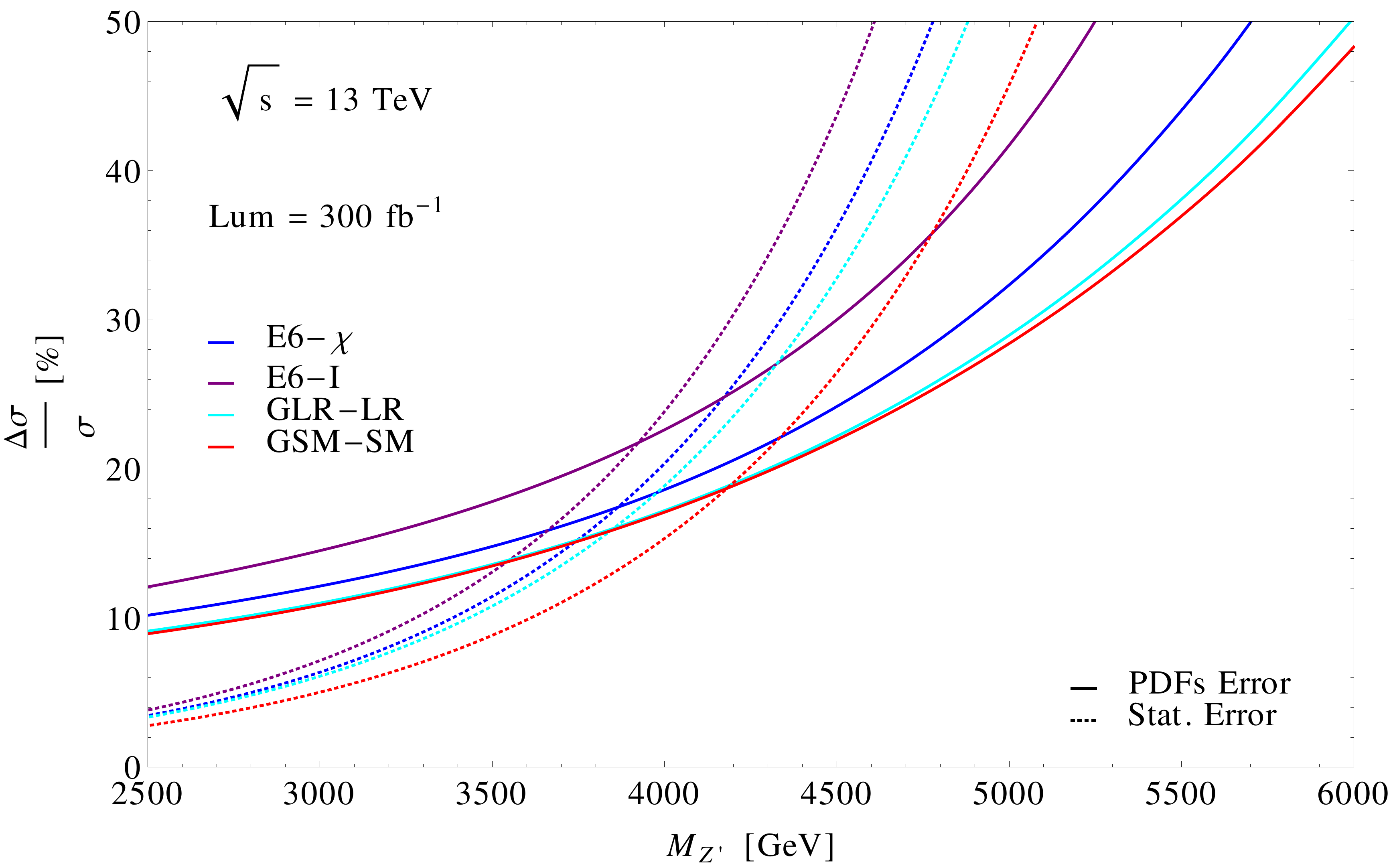}
\label{fig:sigma_pdf_m}
}
\subfigure[]{
\includegraphics[width=7.7cm]{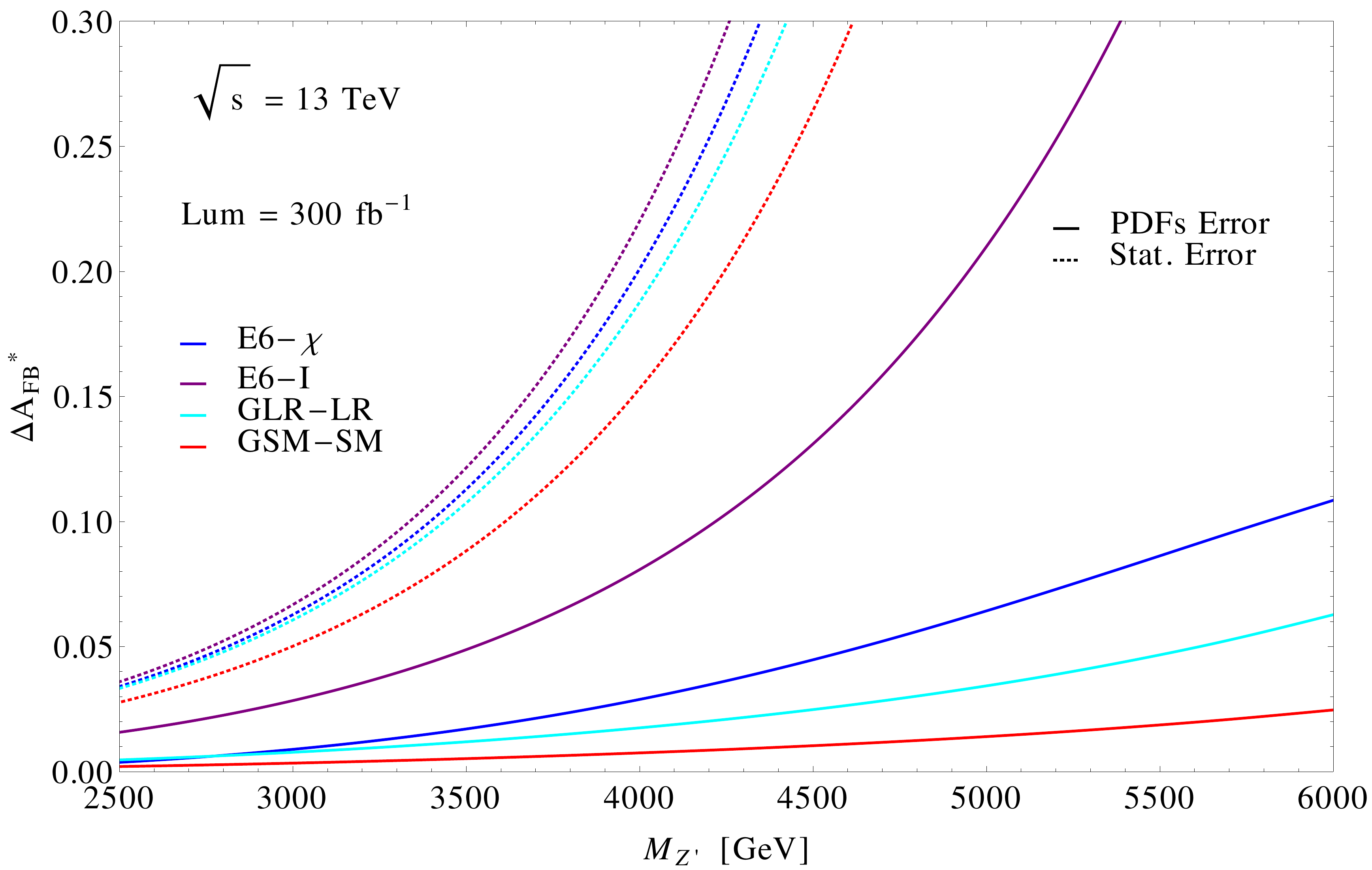}
\label{fig:AFB_pdf_m}
}
\caption{(Colour online)
\subref{fig:sigma_pdf_m} Cross section integrated around the $Z'$-boson mass ($|M_{l\bar l}-M_{Z'}|\le 0.05\times E_{\rm{LHC}}$) as a function of $M_{Z'}$  as predicted by the $E_\chi$, $E_I$, $GLR-LR$ and GSM-SM models. The results are for the LHC at $\sqrt{s}$=13 TeV and $\mathcal{L}=300 fb^{-1}$. Solid lines represent the PDF uncertainty, dashed ones the statistical error. 
\subref{fig:AFB_pdf_m} $A_{FB}^*$ integrated around the $Z'$-boson mass ($|M_{l\bar l}-M_{Z'}|\le 0.05\times E_{\rm{LHC}}$) as a function of $M_{Z'}$  as predicted by the $E_\chi$, $E_I$, $GLR-LR$ and GSM-SM models. The results are for the LHC at $\sqrt{s}$=13 TeV and $\mathcal{L}=300 fb^{-1}$. Solid lines represent the PDF uncertainty, dashed ones the statistical error. 
}
\label{fig:pdferr_energy}
\end{figure}

The strong dependence of the PDF errors on the mass or energy scale also suggests that using as observable the ratio between the $Z'$-boson cross section and the on-peak SM Z-boson cross section $R_\sigma$, might not be entirely PDF safe. The two cross section are indeed a few TeV a part. Given the strong variation of the PDF error with energy, a cancellation analogous to that one  in Eq. \ref{AFBuncertainty} could not happen easily. This is another argument in favour of exploring the AFB as a search variable.


\section{Conclusions}
\label{sec:conclusions}
In this paper we have considered the scope of using AFB, the forward-backward asymmetry, in 
$Z'$-boson searches at the LHC in the di-lepton channel, i.e., via Drell-Yan production and decay.
Such a variable has traditionally been used for diagnostic purposes in presence of a potential signal previously established
through a standard resonance search via the cross section. However, based on the observation that it is affected 
by systematics less than cross sections (being a ratio of the latter), we
have studied the possibility of using AFB for such a purpose for a variety of $Z'$ models, $E_6$, GLR, GSM,
embedding either a narrow or wide resonance. The focus was on determining whether such a resonance could be sufficiently wide and/or weakly coupled such that a normal resonance search may not fully identify it and, further, whether the 
AFB could then provide a signal of comparable or higher significance to complement or even
surpass the scope of more traditional analyses. 

We have found promising results. 
In the case of narrow width $Z^\prime$-bosons, we have proven that 
the significance of the AFB search can be comparable with the usual bump search.  Further, we have
emphasised the fact that the AFB distribution mapped in di-lepton invariant mass can present features amenable
to experimental investigation not only in the peak region but also significantly away from the latter. 
In the case of
wide $Z^\prime$-boson, the AFB search could have a better sensitivity than the cross section studies 
thanks to a more peculiar line-shape and lower systematic and PDF uncertainties.
 In essence, here, AFB in specific regions of the invariant mass of the reconstructed $Z^\prime$-boson could be sensitive to broad resonances much more than the cross section, wherein the broad
distribution of the signal seemingly merges with the background.

We have explored  the above phenomenology  for all the benchmarks under study 
as well as assessed and used  the optimised strategy for  AFB reconstruction. 

\section*{Acknowledgements}
\noindent
We are grateful to Patrik Svantesson for stimulating discussions at the early stage of the project.
This work is supported  by the Science and Technology Facilities Council, grant number  ST/L000296/1.
All authors acknowledge partial financial support through the NExT Institute.


\bibliography{references}


\end{document}